\documentclass[aps,prx,reprint,twocolumn,superscriptaddress,floatfix,nofootinbib]{revtex4-1}
\usepackage{amsmath,amssymb,color,comment,physics}
\usepackage[makeroom]{cancel}
\usepackage[caption=false]{subfig}
\usepackage{mathrsfs}
\usepackage{graphicx}
\usepackage{subfig}
\usepackage[countmax]{subfloat}
\usepackage[english]{babel}
\usepackage{dsfont}
\usepackage[bookmarks=true,colorlinks,linkcolor=OrangeRed,urlcolor=NavyBlue,citecolor=RoyalBlue]{hyperref}
\usepackage[dvipsnames]{xcolor}
\usepackage{braket}
\usepackage{graphicx,color,xcolor}

\definecolor{mygreen}{rgb}{0.25,0.5,0.25}
\definecolor{mygold}{rgb}{0.93,0.59,0.13}
\definecolor{mypurple}{rgb}{0.49,0.18,0.56}

\usepackage[normalem]{ulem} 

\newcommand{\tth}{t_\mathrm{Th}}

\newcommand{\thei}{t_\mathrm{H}}
\newcommand{\tb}{t_\mathrm{b}}

\newcommand{\NN}{D} 

\usepackage{ulem}

\graphicspath{Figures/}

\begin{document}

\preprint{APS/123-QED}

\title{Many-body quantum chaos in stroboscopically-driven cold atoms}

\author{Ceren B.~Da\u{g}}
\email{ceren.dag@cfa.harvard.edu}
\affiliation{ITAMP, Center for Astrophysics $|$ Harvard $\&$ Smithsonian Cambridge, Massachusetts 02138, USA}
\affiliation{Department of Physics, Harvard University, Cambridge, Massachusetts 02138, USA}
\author{S.~I.~Mistakidis}
\affiliation{ITAMP, Center for Astrophysics $|$ Harvard $\&$ Smithsonian Cambridge, Massachusetts 02138, USA}
\affiliation{Department of Physics, Harvard University, Cambridge, Massachusetts 02138, USA}
\author{Amos~Chan}
\affiliation{Department of Physics, Lancaster University, Lancaster LA1 4YB, United Kingdom}
\affiliation{Princeton Center for Theoretical Science, Princeton University, Princeton NJ 08544, USA}
\author{H.~R.~Sadeghpour}
\affiliation{ITAMP, Center for Astrophysics $|$ Harvard $\&$ Smithsonian Cambridge, Massachusetts 02138, USA}

\begin{abstract}

In quantum chaotic systems, the spectral form factor (SFF), defined as the Fourier transform of the two-level spectral correlation function, is known to follow random matrix theory (RMT), namely a ``ramp'' followed by a ``plateau'' in  sufficiently late times. Recently, a generic early-time deviation from the RMT behavior, which we call the ``bump'', was shown to exist in random quantum circuits and spin chains as toy models for many-body quantum chaotic systems. Here we demonstrate the existence of the  ``bump-ramp-plateau'' behavior in the SFF for a number of paradigmatic and stroboscopically-driven 1D cold atom models: (i) Bose-Hubbard model, (ii) spin$-1/2$ Bose-Hubbard model, and (iii) nonintegrable spin-$1$ condensate with contact or dipolar interactions. We find that the scaling of the \textit{many-body Thouless time} $\tth$ --- the onset of RMT ---, and the bump amplitude are more sensitive to variations in atom number than the lattice size regardless of the hyperfine structure, the symmetry classes, or the choice of driving protocol. Moreover, $\tth$ scaling and the increase of the bump amplitude in atom number are significantly slower in spinor gases than interacting bosons in 1D optical lattices, demonstrating the role of locality. We obtain universal scaling functions of SFF which suggest power-law behavior for the bump regime in quantum chaotic cold-atom systems, and propose an interference measurement protocol.
 
\end{abstract}

\maketitle

\section{\label{sec:Intro}Introduction}

Quantum chaos is historically diagnosed with level repulsion: 
A quantum system is considered chaotic if it exhibits spectral statistics given by the random matrix theory (RMT), in sufficiently small energy scales~\cite{bohigas1984characterization}. Such signatures
have been found in a plethora of disciplines including nuclear resonance spectra \cite{Brody,PhysRevLett.48.1086}, mesoscopic physics \cite{meso1997, meso2015,roushan2017spectroscopic}, quantum chaos~\cite{Rigol,cdc2, kos_sff_prx_2018}, black hole physics \cite{gargar2016, Cotler_2017, complexity2017}, and quantum chromodynamics \cite{qcd1995, qcd2018}.
With their unprecedented degree of controllability \cite{bloch2012quantum,browaeys2020many},
trapped cold atoms~\cite{PhysRevLett.75.3969,anderson1995observation,PhysRevLett.55.48,pethick,PhysRevLett.87.160405,Greiner_2002,2009Natur.462...74B}  are excellent platforms to study many-body phenomena, including signatures of quantum chaos in spectral statistics, and for large-scale simulations of quantum systems~ \cite{PhysRevLett.81.3108,weimer2010rydberg,bloch2012quantum,browaeys2020many,greiner2002quantum,kinoshita2006quantum,Rigol2008, Cheneau_2012,tarruell2012creating,islam2015measuring,schreiber2015observation,Kaufman_2016,bernien2017probing,mazurenko2017cold,de2019observation,rispoli2019quantum,semeghini2021probing}.

A particularly intriguing objective of quantum simulations so far has been to understand the underlying mechanisms which lead to thermalization, as in eigenstate thermalization hypothesis (ETH)~\cite{Deutsch, Srednicki, Rigol2008}, or its absence via many-body localization (MBL)~\cite{gornyi2005interacting, BAA, schreiber2015observation,rispoli2019quantum,MBLreview1}, the existence of quantum many-body scars~\cite{bernien2017probing, Turner2018, moudgalya2018}, and Hilbert space fragmentation~\cite{sala2020hsf, khemani2020hsf}. In this vein, as a probe for quantum chaos and ergodicity \cite{PhysRevE.70.016217,KimHuse,PhysRevA.89.053610,hosur2016chaos,MSS,PhysRevX.7.031047,PhysRevX.7.031011,Chen_2016,PhysRevB.96.020406,PhysRevX.9.041017,PhysRevA.99.052322,PhysRevLett.123.230606,xu2020accessing}, observables 
such as entanglement entropy and out-of-time-order correlators (OTOCs) have been intensively studied, and measured with quantum simulators \cite{islam2015measuring,neill2016ergodic,Kaufman_2016,G_rttner_2017,PhysRevX.7.031011,PhysRevLett.120.070501,Landsman_2019,PhysRevA.100.013623,braumuller2022probing}.However, the OTOCs could be susceptible to quantum criticality and order as found by Refs.~\cite{PhysRevB.96.054503,PhysRevLett.121.016801,PhysRevLett.123.140602,https://doi.org/10.1002/andp.201900270}, even in nonintegrable quantum many-body models at infinite temperature \cite{PhysRevB.101.104415}. Moreover, it was demonstrated by Refs.~\cite{PhysRevE.101.010202,PhysRevLett.124.140602} that exponentially fast scrambling detected by OTOCs could be induced at the fixed points in phase space, and is not necessarily the signature of quantum chaos. 
Alternatively, the metrics of spectral statistics can be defined in the energy domain, e.g.~spacing distribution and ratio~\cite{bohigas1984characterization,FRIEDRICH198937,1988PhR...163..205E,OganesyanHuse}, as well as number variance \cite{PhysRevA.47.3571,Rigol}. 
While the spacing distribution can be experimentally obtained for few particles~\cite{PhysRevLett.48.1086,roushan2017spectroscopic}, it is particularly challenging for many-body systems: One must access an exponentially large number of energy levels, and resolve exponentially small many-body energy gaps.
Consequently, a direct experimental probe of level repulsion and, more generally, of spectral correlations in a many-body spectrum remains elusive.

\begin{figure}[t!]
\subfloat{\includegraphics[width=0.49\textwidth]{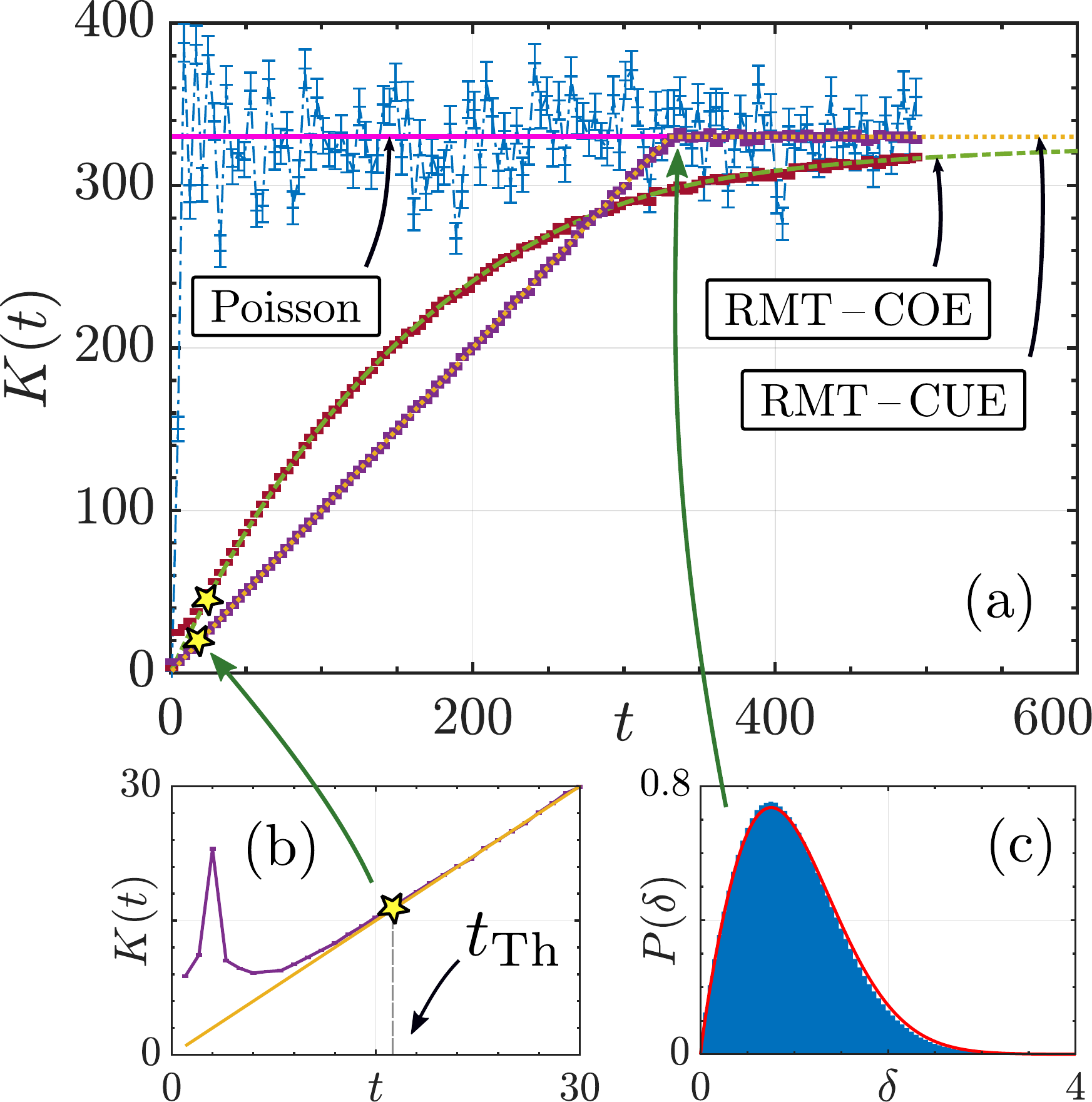}}\hfill
\caption{ (a) Representative 
behavior of the SFF $K(t)$ for the many-body driven quantum chaotic spin-$1/2$ Bose-Hubbard system, e.g., with $u_{\uparrow}=u_{\downarrow}=u_{\uparrow \downarrow}=J$ where $J_{\uparrow}=J_{\downarrow} \equiv J_{\uparrow \downarrow}=J=1$ is the characteristic energy scale. Interacting spin-$1/2$ bosons in a lattice with unit filling factor $\nu=1$, atom number $N=4$, and lattice size $L=4$, are subjected to a two- (red symbols) or three-step (purple symbols) driving protocols where the driving periods are $T=1$ and $T=3$, respectively. The two- and three-step protocols approach the respective RMT behaviors given by COE (green dashed) and CUE (orange dotted).
On the other hand, spin-$1/2$ atoms in a driven single well, $L=1$ (blue line), follows the Poisson statistics given by the flat line (pink), $K_{\mathrm{Poi}}(t)=D$, where $D$ is the Hilbert space dimension. This is a signature of integrability. The Thouless times $\tth$, denoted by stars, and the Heisenberg time $\thei$ determines the onset of the RMT behavior and the appearance of the plateaus respectively. (b) $K(t)$ following the three-step driving protocol in a spin-$1/2$ setting with $\nu = 1/4$, $N=3$, $L=12$. The bump ($t<\tth$) is a signature of \textit{many-body} quantum chaotic systems. (c) The nearest-neighbour spacing distribution $P(\delta)$ of a spin$-1/2$ many-body system with $N=9$ atoms trapped in $M=3$ wells and driven according to two-step protocol with a period of $T=1$ (blue) coincides with the one for the random matrix ensemble (red), and displays level repulsion, which is also captured by the SFF at $\thei$ in (a).
}
\label{Fig1b}
\end{figure}
To circumvent this issue, we consider the \textit{spectral form factor} (SFF), a time-domain observable defined as the Fourier transform of the two-level spectral correlation function~\cite{Mehta,Haake}. For a physical system, the appearance of the RMT-like ``ramp'' and subsequent ``plateau'' in SFF, Fig.~\ref{Fig1b}(a), is a signature of quantum chaos. In particular, the SFF behavior around the Heisenberg time $\thei$ that separates ramp and plateau, is a manifestation of the energy level repulsion, which is often probed with reference to the Wigner-Dyson level spacing distribution~\cite{doi:10.1063/1.1703773} (Fig.~\ref{Fig1b}(c)). SFF has been instrumental in the discovery of a number of novel and  universal signatures of \textit{many-body} quantum chaotic systems in both condensed matter~\cite{bertini2018exact, cdc1, cdc2, kos_sff_prx_2018, friedman2019,flack2020statistics,  PhysRevE.102.062144, PhysRevLett.124.186601,PhysRevLett.125.156601, Liao2020,PhysRevE.102.060202, cdclyap,  moudgalya2021, bertini2021random, garratt2020dw,SUNTAJS2021168469, garratt2020MBL, li2021spectral,prakash2021mblsff, liao2022emergence, chan2021manybody, Cornelius2022, PhysRevLett.129.060602,  Roy_2022,shivam2022ginibre, kurchan2022sff}, and in high energy physics~\cite{Cotler_2017, complexity2017, Gharibyan_2018, ALTLAND2018, Hunter_Jones_2018, winer2020exp, khramtsov2021spectral, Saad2019semiclassical,hepsff2022white}. 
Black holes have been shown to display RMT behavior in late times with SFF \cite{Cotler_2017, Saad2019semiclassical}, and the SFF has been utilized to investigate the ergodic-to-MBL transition in 1D spin chains~\cite{PhysRevE.102.062144, garratt2020MBL,PhysRevLett.124.186601,prakash2021mblsff}.

\begin{figure}[t!]
\subfloat{\label{Fig1a}\includegraphics[width=0.49\textwidth]{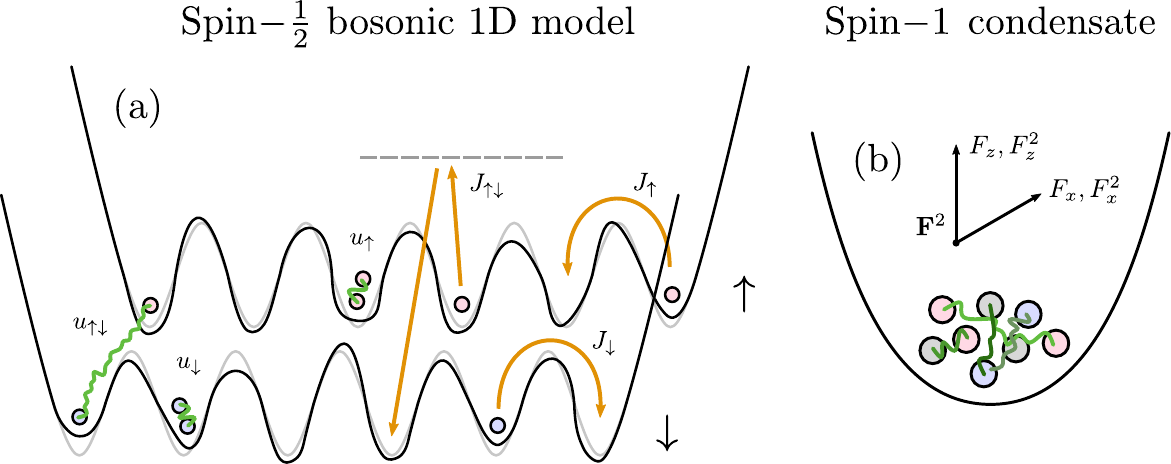}}\hfill
\caption{
(a-b) Schematic representation of periodically driven 1D (a) lattice trapped and interacting spin$-1/2$ bosons and (b) a spin$-1$ condensate in a harmonic potential. The tunnelling ($J_{\uparrow}$, $J_{\downarrow}$) and the spin-mixing $J_{\uparrow \downarrow}$ 
(e.g. facilitated by Raman-assisted tunneling with a certain Rabi frequency, see the dashed line) are depicted. $u_{\uparrow}$, $u_{\downarrow}$ ($u_{\uparrow \downarrow}$) refer to the on-site interactions, while the induced on-site disorder is illustrated by the  irregular wells.}
\end{figure}

For many-body spin models and local random quantum circuits, it has been shown that SFF could deviate from RMT. Specifically a ``bump'' appears at early times $t \ll \thei$ (see Fig.~\ref{Fig1b}(a) and (b)), and RMT is recovered only after the \textit{many-body Thouless time} $\tth$ \cite{cdc2, kos_sff_prx_2018,friedman2019, moudgalya2021, PhysRevE.102.062144, Gharibyan_2018,Roy_2022,PhysRevE.102.060202}, as illustrated by star markers in Fig.~\ref{Fig1b}.
The extension of such results to cold atom systems is nontrivial, since only an approximate mapping between them could be established for the Mott insulating regime, where spinful atoms are confined to lattice sites \cite{PhysRevLett.91.090402}. 
A central result of our work is the existence of the ``bump-ramp-plateau'' behavior in the SFF of stroboscopically-driven interacting (spinless and spin-$1/2$) bosons in optical lattices beyond the Mott insulator regime, as well as spin$-1$ condensates with either short-range contact or long-range dipolar interactions. We account for the trapped atoms' motional degree of freedom in addition to spin degree of freedom. To investigate the dependence of $\tth$ on the locality of the underlying Hamiltonian, we employ spatially-extended Bose-Hubbard (BH) chains \cite{PhysRevB.40.546} (Fig.~\ref{Fig1a}(a)), as opposed to a chaotic spin$-1$ condensate under single-mode approximation (SMA) \cite{PhysRevLett.81.5257} (Fig.~\ref{Fig1a}(b)). The latter refers to an ensemble of pairwise interacting spin$-1$ atoms, and hence effectively a zero-dimensional model \cite{sachdev1993gapless,kitaev2015simple,Swingle_2016,G_rttner_2017}. We find that the Thouless time increases significantly more slowly as a function of atom number in the case of chaotic spin$-1$ condensates as compared to the lattice confined interacting bosons. Furthermore, we reveal that both the bump amplitude and $\tth$ scaling are more sensitive to the atom number than the lattice size. This strong dependence on the atom number is not affected by the atomic hyperfine structure, e.g.,~single- or two-states; the symmetry class, and the choice of driving protocol. 

Our choice of quantum many-body models is also experimentally motivated. The BH model describes the physics of cold atoms trapped in optical lattices \cite{PhysRevLett.81.3108}, as experimentally demonstrated in 
\cite{greiner2002quantum}, while different hyperfine states of bosonic atoms can be routinely used to realize Bose mixtures~\cite{mandel2003controlled,PhysRevLett.105.045303}. The spinor condensates constitute yet another atom-based platform to simulate many-body physics~\cite{RevModPhys.85.1191,PhysRevLett.116.155301,PhysRevA.100.013622,PhysRevLett.126.063401}. We utilize Floquet driving, because the stroboscopically-driven setups have the advantage to produce a uniform density of states, i.e.,~no unfolding~\cite{guhr_random_1998} or filtering~\cite{Gharibyan_2018} procedures are required for the computation of the spectral statistics, unlike in time-independent Hamiltonian systems. Floquet driving of cold atom systems have been realized \cite{wintersperger2020realization,lellouch2017parametric,goldman2014periodically}. 

We empirically obtain a power-law scaling function for the bump regime. This suggests a power-law correction to the SFF ``ramp-plateau'' in cold atom lattices. The deduced scaling function for the bump, in turn, proves to be practically useful to statistically differentiate the bump from ramp in an experiment. In addition, we establish a relation between the SFF and the survival probability thus corroborating the experimental detectability of the former. The latter has been theoretically studied to probe quantum chaos \cite{PhysRevB.99.174313,PhysRevE.102.032208,santos2022boson} and measured in picosecond spectroscopy of molecular dissociation \cite{PhysRevLett.58.475,zewail,bala}. We devise a read-out protocol for the SFF through many-body state interference \cite{PhysRevLett.109.020505}, which measures survival probability and is a state-of-the-art measurement scheme for cold atoms in optical lattices \cite{islam2015measuring}. 

The focus on SFF in driven atomic systems complements the investigations of chaos in atomic models utilizing time-independent setups \cite{PhysRevE.102.032208,santos2022boson,PhysRevA.106.013301}, Rydberg interactions \cite{ZollerSFF2020,ZollerSFF2021}, highly long-range interactions \cite{Roy_2022,PhysRevE.102.060202} or tools such as dynamical fidelity \cite{PhysRevLett.90.054101,Wimberger_2006} and nearest-neighbor energy level statistics \cite{PhysRevA.35.1464,PhysRevLett.91.253002,PhysRevA.72.032511,PhysRevLett.98.130402,OganesyanHuse,Santos2010OnsetOQ,PhysRevA.88.032119,PhysRevE.90.012110,PhysRevA.99.052322,PhysRevA.101.053604,anhtai2023quantum}. Recent studies~\cite{ZollerSFF2020, ZollerSFF2021} have proposed schemes to measure SFF, numerically implement spin-1/2 chains and study the SFF in dipole-blockaded Rydberg atoms, with an emphasis on $\thei$ and the ramp-plateau behavior. In this work, we argue that $\tth$ is not only experimentally more accessible for large systems, because $\tth \ll \thei$ and probing $\tth$ does not require resolving the many-body level spacing, but it is also physically more appealing ---
$\tth$ and the bump regime emerge in {\it many-body} chaotic systems. The onset of the RMT ramp, as quantified by ~$\tth$, provides us with another physically meaningful timescale in quantum many-body systems.

This work is organized as follows. In Secs.~\ref{sec:sff} and~\ref{sec:models}, we define the SFF and the cold atom models under consideration, respectively. In Sec.~\ref{sec:prermt}, we discuss the bump regime in bosonic quantum gases and obtain a scaling function for it. Finally in Sec.~\ref{sec:exp}, we outline our read-out protocol for the SFF before concluding in Sec.~\ref{sec:conc}.

\section{\label{sec:sff}Spectral Form Factor}
SFF is the Fourier transform of the two-level spectral function, and can be defined directly in the time domain as
\begin{eqnarray}
K(t)&=& 
\bigg\langle \left|\Tr [\hat{U}(t)]\right|^2 \bigg\rangle
=
\left\langle \sum_{m,n} e^{i(E_m  - E_n)t} \right\rangle
 \, ,\label{sffDef}
\end{eqnarray}
where $\hat{U}(t)$ is the time evolution operator of a given system with spectrum $\{  E_m\}$, and $\langle \dots \rangle$ denotes averaging over certain ensemble of statistically-similar systems. %
To understand the basic features of SFF for quantum chaotic systems, consider the circular unitary ensemble (CUE) and the circular orthogonal ensemble (COE) in RMT \cite{Haake, Mehta}. CUE is the ensemble of $n$-by-$n$ matrices uniformly distributed in the unitary group $\hat{U}(n)$ according to the Haar measure, and can be used to model quantum chaotic systems without symmetries. In this case, the SFF is given by $K_{\text{CUE}}(t) = t$ for $t\leq \thei $ and $K_{\text{CUE}}(t) = \NN $ for $t > \thei$, where the Heisenberg time $\thei = O(\Delta^{-1}) = O(\NN)$ is proportional to the inverse energy level spacing $\Delta$ and hence to the size of the Hilbert space $D$.
COE is the ensemble of uniformly-distributed unitary symmetric matrices, and can be utilized to model systems with time-reversal symmetry with the antiunitary time reversal operator $\mathcal{T}$ satisfying $\mathcal{T}^2 = 1$~\cite{Haake}. The corresponding SFF behavior is $K_{\text{COE}}(t) = 2t - t \log (1+2t/\thei)$ for $t\leq \thei =  \NN$ and $K_{\text{COE}}(t) = 2\thei -t \log\left[ (2t + \thei)/(2t-\thei)\right]$ for $t > \thei= \NN$. 
The RMT-like ramp in a spin$-1/2$ BH model can be seen in Fig.~\ref{Fig1b}(a) for both symmetry classes.

As mentioned in the Introduction, a deviation  from RMT behavior in SFF, the ``bump'', has been demonstrated to exist at times earlier than the many-body Thouless time $\tth$ in generic many-body quantum systems (see Fig.~\ref{Fig1b}(a) and (b)). 
The temporal region $t < \tth$ of SFF (Fig.~\ref{Fig1b}(b)) will be referred to as the bump regime throughout the text. 
Thouless time is an intrinsic timescale that generically grows with the system size -- with the exception of quantum circuits satisfying the dual unitarity condition~\cite{ bertini2018exact, bertini2021random} -- and can be characterized by a set of Lyapunov exponents \cite{cdclyap}. For systems without conserved quantities, this behavior of $\tth$ originates from the domain walls in emergent statistical mechanical models that separates growing RMT-like regions~\cite{cdc2, garratt2020dw, garratt2020MBL}. For systems with conserved quantities and constraints, the origin of the $\tth$ scaling can be traced back to the diffusive and sub-diffusive modes of the charges~\cite{friedman2019, moudgalya2021}. 

\section{\label{sec:models}Models}

In all of the models, the employed driving protocol is expressed via the unitary operator $\hat{U}(t')\equiv \hat{U}^{t'}$. This refers to the repeated application of the (i) two-step (with period $T=\tau$) and (ii) three-step (alternating from $\tau_1$ to $\tau_2$ and having a period $T=(2\tau_1+\tau_2)/2$) periodic driving scheme defined by the Floquet operator
\begin{eqnarray}
\hat{U}&= & \begin{cases} 
e^{-i\hat{H}_2 \tau} e^{-i\hat{H}_1 \tau} &  \text{two-step}  \\
e^{-i\hat{H}_1 \tau_1} e^{-i\hat{H}_2 \tau_2} e^{-i\hat{H}_1 \tau_1}  \;\; &  
 \text{three-step} 
\end{cases}\;, \label{floquet_operator}
\end{eqnarray}
with  $\tau_1 \neq \tau_2 \neq 0$ which breaks the time-reversal symmetry for the three-step protocol.

In the following, we introduce the spatially extended spinless and spin$-1/2$ BH models, as well as chaotic spin$-1$ condensates. 
The system-specific ingredients of the Floquet protocol are also discussed.

\subsection{\label{sec:BHmodel}Driven Bose-Hubbard chains}

The simplest tight-binding model for interacting bosons, the BH model \cite{PhysRevB.40.546}, can be effectively realized by loading bosonic cold atoms into a one-dimensional optical lattice~\cite{PhysRevLett.81.3108}.
Such a geometry can be readily implemented in an experiment with tightly confined transversal directions so that the excitations along the latter are highly suppressed, and hence the system corresponds to a series of independent 1D tubes across the $x$ direction~\cite{rispoli2019quantum}.  
Initially, we employ a setup where the interacting bosonic atoms have two hyperfine states described by a spin$-1/2$ BH Hamiltonian (Fig.~\ref{Fig1a}(a)),
\begin{eqnarray}\label{eq:bh_mix}
\hat{H}_{\text{BH}} &=& \sum_{r} \left( \sum_{\sigma} J_{\sigma}  \hat{b}_{\sigma,r}^{\dagger} \hat{b}_{\sigma,r+1} + J_{\uparrow \downarrow,r } \hat{b}_{\uparrow,r}^{\dagger} \hat{b}_{\downarrow,r} + \text{h.c.} \right) \notag\\
&+& \frac{1}{2} \sum_{\sigma,r} u_{\sigma} \hat{n}_{\sigma,r} (\hat{n}_{\sigma,r}-1)+u_{\uparrow \downarrow} \sum_r \hat{n}_{\uparrow,r} \hat{n}_{\downarrow,r} \notag \\
&+& \sum_{\sigma,r} \mu_{\sigma,r} \hat{n}_{\sigma,r} \; . \label{hamiltonianMixture}
\end{eqnarray}
Here $\sigma= \, \uparrow, \downarrow$ is the spin index. $J_{\sigma}$, $u_{\sigma}$ and $u_{\uparrow \downarrow}$ denote the tunneling and on-site interaction strengths, that can be experimentally tuned through the laser fields generating the lattice potential \cite{PhysRevLett.81.3108,PhysRevLett.91.090402}.
Additionally, $J_{\uparrow \downarrow}$ refers to the spin-mixing coupling between two hyperfine states.
It can be created and adjusted using Raman assisted tunneling \cite{weimer2010rydberg} or the radio-frequency fields~\cite{lavoine2021beyond}. The position-dependent random potential $\mu_{\sigma,r}$ can be realized with the aid of digital micromirror devices~\cite{rispoli2019quantum,lukin2019probing,bordia2017probing}. 

The Hamiltonians in the Floquet unitaries above are characterized by the fixed parameters $J_{\uparrow}=u_{\uparrow}=u_{\downarrow}=u_{\uparrow \downarrow} = u = J$, while the remaining ones alternate according to
\begin{eqnarray}
\hat{H}_{1} & \rightarrow & \, \hat{H}_{\mathrm{BH}} \text{ with } J_{\downarrow}=J, \; J_{\uparrow \downarrow} = 0, \;  \mu_{\uparrow,r} \in [-J,J] \;, \notag\\
\hat{H}_2 & \rightarrow &  \, \hat{H}_{\mathrm{BH}} \text{ with }
J_{\downarrow}=0, \;  J_{\uparrow \downarrow,r} \in [-J,J], \; \mu=0 \;, \notag 
\end{eqnarray}
with the constraint $\mu_{\uparrow,r}=-\mu_{\downarrow,r}$. As such, within the first step, a random potential is turned on across the chain where the notation $[-J,J]$ means that we randomly choose the site-dependent potential 
values from a uniform distribution. 
However let us note that the strength of the disorder does not affect the physics so long as it is larger than $J$.
Subsequently, in the second step, we switch off this random potential, while the tunneling of the atoms confined in $\Ket{\downarrow}$ hyperfine state vanishes and an onsite random spin-mixing tunneling amplitude between different hyperfine states is turned on. The driving frequency should not be on resonance with the spin-mixing coupling $J_{\uparrow \downarrow}$ not to inadvertently polarize the time-dependent state.

Evolution times for the BH Hamiltonians, e.g.,~$\tau$, are expressed in units of inverse tunneling $1/J$, and to compute $\hat{U}^{t'}$ we apply trotterization. When $\tau \gg 0.1$ holds, the application of the two-step scheme reproduces COE spectral statistics for the spin-$1/2$ bosonic system \cite{supplementary}. 
In contrast, by following the three-step periodic scheme the system transitions from exhibiting COE to CUE spectral statistics as $\tau_1$ and $\tau_2$ increase and has persistent CUE statistics for $\tau_1,\tau_2 \gg 0.1$.  
For convenience, throughout, we define a dimensionless normalized time $t=t'/T$ such that our many-body simulations are directly comparable to the analytic predictions of RMT.

Although our results, to be presented below, are based on the stroboscopically-driven protocol discussed above, there is flexibility in the choice of parameters for establishing a chaotic behavior in the spin-$1/2$ setting. Here, we briefly summarize the range of this flexibility and refer to the SM \cite{supplementary}
for numerical evidence. (i) If desired, the number of simultaneous pulses can be further decreased. (ii) Randomizing either $\hat{H}_1$ or $\hat{H}_2$ is sufficient, e.g.~$J_{\uparrow \downarrow,r}$ does not have to be randomized. (iii) The interaction or tunneling strengths do not need to be equal to each other and one of the three interaction parameters could be set to zero. We observe that the two-step driven Floquet system moves away from COE statistics when the atoms with either spin components transition deep into the superfluid regime, e.g.,~$J_{\uparrow}=20J$. Overall, we find that the necessary parameter to simulate RMT statistics in the spin$-1/2$ BH model is the onsite spin-mixing tunneling $J_{\uparrow \downarrow}$.
The underlying reason can be traced back to the fact that the spin-mixing tunneling term breaks the SU$(2)$ symmetry of the system~\cite{lavoine2021beyond,Mistakidis_2022}. In the presence of this symmetry and two-step driving, the chaotic behavior described by COE statistics is not apparent to the SFF of the entire Hilbert space (i.e.,~unprojected SFF) \cite{supplementary}.

In the absence of spin-mixing coupling $J_{\uparrow \downarrow}=0$, the projection to a spin-preserving subspace could be achieved through transferring all atoms to one hyperfine state. 
This is effectively modeled by the spinless BH model \cite{PhysRevB.40.546}. 
The Hamiltonian of this spinless model could be written in the form of Eq.~\eqref{hamiltonianMixture} with the constraint that all atoms are, for instance, in the spin-$\downarrow$ state. Namely, we consider the parameters $J_{\uparrow \downarrow}= u_{\uparrow \downarrow}=J_{\uparrow}=u_{\uparrow}=\mu_{r,\uparrow}=0$, and assume that $N=N_{\downarrow}$. The remaining parameters are denoted by $J_{\downarrow}=J$, $u_{\downarrow}=u$ and $\mu_{r,\downarrow}=\mu_r$. 
Exploring how to induce chaotic behavior in the spinless BH model and comparing the predictions with the spin-$1/2$ case enable us to determine how the bump regime and the Thouless time depend on the hyperfine structure of the atoms. The periodically driven spinless BH model exhibits chaotic behavior when the Hamiltonians of the Floquet unitary are given by 
$\hat{H}_1 = \hat{H}_\text{BH}$ with $\lbrace J=1,u=J,\mu_{r} \in [-J,J]\rbrace$, and  $\hat{H}_2 = \hat{H}_{\text{BH}}$ with $\lbrace J=0,u=1,\mu_{r}=0\rbrace$.
Note that the first two guidelines stated above, (i) and (ii), for the parameter selection that could lead to RMT behavior, are also valid for the spinless BH model \cite{supplementary}.
Moreover, we utilize an alternative protocol based on Refs.~\cite{PhysRevLett.120.050406,PhysRevA.97.023604} to test whether, and if so how, $\tth$ depends on the choice of the  driving protocol (see Sec.~\ref{sec:BHresults} and SM for further details).

\subsection{\label{sec:spinorModel}Driven short- and long-range interacting spinor condensates}

\begin{figure*}
\subfloat{\includegraphics[width=0.99\textwidth]{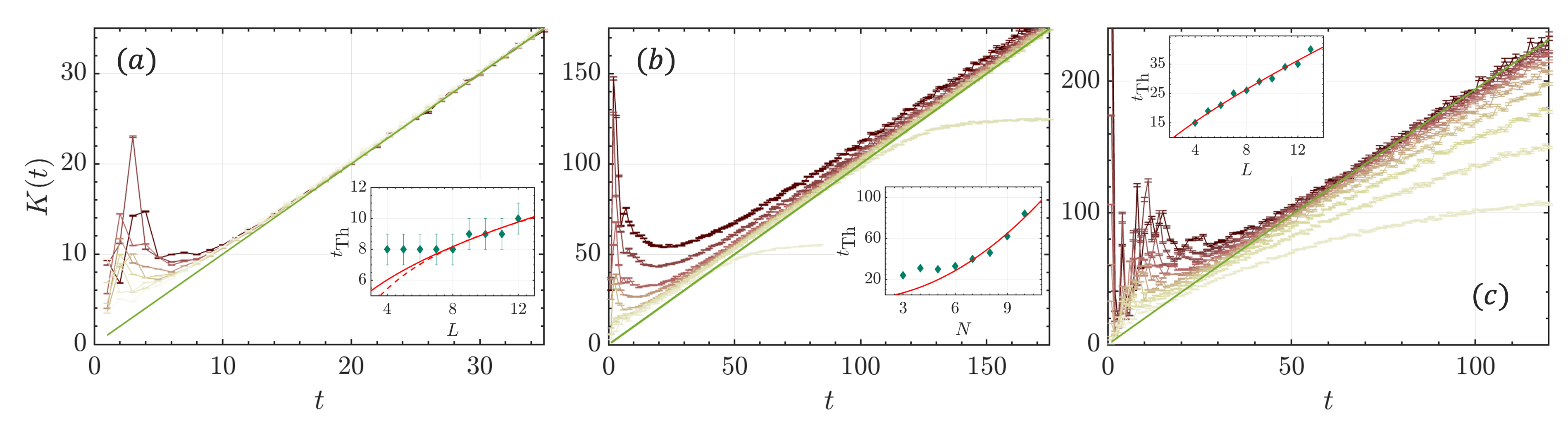}}\hfill
\caption{The bump behavior of spin-$1/2$ BH model when driven according to (a-b) three-step and (c) two-step protocols. (a) SFF for $N=3$ following CUE for a driving period $T=3$ with lattice sizes $L \in [4,12]$. Inset: $\tth$ increases with $L$, however its function cannot be unambiguously determined with the available data. Solid and dashed lines are fit models for power-law and logarithmic, respectively. (b) SFF for a triple well $L=3$ following CUE for a driving period of $T=1.5$ with atom numbers $N\in [3,10]$. Inset: $\tth$ scaling is fitted to a power-law with $\tth(N=3,L)\propto N^{2}$. (c) SFF of three atoms $N=3$ following COE for a driving period $T=1$ with lattice sizes $L \in [4,13]$. The green-solid denote the COE for largest $D$ in the data set. Inset: $\tth$ scaling with $L$ is fitted to a power-law with $\tth(N=3,L)\propto L^{0.78}$. For all cases, the bumps generally increase with $L$ and $N$. Note the $\tth$ is defined as a time at which an error function $\epsilon(t)= |K(t) - K_{\mathrm{RMT}}(t)|/K_{\mathrm{RMT}}(t) $ is smaller than a threshold of choice, where $K_{\mathrm{RMT}}$ is the SFF behavior for RMT of an appropriate symmetry class. See details in \cite{supplementary}.
}
\label{Fig2}
\end{figure*}

Since we aim to identify the role of spatial extendedness in the bump regime, we also employ an $s-$wave interacting spin$-1$ Bose gas within the single-mode approximation (SMA) \cite{PhysRevLett.81.5257,ho1998spinor,yi2002single,mistakidis2022cold} which assumes a separation of the spin and spatial degrees of freedom. In particular, the wave functions for each hyperfine state are described by the same spatial mode $\phi_{m=0,\pm 1}(x)=\phi(x)$ and different spin wave functions such that the decomposition $\hat{\psi}_m (\mathbf{r}) = \hat{a}_m \phi(\mathbf{r})$ holds  where $\hat{a}_m$ is the annihilation operator acting on the spin-$m$. SMA is valid for atom numbers as small as $N=100$ when the condensate is prepared in a tight laser trap \cite{PhysRevLett.126.063401}, and it could break down for long evolution times either due to the build-up of spatial correlations~\cite{RevModPhys.85.1191,mistakidis2022cold} as well as atom loss related processes from the condensate \cite{PhysRevA.100.013622,bookjans2011quantum,RevModPhys.85.1191}. The SMA Hamiltonian for the spin$-1$ condensate, with both linear and quadratic Zeeman fields along the $x-$ and $z-$ spin directions, reads
\begin{eqnarray}
\hat{H}_{\text{SC}} = \frac{c_1}{2N} \hat{\mathbf{F}}^2 + \sum_{i=x,z} \left(p_i \hat{F}_i + q_i \hat{F}_i^2\right). \label{spinorCondensate}
\end{eqnarray}
Here $\hat{\mathbf{F}}^2=\hat{F}_x^2+\hat{F}_y^2+\hat{F}_z^2$ denotes the spin operator, while $c_1>0$ is assumed corresponding to a ferromagnetic spinor gas~\cite{RevModPhys.85.1191}. A magnetic field across the $z-$direction, embodied in the quadratic Zeeman term $q_z \hat{F}_z^2$, breaks the rotational symmetry of the spinor gas \cite{RevModPhys.85.1191}. $c_1 \hat{\mathbf{F}}^2/2N +q_z \hat{F}_z^2$ is still an integrable model \cite{PhysRevA.72.013602,PhysRevA.97.023603}, however, the addition of a linear Zeeman field in the $x-$direction has been recently shown to break its integrability~\cite{PhysRevA.101.053604,PhysRevLett.126.063401}. 
This is because the $p_x \hat{F}_x$ term breaks the SO$(2)$ symmetry of the spinor condensate, giving rise to signatures of level repulsion which are captured by the Brody distribution \cite{Brody}. 

\begin{figure*}
\subfloat{\includegraphics[width=0.99\textwidth]{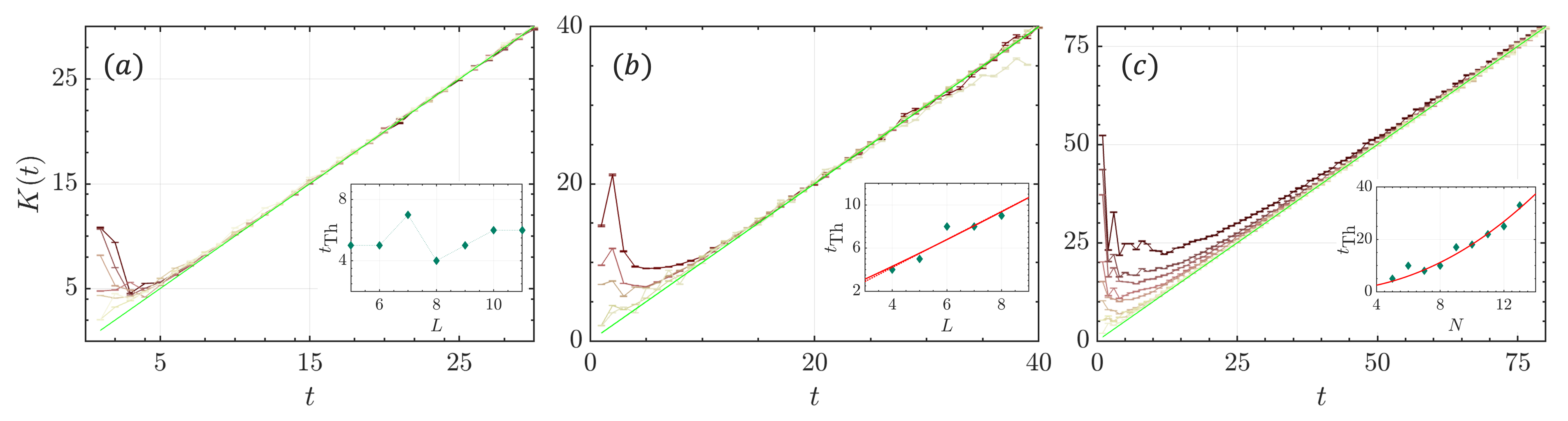}}\hfill
\caption{Bump behavior of spinless BH model in different filling factor regimes with three-step CUE protocol and a driving period of $T=3$. The filling factors are (a) $\nu\rightarrow 0$, (b) $\nu=1$ and (c) $\nu \rightarrow \infty$ as the Hilbert space dimension increases. (a) The atom number is $N=5$ and lattice size changes, $L \in [5,11]$. Inset: $\tth(N=5,L)$ fluctuates with $L$. (b) $N$ and $L$ both change, $L=N\in [4,8]$. Inset: $\tth(N=L,L)$ increases either linearly $1.3L-1$ (dotted) or close to linear $\propto L^{1.12}$ (solid). (c) $L=5$ and  $N \in [5,13]$. The power-law fit to $\tth$ scaling is $\tth(N,L=5)\propto N^{2.16}$.
}
\label{Fig3}
\end{figure*}

We find that spin$-1$ condensates with $p_x \hat{F}_x$ term display stronger level repulsion and spectral rigidity, i.e.,~an extended RMT ramp in SFF, only when a quadratic Zeeman shift in the $x-$direction $q_x \hat{F}_x^2$ or a linear Zeeman shift in the $z-$direction $p_z \hat{F}_z$ are introduced to the system \cite{supplementary}.

There is no spatial-extendedness in this model due to the SMA. All atoms,  regardless of the physical distance between them, could interact with one another equally likely through contact interaction, i.e. such a spinor Hamiltonian is effectively zero-dimensional. The spinor Hamiltonian described in Eq.~\eqref{spinorCondensate} can also be expressed in the Fock basis $\Ket{n_{-1},n_{0},n_1}$, where $n_i$ is the population in each hyperfine state. By utilizing the SU(3) Lie algebra operators \cite{PhysRevA.88.033629},
\begin{eqnarray}
\hat{H}_{\text{SC}} &=& \frac{c_1}{2N} \bigg[ \hat{a}_0^{\dagger} \hat{a}_0^{\dagger} \hat{a}_1 \hat{a}_{-1} + \text{h.c.} + \hat{N}_0(\hat{N}_1+\hat{N}_{-1}) \notag \\
&+& \frac{1}{2} \left(\hat{N}_1-\hat{N}_{-1}\right)^2\bigg] + \frac{p_x}{\sqrt{2}}\bigg [(\hat{a}_1^{\dagger}+\hat{a}_{-1}^{\dagger})\hat{a}_0 + \text{h.c.}\bigg ] \notag \\
&+& p_z \left(\hat{N}_1-\hat{N}_{-1}\right) + \frac{q_x}{2} \bigg[\hat{a}_1^{\dagger}\hat{a}_{-1} + \text{h.c.} \bigg ] - q_z \hat{N}_0, \label{SCinFock}
\end{eqnarray}
where $\hat{a}_{\nu}$ ($\hat{a}_{\nu}^{\dagger}$) and $\hat{N}_{\nu}$ refer to the annihilation (creation) and the number operators of an atom occupying the spin$-\nu$ hyperfine level, respectively. Note also the presence of spin-mixing collision interactions \cite{PhysRevLett.81.5257} in the first two terms of Eq.~\eqref{SCinFock}. Let us emphasize that without $p_z \hat{F}_z$ term in Eq.~\eqref{SCinFock}, the condensate preserves the inversion symmetry. Therefore, in order to capture the bump-ramp-plateau behavior in SFF, we either perform projection onto the symmetry subspace with even parity $\Ket{N_0,\mathcal{M}} \rightarrow \left(\Ket{N_0,\mathcal{M}}+\Ket{N_0,-\mathcal{M}}\right)/\sqrt{2}$ where $\mathcal{M}=N_1-N_{-1}$ is the magnetization, or we add linear Zeeman field in the $z-$direction, as in Eq.~\eqref{SCinFock}, and compute the unprojected SFF. The latter approach is experimentally feasible, see~\cite{supplementary} for details on the inversion symmetry subspace projection.

Furthermore, we consider the effect of long-range interactions in the spin$-1$ condensate and study the bump regime in the even parity subspace when there are dipolar magnetic interactions between the spinor atoms \cite{RevModPhys.85.1191,PhysRevA.73.023602}. For an anisotropic harmonic trap, it is possible to arrive at the so-called two-axis counter-twisting (TACT) Hamiltonian 
\cite{PhysRevA.93.023627}, namely
\begin{eqnarray}
\hat{H}_{\text{TACT}} &=& \chi \left(\hat{F}_x^2-\hat{F}_y^2-\hat{D}_{xy} \right), \label{TACT}\\
&=& \chi \bigg[2\hat{N}_0 \hat{a}_{-1}^{\dagger}\hat{a}_1 + \hat{a}_0^{\dagger}\hat{a}_0^{\dagger}(\hat{a}_1^2+\hat{a}_{-1}^2) + \text{h.c.}\bigg]. \notag
\end{eqnarray}
The latter term in Eq.~\eqref{TACT} is one of the eight SU$(3)$ Lie algebra operators $\hat{D}_{xy} = \hat{a}_1^{\dagger}\hat{a}_{-1} + \text{h.c.}$. Accordingly, we write the Hamiltonian for the dipolar spin$-1$ condensate as
\begin{eqnarray}
\hat{H}_{\text{DC}} = \frac{c_1}{2N} \hat{\mathbf{F}}^2 + p_x \hat{F}_x + q_z \hat{F}_z^2 + \hat{H}_{\text{TACT}}. \label{dipolarCondensate}
\end{eqnarray}

Similar to the BH model protocols described above, the two- and three-step stroboscopic protocols of Eq.~(\ref{floquet_operator}) lead to COE and CUE statistics with the following Hamiltonians in the Floquet unitary,
\begin{eqnarray}
\hat{H}_1 & \rightarrow & \,  \hat{H}_{\text{SC/DC}}   \text{ with }q_z \in \left[0,\frac{c_1}{N} \right], \; \; p_x=\frac{c_1}{N} \, , \notag \\
\hat{H}_{2} & \rightarrow &  \,  \hat{H}_{\text{SC/DC}}   \text{ with } q_z \in \left[-\frac{c_1}{N},0 \right], \; \; p_x=-\frac{c_1}{N} \,.
\end{eqnarray}
For concreteness throughout the text, we will assume a spinor condensate in the Thomas-Fermi limit and confined in a 1D trap \cite{PhysRevA.60.1463}, such that $c_1/N \propto N^{-1/3}$ holds.

\section{\label{sec:prermt}Bump regime  and Thouless Time}

Now we discuss the behavior of the bump regime and the scaling of $\tth$ for the spatially-extended driven BH models, and the spinor condensates in subsections~\ref{sec:BHresults} and~\ref{sec:spinorResults}, respectively. To summarize, we find that the bump regime and $\tth$ scaling are more sensitive to variations in the atom number than in the chain size. Moreover, we observe that interacting bosons trapped in a single well display RMT behavior in SFF much faster than the bosonic system in spatially extended potentials. We conclude this section by empirically deducing a scaling function for the bump regime of the SFF for the interacting trapped bosons. Our results discussed below are based on 10000-50000 realizations of $\hat{U}$.

\subsection{\label{sec:BHresults}Dependence on atom number and lattice size}

\begin{figure*}
\subfloat{\includegraphics[width=0.99\textwidth]{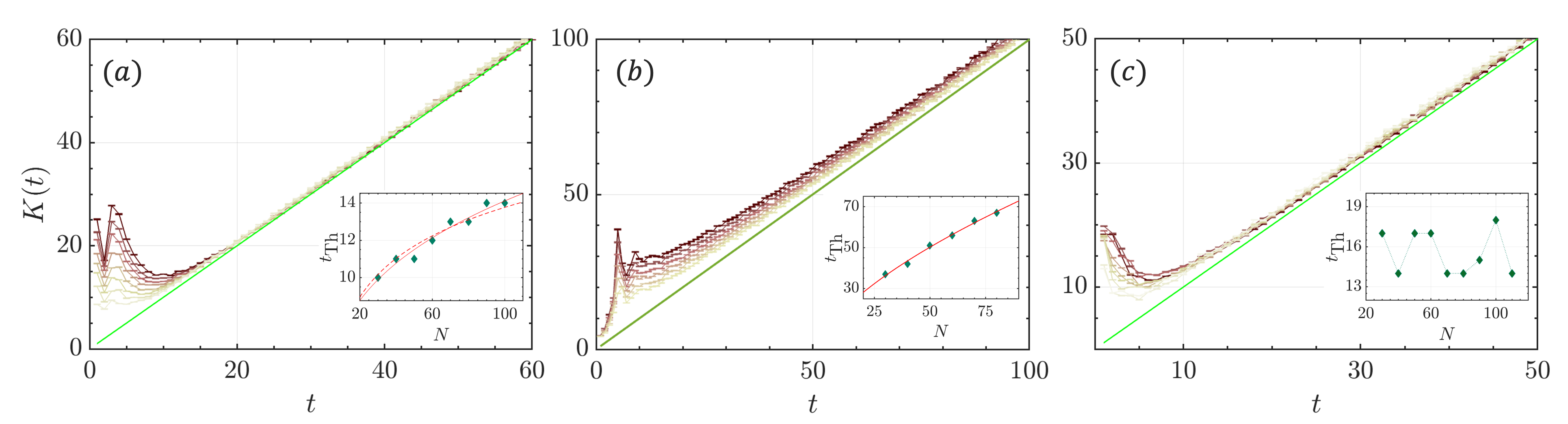}}\hfill
\caption{(a) Projected SFF of the spinor condensate $H_{\text{SC}}$ on even parity sector driven according to three-step protocol and with $p_z=0$, $q_x=2$. The atom number changes between $N \in [30-100]$ from yellow to red. Inset: The Thouless timescales with the atom number, either in a slow power-law ($\propto N^{0.3} N/|c_1| \propto N^{0.63}$, solid) or logarithmic ($\propto \log (N)$, dashed). (b) SFF of the entire Hilbert space of the spinor condensate $H_{\text{SC}}$ driven according to three-step protocol, with $p_z=2$, $q_x=0$ and the atom number changing between $N \in [30,80]$. Inset: The Thouless timescales with the atom number in a slow power-law $\propto N^{0.63} N/|c_1| \propto N^{0.96}$. (c) Projected SFF of the spinor condensate with dipolar interactions $H_{\text{DC}}$ on even parity sector driven according to three-step protocol and with $\chi=2$. The dipolar interaction strength is $\chi=2$ for atom numbers $N \in [30,110]$ from yellow to red. Inset: The Thouless time fluctuates with the atom number, resulting in $\tth(N) \propto N/|c_1| \propto N^{1/3}$ scaling in atom number. In all panels, we set the driving period as $T=3$.}
\label{Fig4}
\end{figure*}

In contrast to spin systems and quantum circuits, the size of the Hilbert space in BH models is determined by two parameters: the chain size $L$ and the atom number $N$. These parameters dictate the size of the Hilbert space as $D=(N+\alpha L-1)!/(N!\times (\alpha L-1)!)$, where $\alpha$ is the number of hyperfine levels. Therefore, we can explore the dependence of the bump regime and $\tth$ scaling not only on $L$ but also on $N$, and hence the filling factor $\nu$ of the lattice. 
We confirm that the bump-ramp-plateau behavior persists in the Mott insulator regime, where $u \gg J$ holds for the spin$-1/2$ bosonic system \cite{supplementary}, and further explore the bump and the emergence of RMT beyond the Mott insulator regime below.

Figs.~\ref{Fig2}a and~\ref{Fig2}b demonstrate that the early time behavior of SFF in a spin$-1/2$ BH model, converges to the CUE behavior, as a function of $L$ ($N$) with fixed $N=3$ ($L=3$). Fig.~\ref{Fig2}(c), on the other hand, shows the presence of the bump when the system is driven according to the two-step protocol, and thus exhibits COE statistics as a function of $L$ with three atoms.
The amplitude of the bump increases with $L$ at fixed $N$ for both protocols that simulate CUE and COE symmetry classes as illustrated in Figs.~\ref{Fig2}a and~\ref{Fig2}c.

An increase of $L$ ($N$) while keeping all other parameters fixed results in a more dilute (denser) lattice trapped gas. Interestingly, the bump is significantly enhanced for a denser gas as seen by comparing Fig.~\ref{Fig2}(b) and Fig.~\ref{Fig2}(a). In all cases of spin$-1/2$ BH model, $\tth$ increases either as a function of $L$ or $N$, but it is difficult to determine the functional form of $\tth(N,L)$ based on the accessible finite-size data. To compare the scaling behavior among different cases, power-law fits are provided in the insets of Fig.~\ref{Fig2}. We observe that the scaling exponent of $\tth$ is larger in $N$ for filling factors $\nu \rightarrow \infty$ ($N \rightarrow \infty$, fixed $L$) e.g.,~$\tth(N,L=3) \propto N^{2}$, than in $L$ for $\nu \rightarrow 0$ ($L \rightarrow \infty$, fixed $N$) e.g.,~$\tth(N=3,L) \propto L^{0.44}$. 

For the two-step protocol simulating COE symmetry class presented in Fig.~\ref{Fig2}(c), the bump is more prevalent in $L$ compared to the three-step protocol simulating CUE class shown in Fig.~\ref{Fig2}(a). Consistently, $\tth(N=3,L)\propto L^{0.78}$ scaling is slightly faster. For the corresponding scaling behavior at $N > 3$ or $L>3$, see SM. A similar stark difference between the scaling in $N$ and $L$ persists also for the two-step protocol under COE symmetry class with $\tth(N,L=3) \propto N^{1.95}$ and $\tth(N=3,L) \propto L^{0.78}$, respectively \cite{supplementary}. 

These observations suggest that $\tth(N,L)$ is more sensitive to $N$ than to $L$ for spin$-1/2$ interacting bosons in optical lattices. This dependence on atom number, in turn could be traced back to the interactions between lattice confined atoms.
In fact, increasing the interaction strengths $u=u_{\uparrow}=u_{\downarrow}=u_{\uparrow \downarrow}$ so that the system transitions to a Mott insulator, increases the Thouless time and results in a larger bump amplitude \cite{supplementary}. 

To further test the generality of the argument given above, we compare the SFF of the spinless BH model with three different filling factor regimes, namely $\nu \rightarrow 0$, $\nu =1$ and $\nu \rightarrow \infty$ as $D$ increases. As depicted in Fig.~\ref{Fig3}(c), the bump signature is more pronounced for increasing $\nu$, and $\tth$ scales faster ($\tth(N,L=5)\propto N^{2.16}$) than in Fig.~\ref{Fig3}(b), where $\nu=1$ holds ($\tth(N,L=N)\propto N^{1.12}$).  
We observe that this difference also depends on the choice of the driving protocol. Indeed, the $\tth(N,L)$ scaling turns out to be the same for the filling factors $\nu = 1$ and $\nu \rightarrow \infty$ both leading to a scaling $\tth(N,L)\propto N^4$ when the stroboscopic driving scheme is set to be an alternative protocol based on Ref.~\cite{PhysRevA.97.023604} which we refer to as the VE-protocol. 

The VE protocol is defined by the Floquet unitary, $\hat{U}_{\text{VE}}=\prod_m^M e^{-i \hat{H}_m \tau}$ where $\hat{H}_m$ is a spinless BH Hamiltonian with a different random potential landscape at each Floquet step $\mu_{r,m}\in [-J,J]$ for $J=1$ and $u=J$. This protocol reproduces COE symmetry class for a two-step $M=2$ Floquet unitary and smoothly transitions to simulating CUE symmetry class as the number of steps in the Floquet unitary increases to $M > 5$ \cite{supplementary}. The comparison of the SFF with VE-protocol suggests that the onset of RMT behavior occurs earlier in the three-step protocol described by Eq.~\eqref{floquet_operator}, giving rise to CUE in cold atoms earlier than the VE-protocol does. Nevertheless, for both protocols we observe a significant difference in the bump regimes and $\tth$ scaling between $\nu \rightarrow \infty$ and $\nu \rightarrow 0$. This manifests again the sensitivity of $\tth(N,L)$ to $N$. 

Differently than the spin-$1/2$ BH model, $\tth$ fluctuates around a fixed value instead of exhibiting a monotonic increase with $L$ for $\nu \rightarrow 0$ as illustrated in the inset of Fig.~\ref{Fig3}(a). We confirm the independency of this behavior from the protocol choice \cite{supplementary}. In SM, we show a slowly increasing trend in $L$ with fluctuations for $\tth(N=4,L)$, so that we can simulate up to $L=15$. Therefore, observing no scaling for $\tth(N=5,L)$ in $L$ might be a finite-size effect.

\subsection{\label{sec:spinorResults}Role of locality}

For a many-body system that does not extend spatially in a 1D chain, the Thouless time scaling is significantly slower due to lack of locality \cite{Gharibyan_2018,cdc1}. As we show in Fig.~\ref{Fig1b}(a), a spin$-1/2$ gas trapped in a single well does not exhibit RMT statistics (blue markers). 
For this reason below we employ a spin$-1$ condensate within a harmonic trap, and explore the Thouless time scaling with respect to $N$ as the dimension of the Hilbert space is $D=(N+2)(N+1)/2$. Fig.~\ref{Fig4}(a) shows the presence of the bump regime for $H_{\text{SC}}$ driven according to the three-step protocol and projected to even parity subspace when $p_z=0$ and $q_x=2$ hold. Although this spin$-1$ gas is a zero-dimensional model, we observe a bump in the SFF whose amplitude increases with $N$, i.e.~the bump feature is not a finite-size effect. However, $\tth$ scaling with $N$ is significantly slower compared to 1D BH models. For instance, assuming a power-law scaling function, we obtain $\tth(N)\propto N^{0.3} N/|c_1| \propto N^{0.63}$ when we take $N$ dependence of $c_1$ into account (see Sec.~\ref{sec:models}). 

An experimental operation to achieve the subspace projection is through introducing a linear Zeeman term along the spin $z-$direction, i.e.,~$p_z \neq 0$ breaking the inversion symmetry. The resulting SFF when taking into account the entire Hilbert space is presented in Fig.~\ref{Fig4}(b), where $p_z=2$ and $q_x=0$ were set. The bump gets sharper with increasing atom number, and $\tth$ scaling can be described with a power-law increase in $N$ with an exponent that is still smaller than the spatially extended BH models with $\gamma \sim 1$.

In order to generalize our results beyond the short-range $s-$wave interactions, we next introduce dipolar interactions into the spin$-1$ condensate, e.g.,~$\chi=2$ in Eq.~\eqref{dipolarCondensate}, and drive it according to the three-step protocol. The projected SFF on the even parity sector is presented in Fig.~\ref{Fig4}(c). A tendency for suppression in the amplitude of the bump takes place compared to Fig.~\ref{Fig4}(a), while $\tth$ does not exhibit any clear scaling trend with $N$. When $N/|c_1|$ rescaling is taken into account, we find $\tth(N)\propto N^{1/3}$. Therefore, our findings corroborate the idea that the locality of the underlying Hamiltonian is indeed reflected on the bump regime and $\tth$ scaling.

\subsection{\label{sec:crossover}
Scaling function of the bump regime}
\begin{figure}
\subfloat{\includegraphics[width=0.49\textwidth]{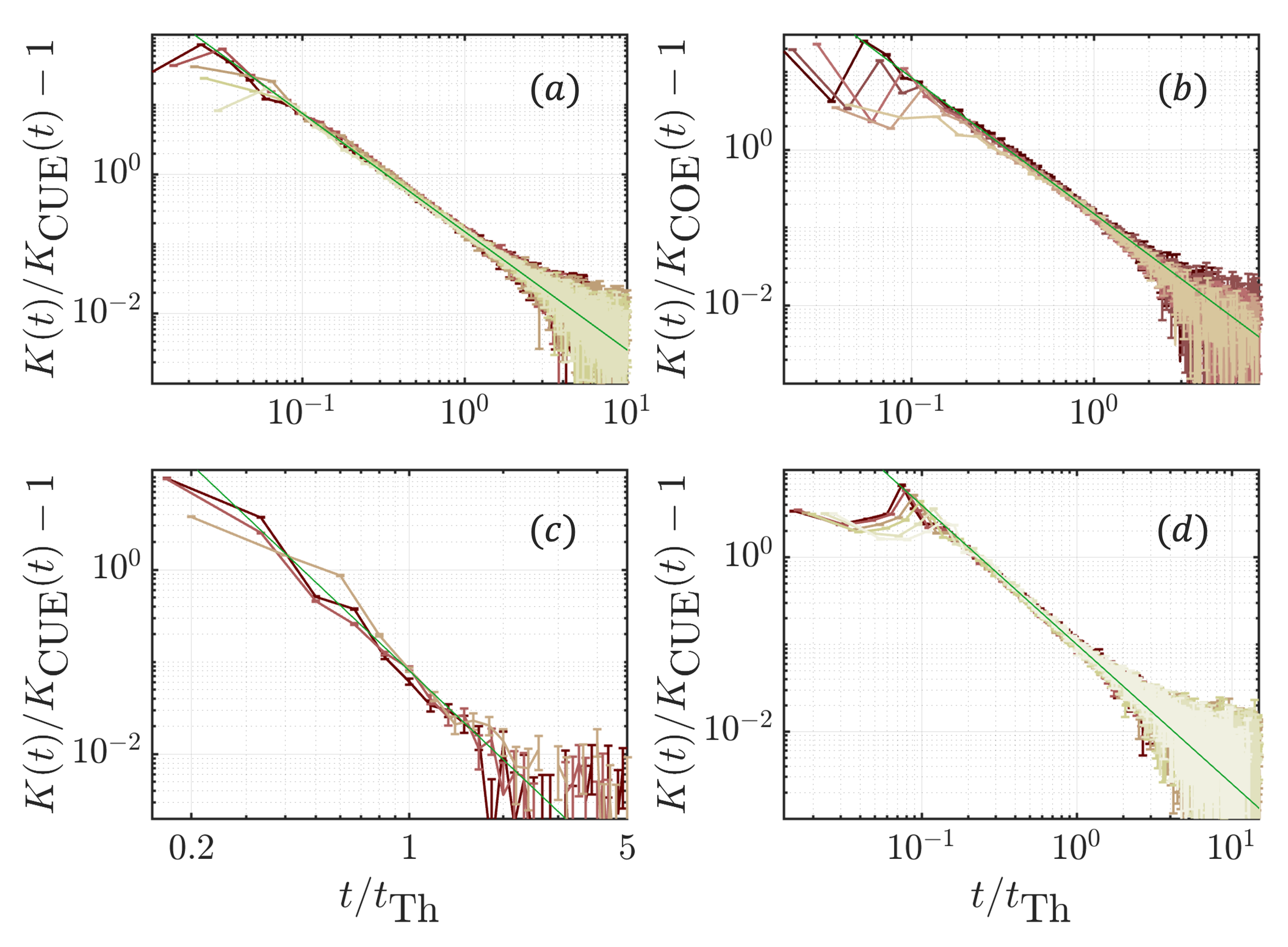}}\hfill
\caption{The collapse of data around $\tth$. (a) Spin$-1/2$ BH model for atom numbers ranging between $N \in [6,10]$ from yellow to red in a triple well following the CUE statistics and giving $\xi=-1.75$ in Eq.~\eqref{bumpEq}. (b) Spin$-1/2$ BH model for atom numbers ranging between $N \in [4,9]$ from yellow to red in a triple well following the COE statistics and giving $\xi=-1.75$. (c) Spinless BH model of five atoms for lattice sizes ranging between $L \in [9,11]$ from yellow to red in a triple well following the CUE statistics and giving $\xi=-3.25$. (d) Spinor condensate with $p_z=2$ and $q_x=0$ for atom  numbers ranging between $N \in [30,80]$ from yellow to red following the CUE statistics and giving $\xi=-1.6$. The $x-$axes are scaled with the numerically extracted Thouless times. The $y-$axes are scaled with the corresponding RMT behavior.}
\label{Fig5}
\end{figure}

Figs.~\ref{Fig2},~\ref{Fig3} and~\ref{Fig4} suggest the existence of rescaling parameters in both $t-$ and $y-$axes where $K(t)$ for different $D$ could collapse on each other in the  bump region. We rescale time with $\tth$ and $K(t)$ with the analytical ramp expression to spotlight the bump region: $K(t)/K_{\text{rmt}}(t)$ which should saturate at unity for $t> \tth$. To determine how the bump approaches unity and whether there is a universal behavior for all $D$, we plot $K(t)/K_{\text{rmt}}(t)-1$. This rescaling analysis reveals a collapse of data around $\tth$. The data collapses well in the RMT regime $t/t_{\text{th}} \gtrsim 1$ as expected, and in the bump region $t < \tth$. We observe no data collapse for $t/\tth \rightarrow 0$. Interestingly for all models, a collapse in the rescaled SFF $K(t)/K_{\text{rmt}}(t)-1$ for sufficiently large $D$ is present, and $K(t)/K_{\text{rmt}}(t)-1$ approaches the saturation value $\sim 0$ as a power-law in the rescaled time $t/t_{\text{Th}}$. We notice that this power-law approach to RMT in fact describes the region where $K(t)$ decreases from certain time that we call $\tb$, reaches a minimum and increases towards RMT. We can then write the following empirical expression for the SFF which also characterizes the bump region,
\begin{eqnarray}
K(\tb < t < \thei) \approx K_{\textrm{RMT}}(t)\left[\beta\left(\frac{t}{\tth}\right)^{\xi}+1\right], \label{bumpEq}
\end{eqnarray}
where $\beta$ and $\xi < 0$ are scalar. Based on our analysis, we observe $\beta \ll 1$ is a small parameter, and the exact value of $\xi$ could depend on the microscopics, e.g.,~$-4 < \xi < -1.4$. We note that $\tb$ coincides with time where the bump exhibits a local peak  before reaching a local minimum for all cases.

We depict two cases of spin$-1/2$ BH chain in a triple well with different atom numbers in Figs.~\ref{Fig5}(a) and~\ref{Fig5}(b) where (a) data is for CUE and (b) for COE. The associated bump region and $\tth$ scaling of (a) are already given in Fig.~\ref{Fig2}(b). 
Hence Fig.~\ref{Fig5}(b) suggests that the physics discussed earlier also holds for COE. We plot the rescaled SFF in Fig.~\ref{Fig5}(c) for the spinless BH model of five atoms trapped in optical lattices of varying sizes. The associated bump region is given in Fig.~\ref{Fig3}(a) whose inset actually shows no $\tth$ scaling with lattice size. Nevertheless the bump regime is present, and consistently with the rest of the observations, it introduces a power-law correction to the RMT. Finally in Fig.~\ref{Fig5}(d), we demonstrate the bump scaling for a chaotic spinor condensate and again find that the expression in Eq.~\eqref{bumpEq} holds. 

The observation that a consistent scaling, Eq.~\eqref{bumpEq}, exists for the different atomic systems and parameter values suggests universality beyond the RMT regime. This means that the physical signature of quantum chaos in many-body systems precedes the ramp and the Thouless time, and includes the bump region $\tb  < t < \tth$. 
An indirect physical evidence in support of this argument can be found in the heating times of local observables \cite{PhysRevE.90.012110}. For the considered lattice sizes, we find that the heating times of density populations in the BH models are less than the corresponding $\tth$, and in fact sometimes comparable to $\tb $, i.e.~the time when the bump regime collapse starts \cite{supplementary}. Hence, the knowledge of relevant timescales of quantum chaos in an experiment may have practical consequences for information acquisition.

\section{\label{sec:exp} Experimental protocol}

Next we discuss an experimental protocol applicable to cold atoms trapped in optical lattices. For concreteness, we will focus on the BH model, however the arguments are valid for the spin$-1/2$ BH model which requires an additional hyperfine state. Later, we comment on the case of spin$-1$ condensates.

SFF is defined through a trace operation (see Eq.~\eqref{sffDef}) meaning that the system should be initialized in an infinite-temperature state whose controlled preparation in an experiment is not a simple task. On the other hand, the BH model in an optical lattice can be routinely initialized in the Mott insulator regime, i.e.,~a product state \cite{islam2015measuring}. The VE-protocol could be utilized to generate a random state $\Ket{\psi}$ from a Mott insulating product state $\Ket{\mathbf{1}}$, as a variant of this protocol has been shown to generate unitary$-2$ design \cite{PhysRevA.97.023604}. 

\begin{figure}[h]
\includegraphics[width=0.45\textwidth]{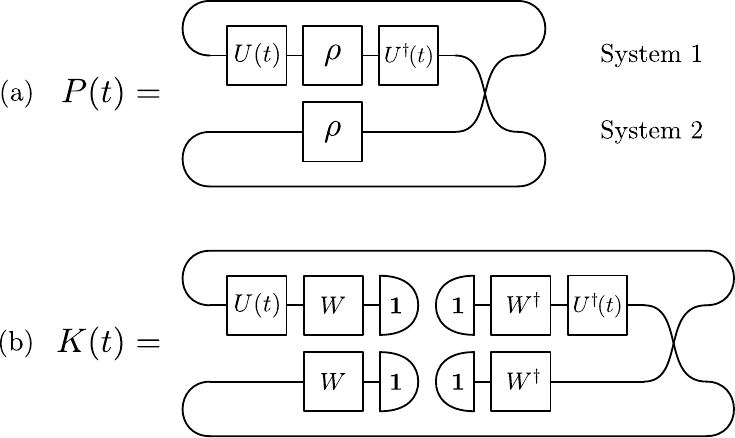}
\caption{Diagrammatical tensor network representation of (a) the survival probability, which is a quantum circuit of a pair of replicated systems, a time evolution operation on system 1 and a swap operator; and (b) the SFF, in terms of survival probability with initial state chosen to be a certain pure state, e.g. a Mott insulating product state $\ket{\mathbf{1}}$, and a set of global rotations $W$ which form a 2-design.}
\label{fig:surv_prob}
\end{figure}

An important ingredient underlying our protocol is the relation between the survival probability and SFF. The survival probability for an initial state $\ket{\psi}$ reads,
\begin{eqnarray}
 P(t, \{U, \ket{\psi} \}) &=&|\Bra{\psi}\hat{U}(t) \Ket{\psi}|^2 \;. \label{survProb}
\end{eqnarray}
Consider replicating the system with the pure initial state $\ket{\psi}$, so that the density matrix of both systems is $\rho = (\ket{\psi} \otimes \ket{\psi})  (\bra{\psi} \otimes \bra{\psi})$. Defining the swap operator $\hat{V}_2$ by the action $\hat{V}_2 \ket{\psi_1} \otimes \ket{\psi_2} = \ket{\psi_2}   \otimes \ket{\psi_1}$, it is possible to express the survival probability in the form
\begin{eqnarray}\label{eq:sp}
 P(t) =\text{Tr}\left[\left(\hat{U}(t) \otimes \mathds{1}\right) \rho 
 \left(\hat{U}^{\dagger}(t) \otimes \mathds{1}\right)
  \hat{V}_2  \right] \;, 
\end{eqnarray}
which has a diagrammatical tensor network representation shown in Fig.~\ref{fig:surv_prob}(a). 

SFF can be expressed as the survival probability for certain initial states.  For instance, $P(t, \{\hat{U}, \ket{\psi}\}) = K(t)$ if 
$\ket{\psi}= D^{-1}\sum_a \ket{E_a} $ is chosen to be the equal-amplitude superposition of the eigenstates $\ket{E_a}$ of $\hat{U}$ or $\hat{H}$ with Hilbert space dimension $D$.
Here, instead, we use an experimentally realizable initial state given by
\begin{equation}
\Ket{\psi}_{\mathrm{CUE}} = \hat{W} \Ket{\mathbf{1}} , \quad \; \; \hat{W}  \in \textrm{CUE}(D) \;.
\end{equation}
so that 
\begin{equation}
 K^{\text{ex}}(t) =   \left\langle  
   \mathbb{E} \left[ 
P(\{\ket{\psi}_{\mathrm{CUE}}, \hat{U}(t) \})
\right]_W
\right\rangle D (D+1) - D \;. \label{Kex}
\end{equation}
where we have used $  \mathbb{E}\left[ \dots \right]_W$ to denote averages over $W$. The circuit to measure $K(t)$ in terms of $P(t)$ is sketched in Fig.~\ref{fig:surv_prob}(b).

Now we show that the survival probability and consequently, the SFF of a many-body system can be experimentally measured utilizing Eq.~\eqref{eq:sp}. A feasible read-out protocol for probing SFF of cold atoms in optical lattices is the interference measurement through the measurement of swap operator $\hat{V}_2$ \cite{PhysRevLett.109.020505,islam2015measuring} which is outlined below:

(i) Initiate two BH chains in a Mott insulator state and apply VE-protocol to both copies with global rotations $\hat{W}$ to obtain the same random state $\Ket{\psi}$. In other words, $\hat{W}=\prod_m^M e^{-i \hat{H}_m \tau}$ where $M$ should be chosen $M\geq 6$ \cite{supplementary}. 

(ii) Freeze one of the copies, and evolve the other one with the Floquet unitary of either two-step or three-step protocol, $\hat{U}$ for an evolution time $t$.

(iii) Switch off the hoping parameters across the 1D chains, and subsequently lower the potential between the two chains to interfere the respective many-body states. This corresponds to beam-splitting operation \cite{islam2015measuring}.

(iv) Measure the parity of each site in the first chain, e.g.,~system 1 in Fig.~\ref{fig:surv_prob}(a) with a quantum gas microscope \cite{2009Natur.462...74B}, and compute $P(t)=\prod_{j \in R}e^{i\pi n_j}$ where $R$ is the entire chain.

(v) Repeat steps (i)-(iv) for a number of different realizations of $\hat{U}$ and $\hat{W}$ at each time $t$ to compute $\left\langle  \mathbb{E} \left[  P(\{\ket{\psi}_{\mathrm{CUE}}, \hat{U}(t) \}) \right]_{\hat{W}} \right\rangle$. 

(vi) Compute SFF via Eq.~\eqref{Kex}.

An alternative proposal for the experimental detection of SFF has been recently introduced in Ref.~\cite{ZollerSFF2020}, using interferometry with a control atom \cite{PhysRevA.68.022302,PhysRevLett.102.170502} for Rydberg gases in tweezer arrays which realize spin models. We comment on the application of interferometry to our systems in the SM, which could be realized by introducing a cavity to implement the controlled coupling of the auxiliary atom to the many-body quantum simulator \cite{Jiang_2008}. Another alternative protocol is proposed for Rydberg gases by utilizing the so-called randomized measurement toolbox in Ref.~\cite{ZollerSFF2021} where measurements are performed on the system of interest after a time evolution and random on-site rotations. Our interference measurement complements this randomized measurement circuit described in Ref.~\cite{ZollerSFF2021}: By introducing a replica of the chain and many-body interfering the copies, we remove the requirement of applying $\hat{W}^{\dagger}$ before read-out. This potentially helps with reducing the total time of a single experimental run.

\begin{figure}[h]
\subfloat{\includegraphics[width=0.49\textwidth]{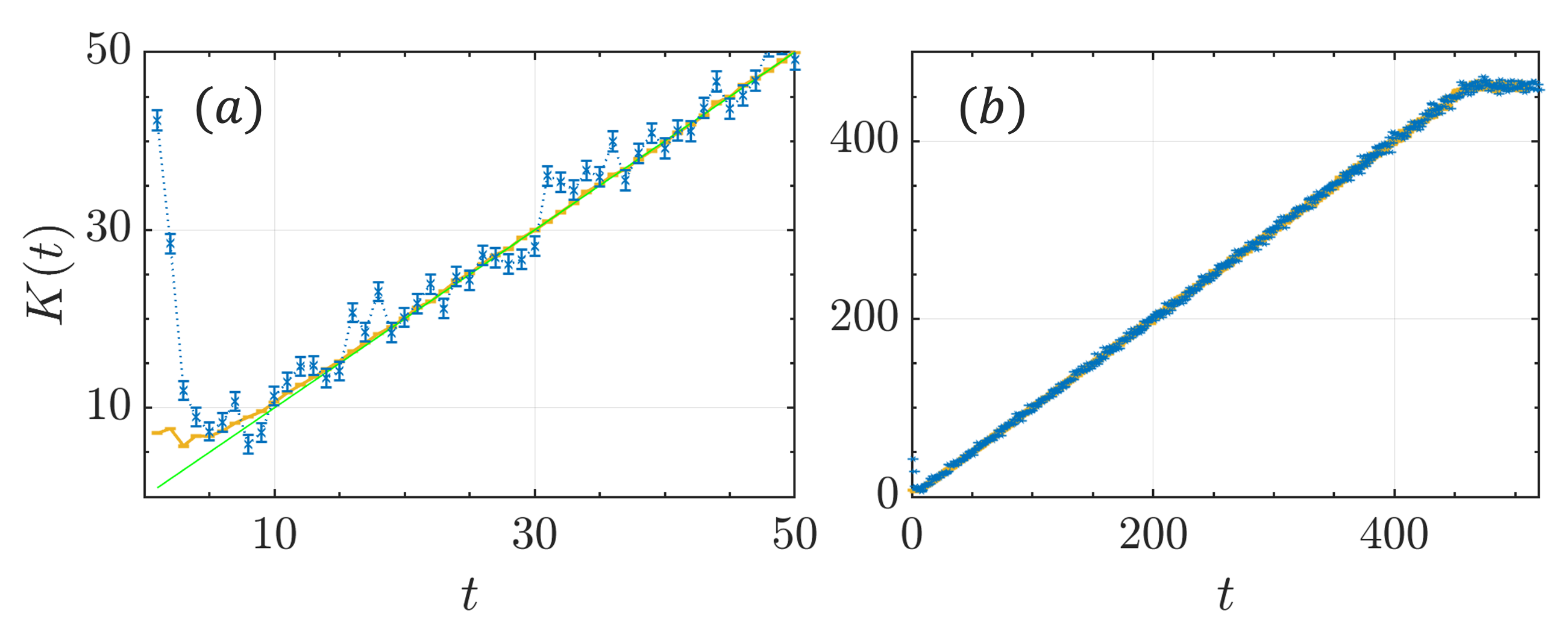}}\hfill
\caption{Eq.~\eqref{Kex}, interference measurement result, (blue, dotted-crosses) is compared to Eq.~\eqref{sffDef}, the theoretical prediction (yellow-solid) for a three-step [CUE] protocol applied to spinless BH model at $N=L=6$. To prepare $W$ operators, we used the VE-protocol with $M=8$. (a) Focus on the early-time SFF where the experimental protocol with $50000$ realization number captures $\tth=8$ and most of early-time SFF well except $t \leq 3$. The green line is the CUE. (b) The experimental protocol captures the ramp and plateau well.} 
\label{Fig6}
\end{figure}

We emphasize that it is important to measure the SFF with its standard normalization, e.g.~Eq.~\eqref{Kex}, instead of a normalization as $0 < K^{\text{ex}}(t)/D^2 \leq 1$ as in \cite{ZollerSFF2020,ZollerSFF2021}. This ensures a sufficient resolution for the bump regime and not to inadvertently suppress the bump amplitude as $D$ increases.

An experimental implication of our results concerns the resolution of $K(\tb < t \lesssim \tth)$ where we would aim to differentiate the bump from the ramp. Eq.~\eqref{bumpEq} provides the difference between bump and ramp amplitudes as
$K(\tb < t < \thei ) - K(t \geq \tth ) \sim \beta t^{1+\xi} \tth^{|\xi|}$. For the bump to be measured, this difference should be greater than the standard deviation of $\mathcal{A}$ measurements. 
Denoting $\mathcal{K}\equiv |\text{Tr}[\hat{U}(t)]|^2$, and the variance  $\sigma_\mathcal{K}(t) \equiv \big[\Braket{\mathcal{K}^2}-\Braket{\mathcal{K}}^2\big]^{1/2}$, an upper bound for $\sigma_{\mathcal{K}}(t)$ must read
\begin{eqnarray}
\frac{\sigma_{\mathcal{K}}(t)}{\sqrt{\mathcal{A}}} &<& \beta t^{1+\xi} \tth^{|\xi|} = \beta t^{1+\xi} N^{\gamma |\xi|}. \label{error}
\end{eqnarray}
The equality follows from assuming a power-law form for the Thouless time scaling in atom number, $N^{\gamma}$. For spatially extended models $\gamma > 1$, whereas for spinor condensates $\gamma < 1$, and $\xi < -1$ based on our analysis in the previous subsections resulting in $\gamma |\xi| > 1$. Although setting $D$ sufficiently small is crucial to be able to resolve $\thei$ and recover the ramp-plateau behavior, we observe that larger atom numbers increase the upper bound for the error in measuring the bump. For the data set depicted in Fig.~\ref{Fig6} for $\nu=1$, Eq.~\eqref{error} requires $\mathcal{A} \sim 5000$ runs of the experiment for a single time point to differentiate the bump from ramp. On the other hand, $\mathcal{A} \sim 2000$ is sufficient for the same system with $N=8$ and $L=5$, which has a similar Hilbert space dimension and $\tth=10$. In passing, let us mention that the error bars in all figures on SFF are found by calculating $\sigma_{\mathcal{K}}(t)/\sqrt{\mathcal{A}}$.

Atomic systems in optical lattices are highly flexible platforms~\cite{bloch2012quantum}. 
Indeed, their parameters, including the on-site interaction and hopping terms, can be readily tuned in the experiment over a wide range of values. 
For instance, it is possible to adjust the hopping parameter $J$ from $J/h \propto2$ Hz preparing a shallow lattice to $J/h \propto150$ Hz constructing a deep lattice~\cite{sherson2010single} where $h$ is the Planck's constant. 
By considering a hopping amplitude of the order of $100$ Hz for the spin$-1/2$ BH model with $N=L=4$ as depicted in Fig.~\ref{Fig1b}(a), typical total evolution times correspond to $\propto 10$~s [$\propto 30$~s] for two [three] step driving protocol. For this system, Thouless time is around $\tth \propto 0.6$~s in both driving protocols which should be experimentally feasible to monitor. 
This is in contrast to the respective Heisenberg times that are attained for $\thei \propto6.6$~s [$\thei \propto 19.8$~s] in the case of a two-[three-]step driving protocol.

Turning to the short-range interacting spin$-1$ condensate and using a spin-spin interaction coefficient $c_1=-2 \pi \times 9$~Hz~\cite{RevModPhys.85.1191} the Thouless time for a three-step driving protocol corresponds to $\tth \approx 3.93$~s and $\tth \approx 2.17$~s in the case of $N=100$ and $N=40$ respectively. 
Along the same lines, the Thouless time for a dipolar spin$-1$ bosonic gas is $\tth \approx 4.43$~s [$\tth \approx 2.54$~s] for $N=100$ [$N=40$]. For a spinor condensate trapped in a single well, a single copy randomized measurement \cite{ZollerSFF2021} might be more feasible than two-copy beam-splitting operation. The condensate could be initialized in a state where all atoms occupy the spin-$0$ hyperfine level and a measurement on the number of atoms in each hyperfine level can be made at the end of the protocol to estimate Eq.~\eqref{survProb}.

\section{\label{sec:conc}Conclusions}

We investigate signatures of quantum many-body chaos in various stroboscopically-driven atomic setups ranging from lattice trapped spinless or spin$-1/2$ bosons to harmonically confined spin$-1$ condensates. 
A particular focus is placed on the behavior of the spectral form factor which was found to feature a universal bump regime and for longer evolution times, the ramp and plateau regimes. The latter is a measure of spectral rigidity, and hence quantum chaos. Our many-body cold atom models, regardless of the locality of the underlying Hamiltonian, hyperfine structure, driving protocols or the symmetry classes, exhibit a universal bump, and suggest a power-law correction to the RMT prediction of the SFF at early experimentally accessible times. Extension of universality from the RMT to the bump regime and the observation that heating times of density populations are significantly shorter than the corresponding $\tth$ highlight the role of $t_b$, the start time of the bump regime collapse. The practical consequences of such a separation in the early timescales in many-body quantum chaos, e.g.,~for information retrieval from a heating system, is an intriguing endeavor. 

We study $\tth$ scaling with respect to relevant system parameters, the atom number and chain size. The Thouless time scaling with atom number is significantly slower in spin$-1$ gases trapped in a single well than the 1D-lattice trapped BH models with $L \geq 3$. This suggests that the spatially-extended cold atom systems take longer to exhibit RMT.
Consequently, the locality of the underlying Hamiltonian is a key condition for the Thouless time scaling and the bump amplitude. Importantly, we find that $\tth$ of spatially-extended systems is more sensitive to the atom number than the lattice size. Namely, the scaling in atom number is larger than in lattice size, and subsequently it takes a denser gas longer to reach RMT. These conclusions highlight the role of atom number for determining the onset of RMT in many-body systems, and therefore the impact of interactions. Consistently, the bump regime is enhanced as the lattice-trapped gas becomes denser. These observations hold true regardless of the hyperfine structure, symmetry classes or the choice of driving protocol. 

Our results suggest the existence of a universal scaling function for the Thouless time depending on both the lattice size and the atom number $\tth(N,L)$. Determining the exact form of this scaling function is an important next step to better understand the nature of many-body quantum chaos in experimentally realizable systems. Constructing \textit{dual} cold atom circuits for this purpose could be a fruitful direction.
Alternatively, one could utilize advanced numerical techniques such as time-dependent density matrix renormalization group \cite{Daley_2004} or multiconfiguration approaches~\cite{cao2017unified} in order to access larger lattice sizes and atom numbers while taking into account relevant many-body correlation effects.

As proposed, the bump regime and Thouless time can be measured with cold atoms in optical lattices with quantum gas microscopy. How the bump regime changes as one deviates from random initial states is an intriguing and potentially useful prospect for experiments. Exploring the signatures of the bump regime and the Thouless time in infinite-temperature correlators and entanglement entropy is an interesting future direction, and can provide a route to understand connections between spectral and spatio-temporal correlations. 
Finally, devising spectroscopic probes of the bump and the Thouless energy in many-body systems is also a fruitful direction, as these features do not require resolving the many-body level spacing.

\section*{Acknowledgements} 

We thank Soonwon Choi, Bertrand Evrard, Wen-Wei Ho, David Huse, Di Luo, Minh Tran, Norman Yao, Sina Zeytinoglu and Peter Zoller for stimulating discussions. 
C.B.D., S.I.M., and H.R.S. acknowledge support from the NSF through a grant for ITAMP at Harvard University. A.C. acknowledges funding support from the PCTS at the Princeton University.

\bibliographystyle{apsrev4-1}

\end{document}


\preprint{APS/123-QED}

\title{Many-body quantum chaos in stroboscopically-driven cold atoms}

\author{Ceren B.~Da\u{g}}
\email{ceren.dag@cfa.harvard.edu}
\affiliation{ITAMP, Center for Astrophysics $|$ Harvard $\&$ Smithsonian Cambridge, Massachusetts 02138, USA}
\affiliation{Department of Physics, Harvard University, Cambridge, Massachusetts 02138, USA}
\author{Simeon I. Mistakidis}
\affiliation{ITAMP, Center for Astrophysics $|$ Harvard $\&$ Smithsonian Cambridge, Massachusetts 02138, USA}
\affiliation{Department of Physics, Harvard University, Cambridge, Massachusetts 02138, USA}
\author{Amos Chan}
\affiliation{Department of Physics, Lancaster University, Lancaster LA1 4YB, United Kingdom}
\affiliation{Princeton Center for Theoretical Science, Princeton University, Princeton NJ 08544, USA}
\author{Hossein R. Sadeghpour}
\affiliation{ITAMP, Center for Astrophysics $|$ Harvard $\&$ Smithsonian Cambridge, Massachusetts 02138, USA}

\maketitle
\onecolumngrid

\setcounter{page}{1}
\setcounter{equation}{0}
\setcounter{figure}{0}
\renewcommand{\thetable}{S\arabic{table}}
\renewcommand{\theequation}{S\arabic{equation}}
\renewcommand{\thefigure}{S\arabic{figure}}
\setcounter{secnumdepth}{2}

\section*{\label{AppD}Supplementary Note 1: Alternative driving protocol and error function}

\begin{figure}[h]
\subfloat[]{\label{SFig2a}\includegraphics[width=0.33\textwidth]{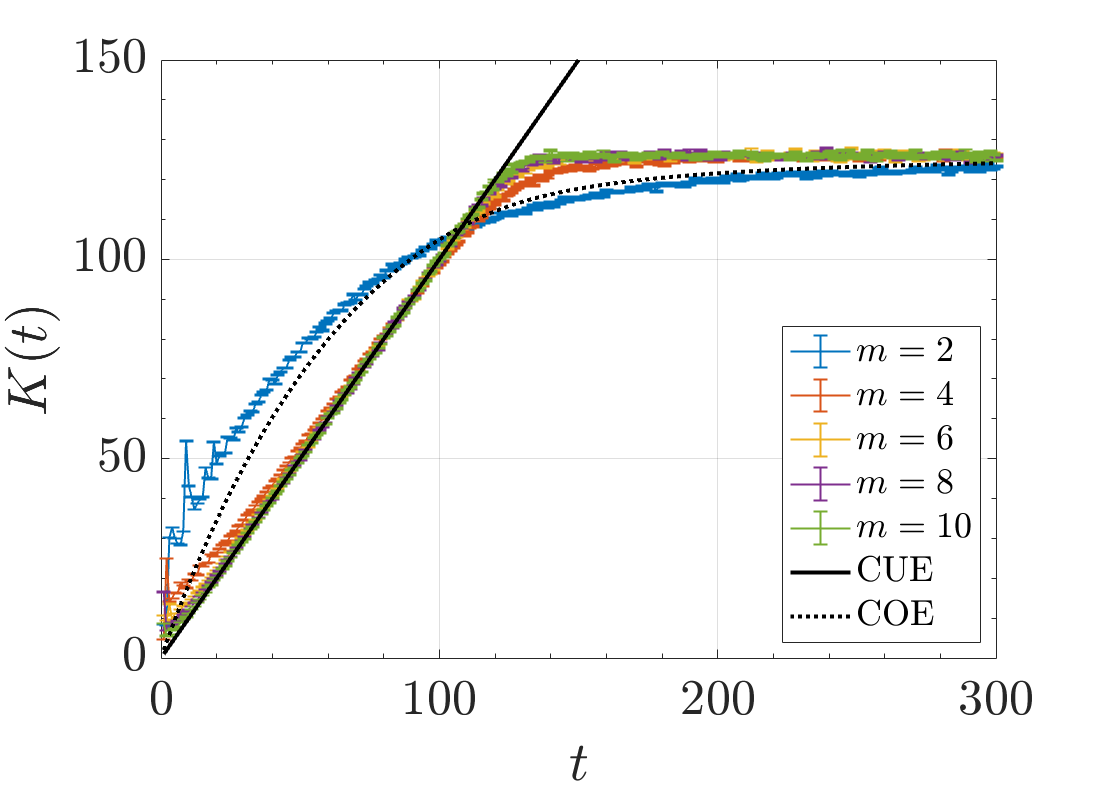}}
\subfloat[]{\label{SFig2b}\includegraphics[width=0.33\textwidth]{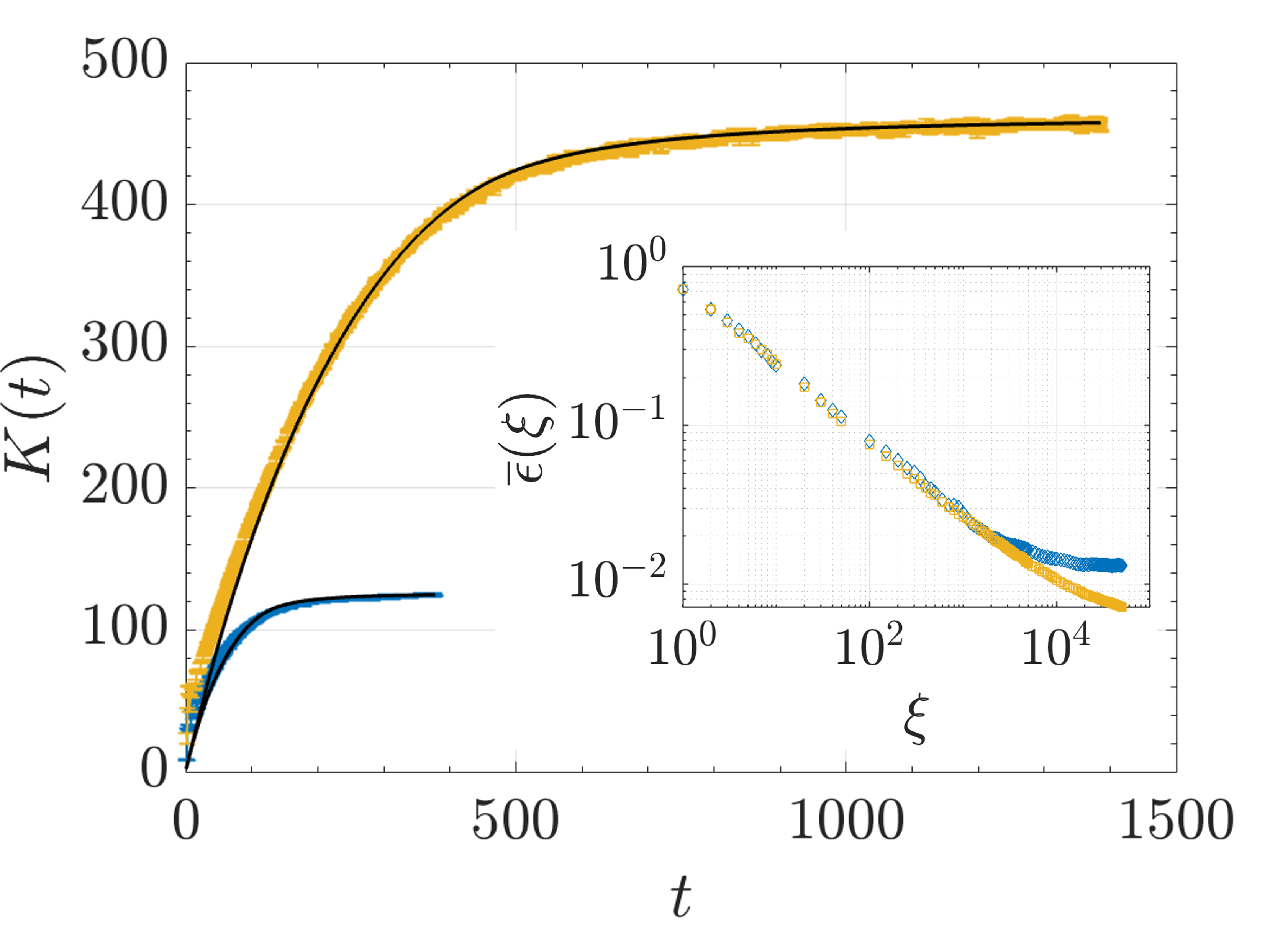}}
\subfloat[]{\label{SFig2c}\includegraphics[width=0.33\textwidth]{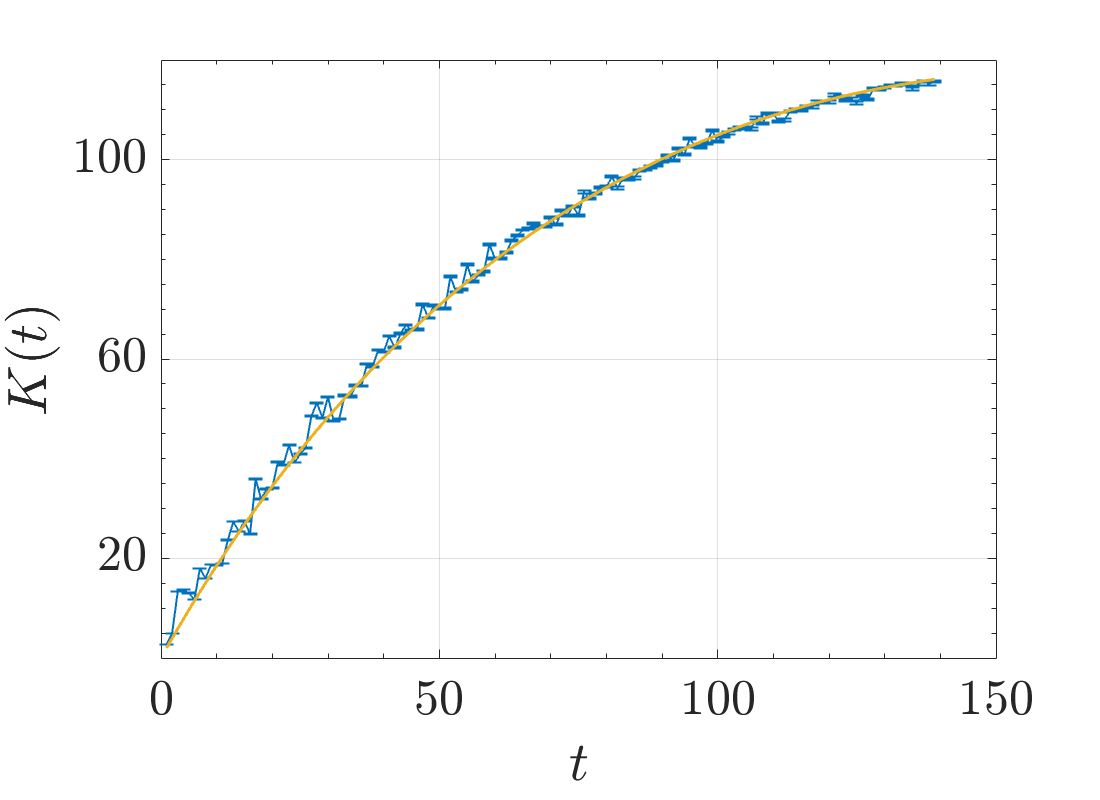}}
\caption{(a) SFF of spinless BH model with $N=L=5$ based on VE-protocol with $m$ specifying the number of quenches with randomized potential. We observe COE for $m=2$ and as $m$ increases the ensemble changes to CUE. (b) The imperfect match for $m=2$ in (a) is a finite-size effect, which is compared here with a larger system $N=L=6$ (yellow). The inset shows the time-averaged error between the data and the COE analytics. The error decreases as $D$ increases. (c) A better match with COE could still be achieved for small lattice sizes, $N=L=5$, if one introduces an additional quench to the two-step ($m=2$) protocol. SFF data is in blue, and the COE prediction is in orange. See the text for details.}
\label{SFig2}
\end{figure}

The Floquet unitary for the VE-protocol reads $\hat{U}_{F,n} = \prod_{m}^M e^{-i \hat{H}_m \tau}$. Here the Hamiltonian at each step $m$ is randomized in the chemical potential with $\mu_{r}^{(m)}=[-J,J]$, while the rest of the system parameters are held constant $u=J$. We find that for a fixed  driving period $\tau$ at each step, having two quenches ($M=2$) in the driving protocol gives rise to circular orthogonal ensemble (COE), and as one increases the number of quenches we obtain circular unitary ensemble (CUE), Fig.~\ref{SFig2a}. One could notice that the spectral form factor (SFF) behavior for $m=2$ does not completely match with COE in Fig.~\ref{SFig2a}. This is a finite-size effect. We calculate the SFF for exactly the same parameters but different lattice size, $N=L=6$ in Fig.~\ref{SFig2b} (yellow). As it can be seen, SFF now approaches the COE prediction. To quantify this approach, we calculate the error between the data and the COE analytics, see the inset of Fig.~\ref{SFig2b}. 
Throughout the study we use an error function that depends on time and the realization number,
\begin{eqnarray}
\epsilon(t,\xi) =\frac{ |K(t)-K_{\text{RMT}}(t)|}{K_{\text{RMT}}(t)}.\label{errorFunc}
\end{eqnarray}
This error function is also utilized to approximately determine the Thouless time at a set $\xi$. When $\epsilon(t^*,\xi) \leq \epsilon$ where $\epsilon$ is a set threshold, $t^*=t_{\textrm{Th}}$. The error calculated in the inset of Fig.~\ref{SFig2b} is the time-averaged error. We define it as
\begin{eqnarray}
\overline{\epsilon}(\xi) = \frac{1}{t_L-t_{\textrm{Th}}} \int_{t_{\textrm{Th}}}^{t_L} \epsilon(t,\xi) dt. \label{timeAveragedError}
\end{eqnarray}
Particularly for the inset of Fig.~\ref{SFig2b}, we use an error threshold of $\epsilon=0.05$ in determining the Thouless time.

It is still possible to see a better match with COE for $m=2$ while keeping the lattice size small at $N=L=5$. This could be achieved by introducing another quench to the driving protocol where the alternating Hamiltonians are
\begin{eqnarray}
\hat{H}_{\text{bh}}^{(1)}\lbrace J=1,u=J,\mu=[-J,J]\rbrace\rightarrow \hat{H}_{\text{bh}}^{(2)}\lbrace J=0,u=1,\mu=[-u,u]\rbrace.
\end{eqnarray}
Here the chemical potential is randomized differently at each step. Alternatively one could use
\begin{eqnarray}
\hat{H}_{\text{bh}}^{(1)}\lbrace J,u=J,\mu=[-J,J]\rbrace \rightarrow \hat{H}_{\text{bh}}^{(2)}\lbrace J=0,u,\mu=0\rbrace,
\end{eqnarray}
which is the two-step protocol introduced in the main text. The SFF for the latter can be seen in Fig.~\ref{SFig2c}.

\subsection{Different filling factors with VE-protocol}

\begin{figure}[h]
\subfloat[]{\label{SFig5a}\includegraphics[width=0.24\textwidth]{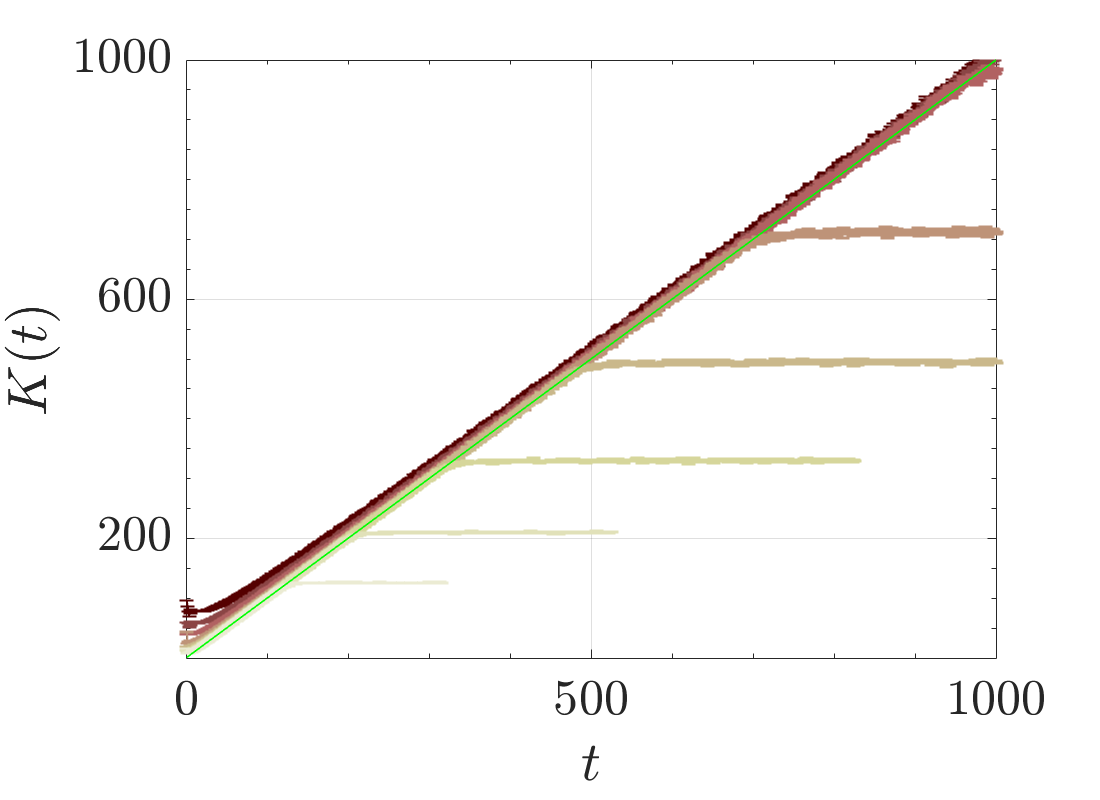}}
\subfloat[]{\label{SFig5b}\includegraphics[width=0.24\textwidth]{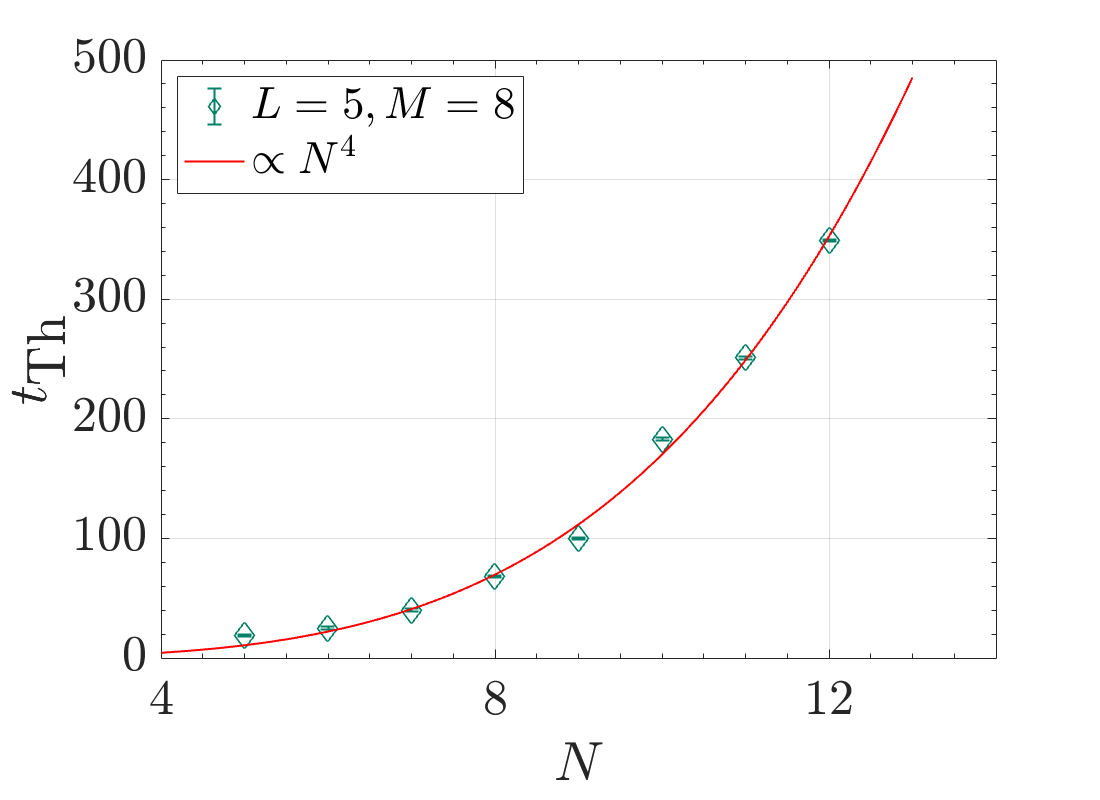}}
\subfloat[]{\label{SFig5c}\includegraphics[width=0.24\textwidth]{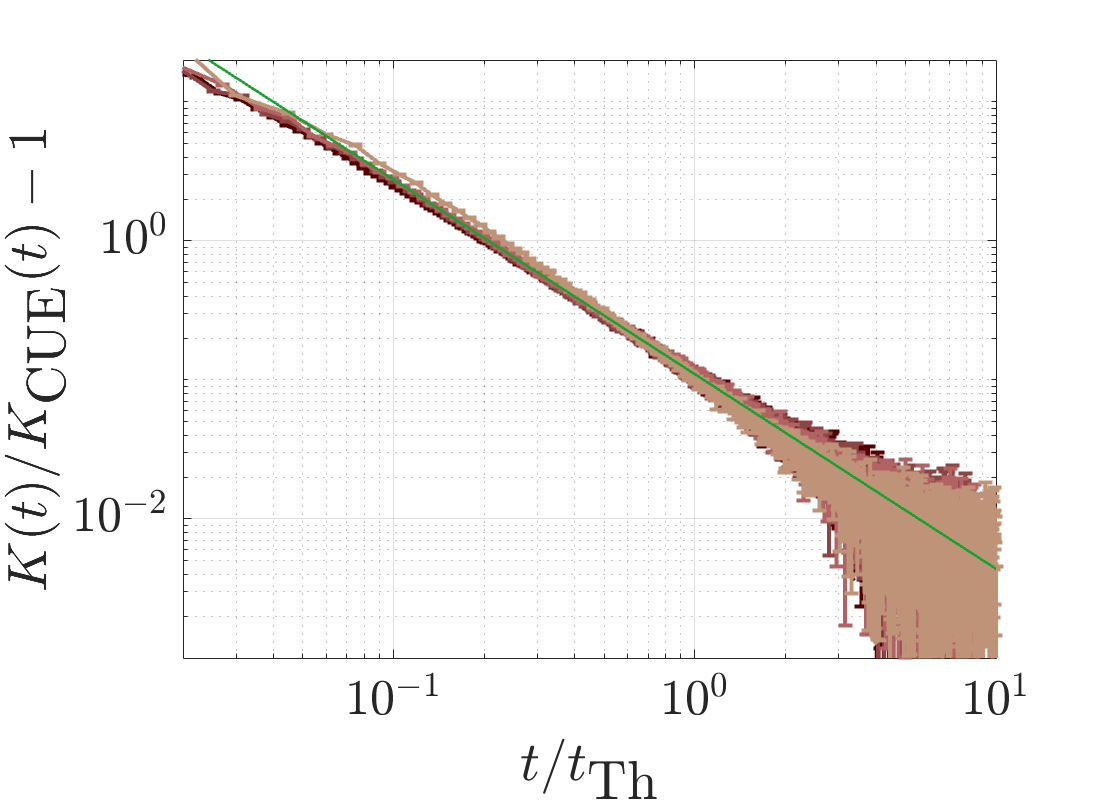}}
\subfloat[]{\label{SFig5d}\includegraphics[width=0.24\textwidth]{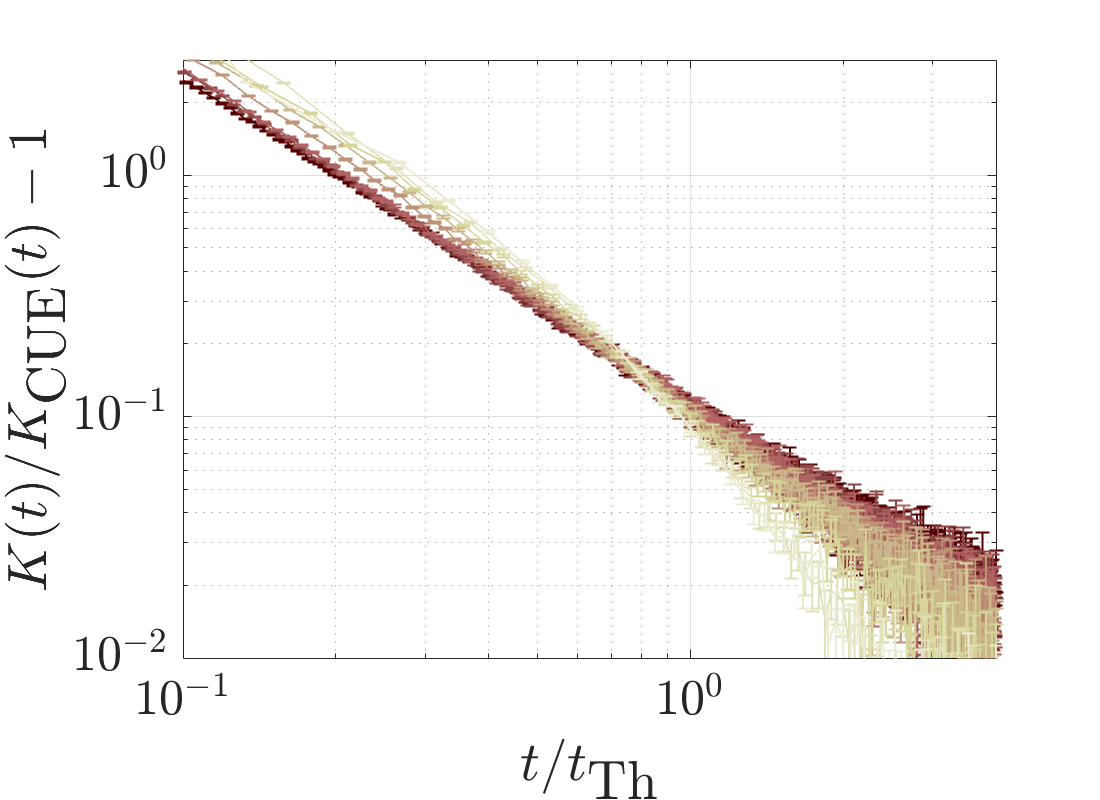}}
\caption{Results for VE-protocol applied to spinless BH model in an optical lattice with $L=5$ wells and $M=8$ number of random quenches. (a) SFF for various $N=5-12$ from light to dark red giving $\nu \geq 1$. (b) Thouless time scaling with atom number $N$ for $\epsilon=0.1$ fitted with a power law with $N^4$. We exclude the first two data points in fits. (c) The scaled SFF with scaled time $t/t_{\textrm{Th}}$ for various different sizes $N=9-12$ from lightest to darkest red. The green line is a power-law fit with $\xi=-1.4$. (d) Focus around $t/t_{\textrm{Th}}\sim 1$ with $L=5-12$, which shows how sufficiently large lattice size data collapse on each other, whereas the data for smallest lattice sizes (yellow) cross the others.}
\label{SFig5}
\end{figure}

\begin{figure}[h]
\subfloat[]{\label{SFig10a}\includegraphics[width=0.24\textwidth]{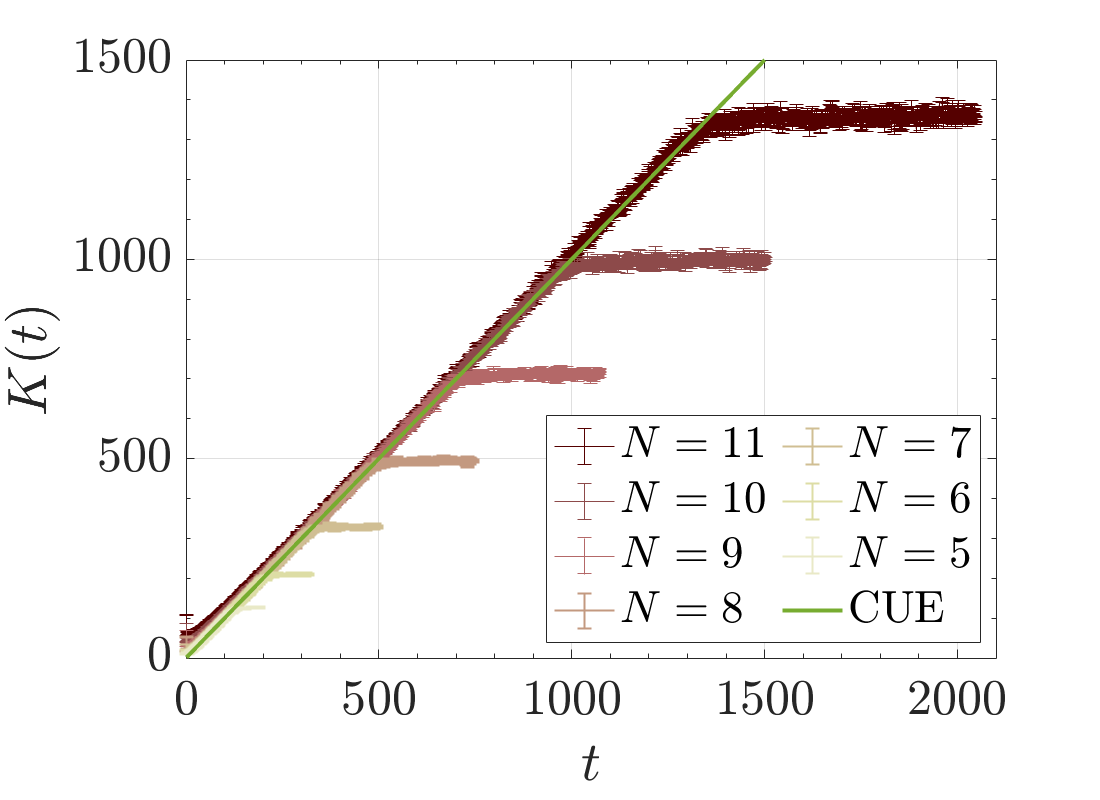}}
\subfloat[]{\label{SFig10b}\includegraphics[width=0.24\textwidth]{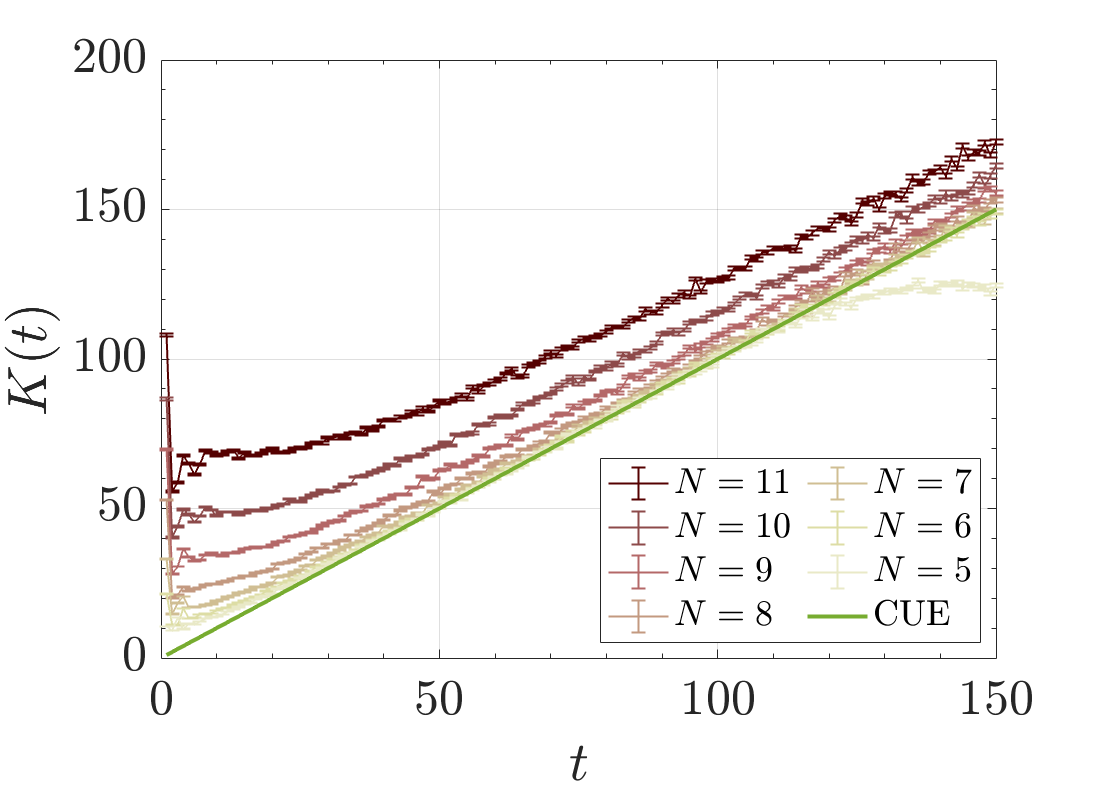}}
\subfloat[]{\label{SFig10c}\includegraphics[width=0.24\textwidth]{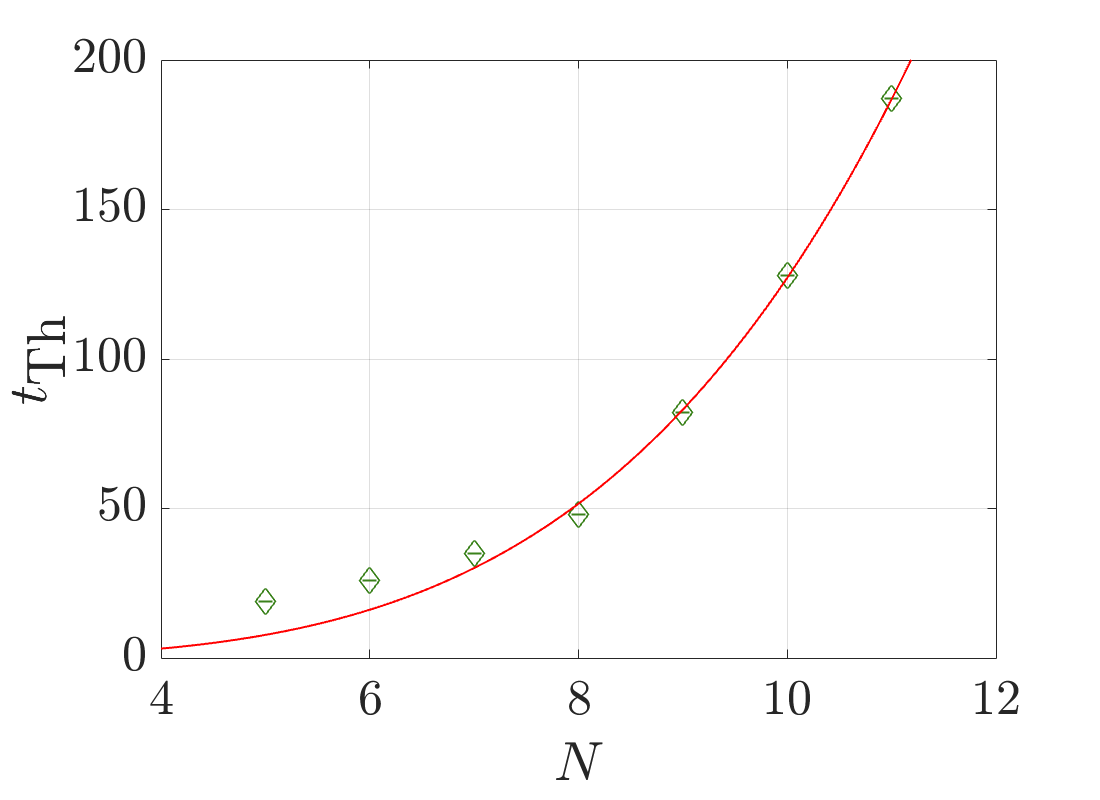}}
\subfloat[]{\label{SFig10d}\includegraphics[width=0.24\textwidth]{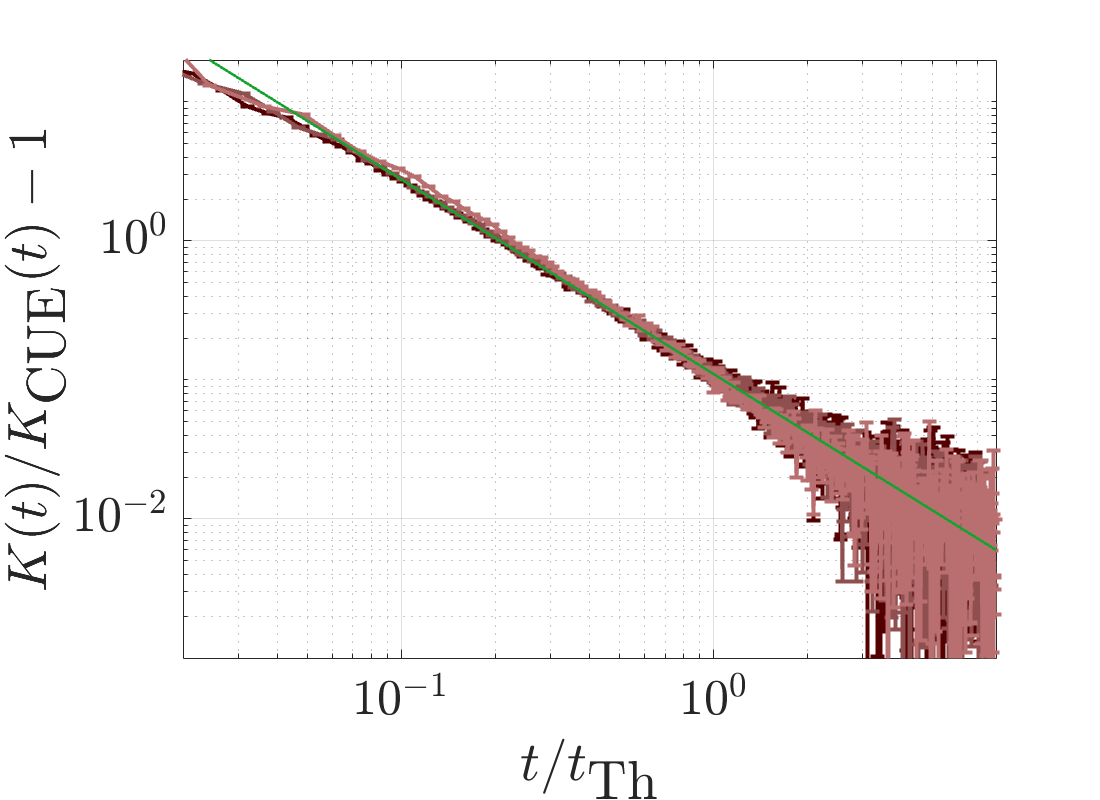}}
\caption{Results for VE-protocol applied to spinless BH model in an optical lattice with $L=5$ wells and $M=6$ number of random quenches. (a) SFF for various $N=5-11$ from light yellow to dark red giving $\nu \geq 1$. (b) SFF focused on the early time to show the bump. (c) Thouless time scaling with atom number $N$ for $\epsilon=0.1$ fitted by a power law with $N^4$. We exclude the first two data points in fit. (d) The scaled SFF with scaled time $t/t_{\textrm{Th}}$ for atom numbers $N=9-11$ from light to dark red. The green line is a power-law fit with $\xi=-1.4$.}
\label{SFig10}
\end{figure}

Next, we investigate the behavior of the SFF of the spinless BH model which is driven according to the VE-protocol. Fig.~\ref{SFig5} presents the case for a fixed lattice size with $L=5$ and increasing atom number $N=5-12$. The number of random quenches is set to $M=8$. Fig.~\ref{SFig5a} shows ``ramp-plateau'' emerging and for early times, one could even notice the bump. $\tth$ scales as $N^4$ which remains to be the same when we change the number of quenches to $M=6$ in Fig.~\ref{SFig10c}. Since the exact scaling function of $\tth$ is not yet obvious, we also perform fitting of the data with other alternative functions, such as an exponential in Figs.~\ref{SFig5b} and~\ref{SFig10c}. Let us note that the VE-protocol results in a particularly enhanced bump regime as can be seen in focus in Fig.~\ref{SFig10b}. This is consistent with the observation of $\tth \propto N^4$. Figs.~\ref{SFig5d} and~\ref{SFig10d} show the rescaled SFF for these cases. In both of them, we see a collapse before $\tth$ too for large enough Hilbert space dimension. Let us also note that as we decrease the number of quenches from $M=8$ to $M=6$ the sharpness at $\thei$ decreases.

\begin{figure}[h]
\subfloat[]{\label{SFig6a}\includegraphics[width=0.33\textwidth]{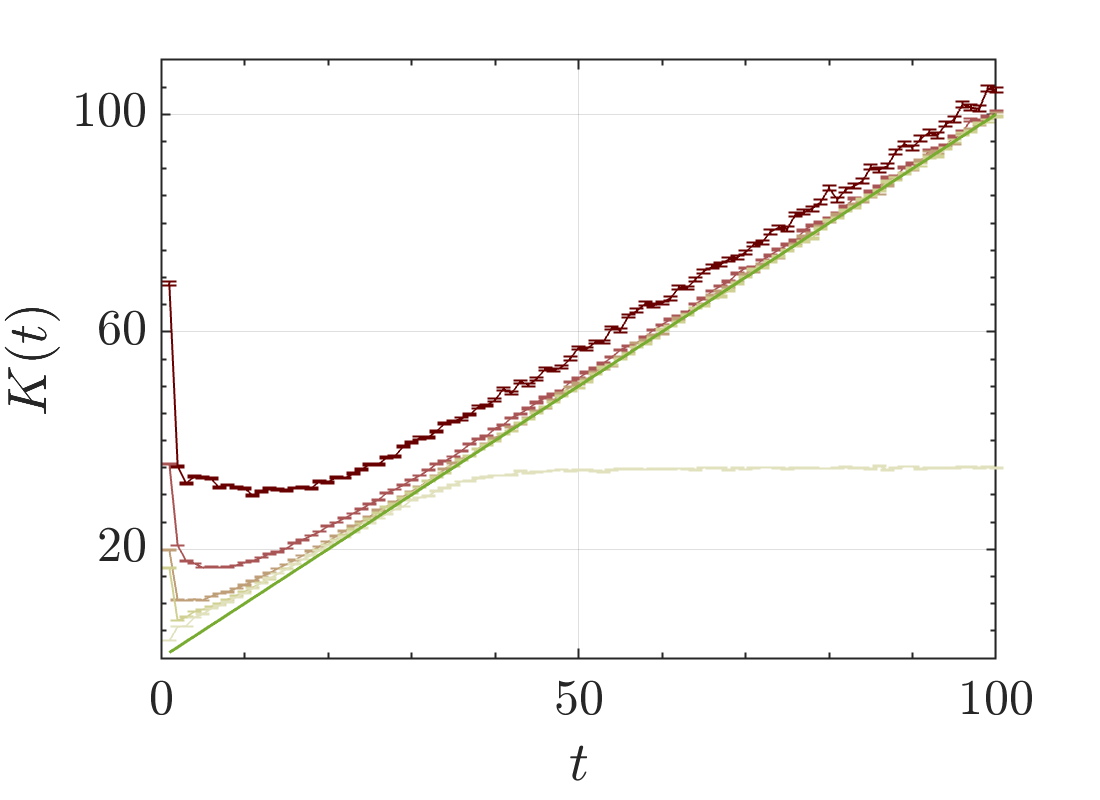}}
\subfloat[]{\label{SFig6b}\includegraphics[width=0.33\textwidth]{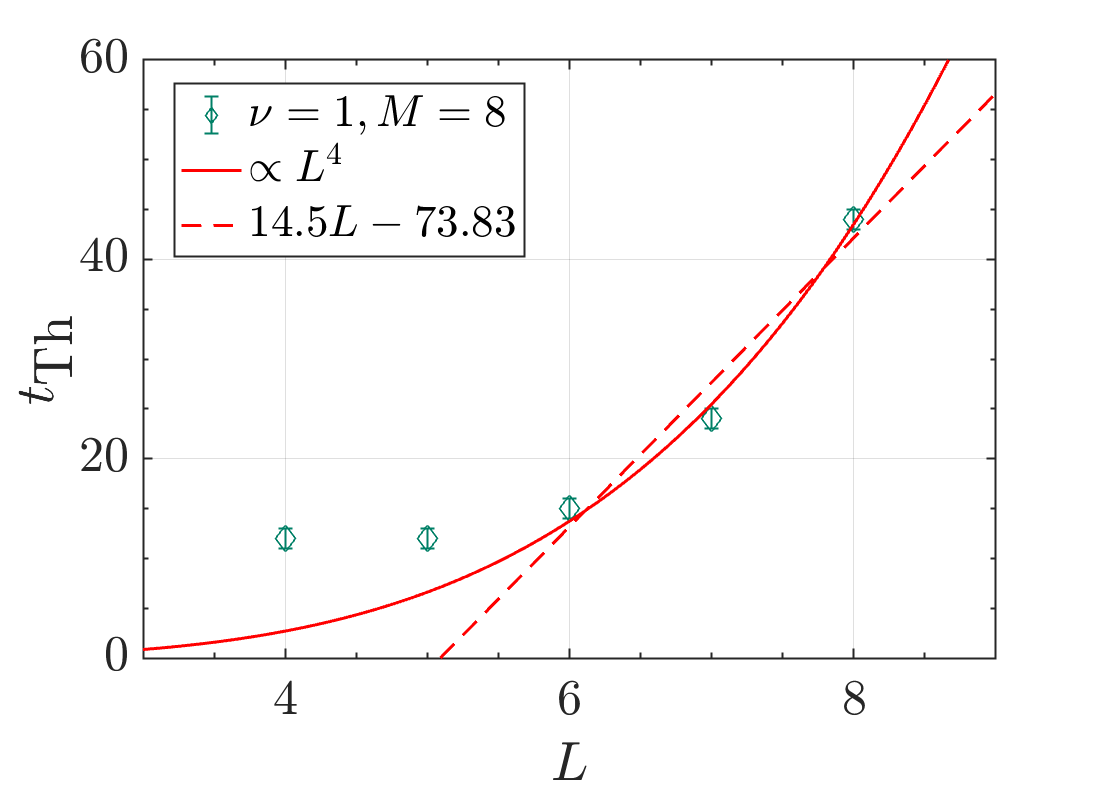}}
\subfloat[]{\label{SFig6c}\includegraphics[width=0.33\textwidth]{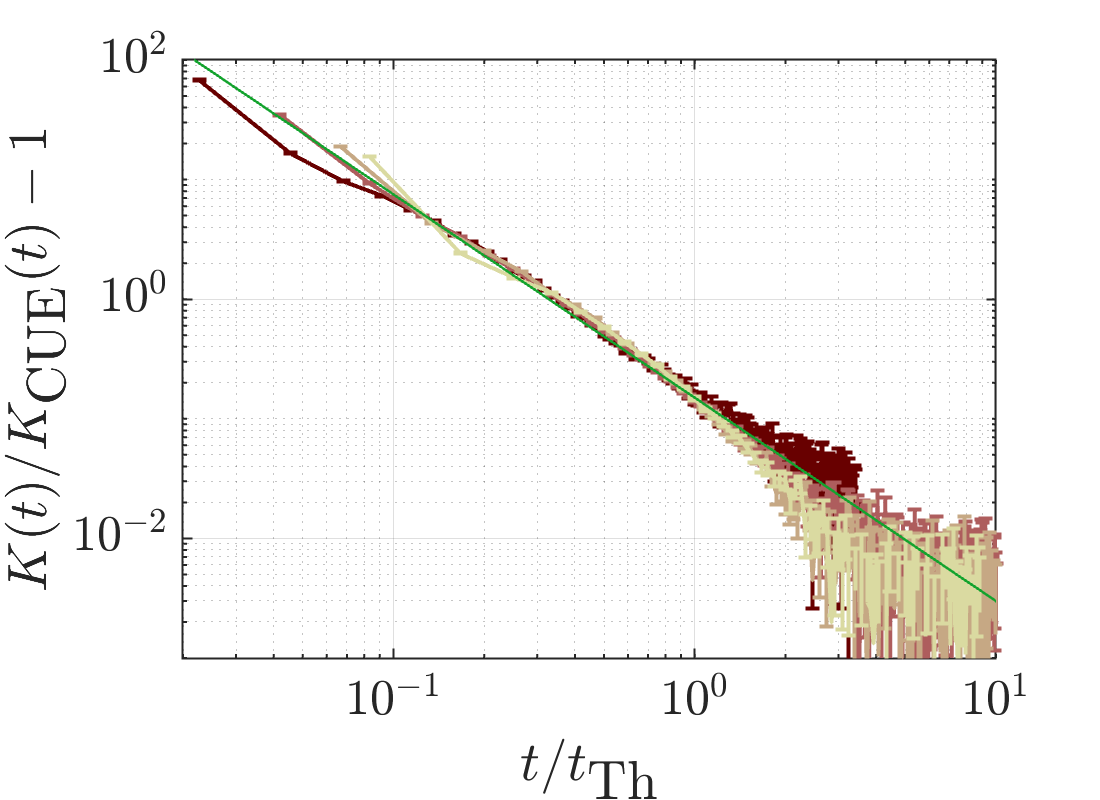}}
\caption{Results for VE-protocol applied to spinless BH model in an optical lattice with unit filling $\nu=1$ and $M=8$ number of random quenches (a) SFF for various $L=N=4-8$ from light to dark red. (b) Thouless time scaling with both the lattice size $L$ and the atom number $N$ for $\epsilon=0.15$ fitted with possible models: power law with $L^4$ and linear. We exclude the first two data points in fits. (c) The scaled SFF with scaled time $t/t_{\textrm{Th}}$ for various different sizes $L=N=5-8$ from yellow to dark red. The green line is the power-law fit with $\xi=-1.7$.}
\label{SFig6}
\end{figure}

Now we apply the VE protocol to a case where the filling factor is set to $\nu=1$. Fig.~\ref{SFig6a} shows a clear signature of bump that becomes more pronounced as $L$ and $N$ increase proportionally. Fig.~\ref{SFig6b} shows possible $\tth$ scaling functions: power-law with $N^4$, exponential $\exp(0.56L)$ or linear. Due to the finite sizes, it is not possible to pin down the exact scaling function. Nevertheless, we observe the proposed power-law correction to the random matrix theory (RMT) for this set of parameters too in Fig.~\ref{SFig6c}. 

\begin{figure}[h]
\subfloat[]{\label{SFig7a}\includegraphics[width=0.24\textwidth]{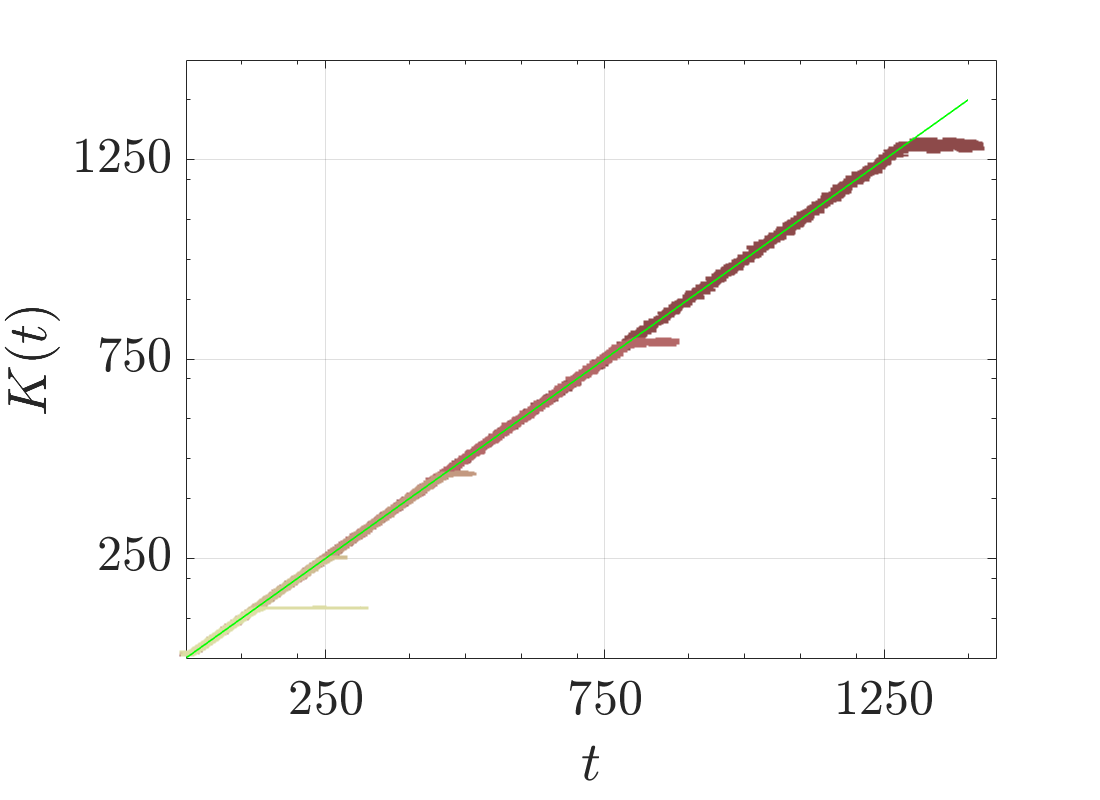}}
\subfloat[]{\label{SFig7b}\includegraphics[width=0.24\textwidth]{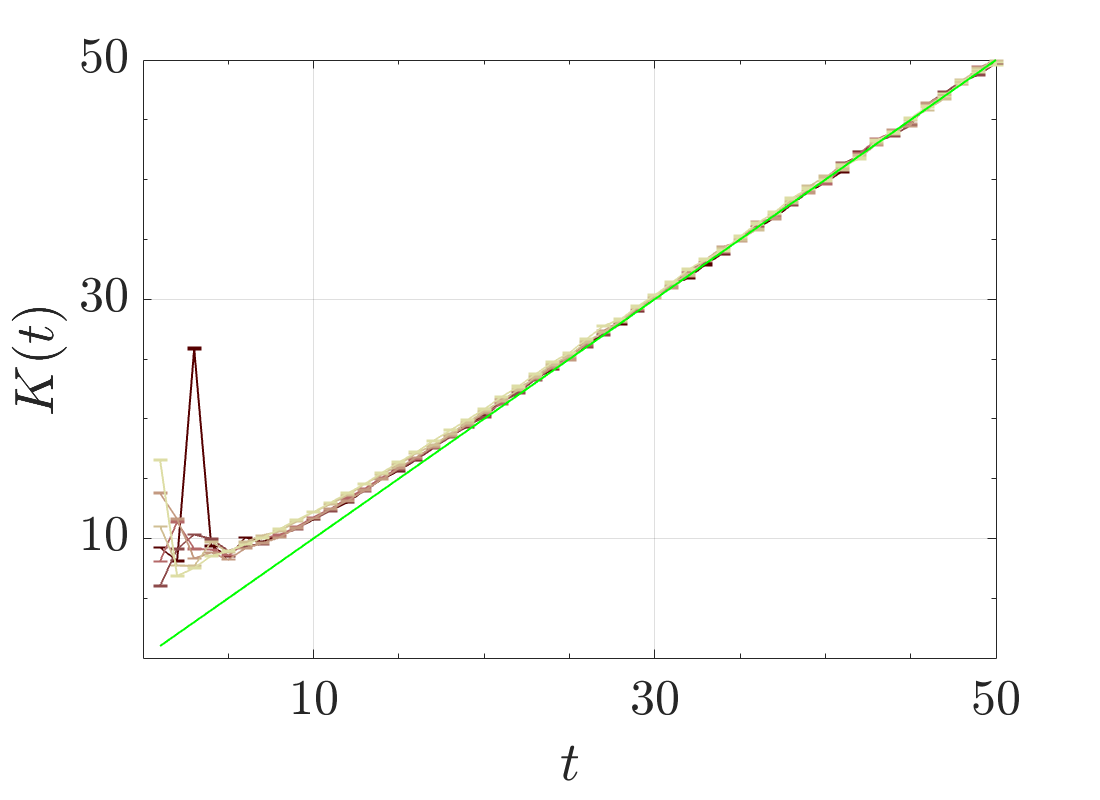}}
\subfloat[]{\label{SFig7c}\includegraphics[width=0.24\textwidth]{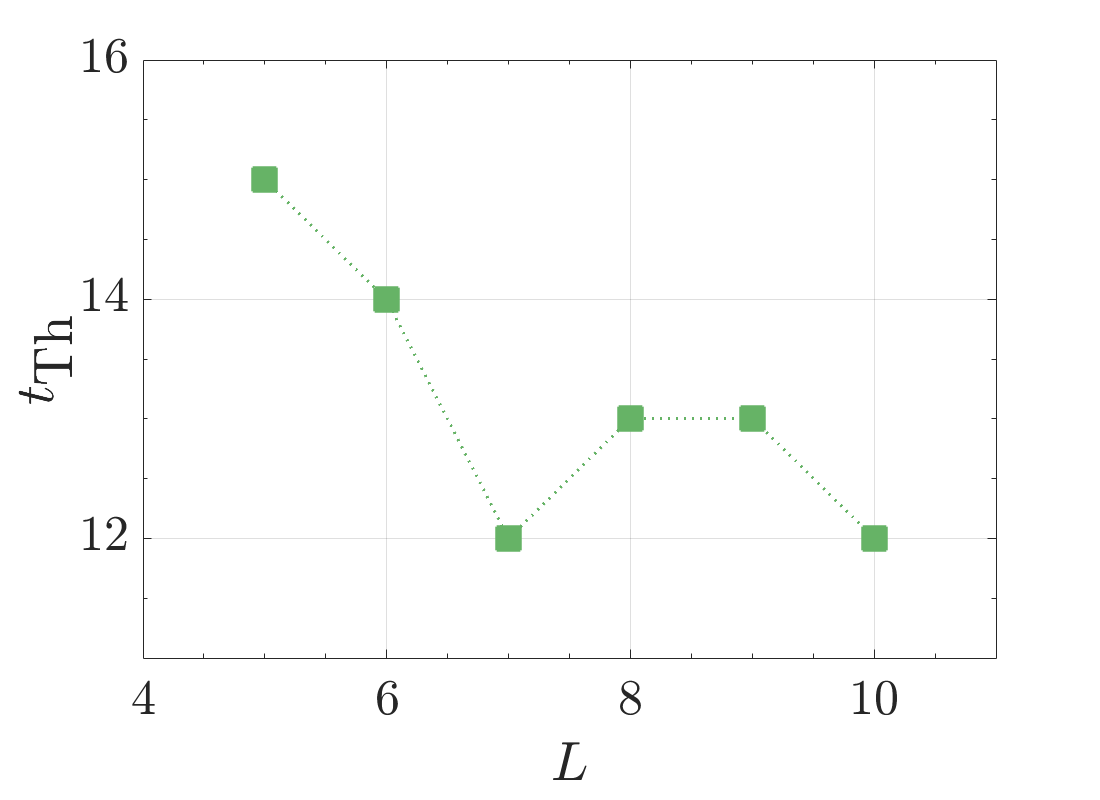}}
\subfloat[]{\label{SFig7d}\includegraphics[width=0.24\textwidth]{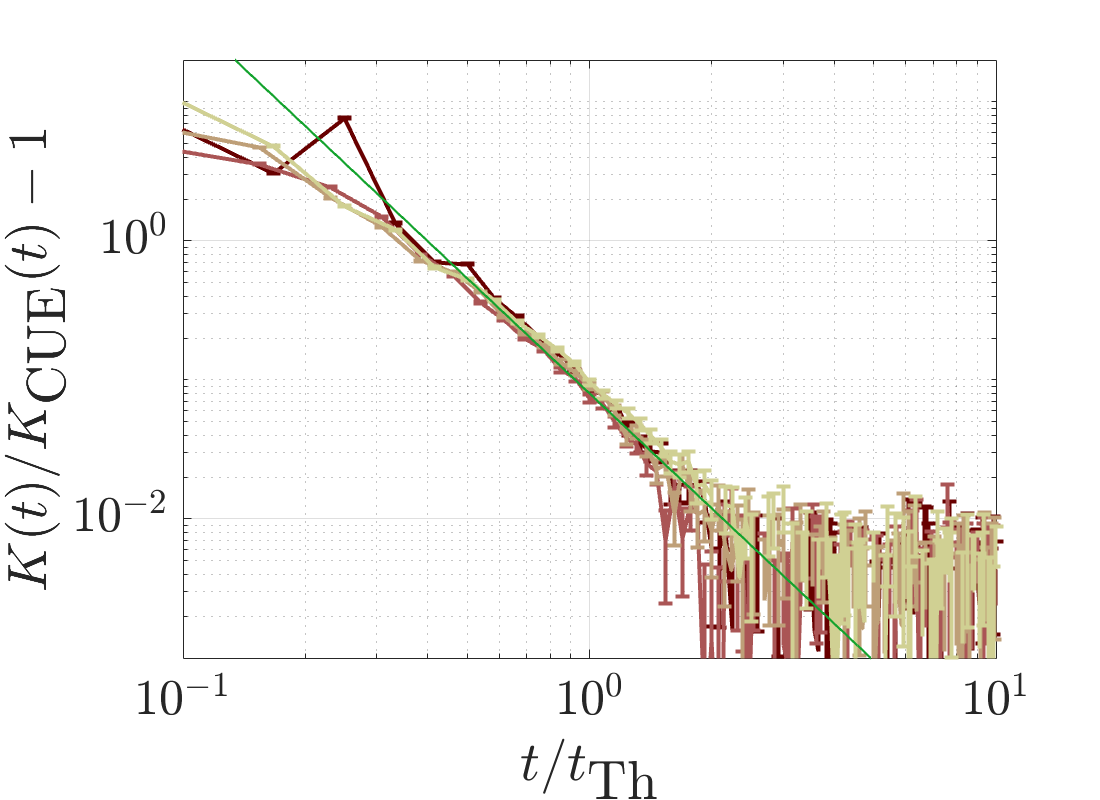}}
\caption{Results for VE-protocol applied to spinless BH model in an optical lattice with $N=5$ number of atoms and $M=8$ number of random quenches. (a) SFF for various $L=5-10$ from light to dark red exhibiting the SFF of CUE (green-solid) and giving $\nu \leq 1$. (b) Focus on bump regime and Thouless time on SFF. (c) Although the bump is enhanced as the lattice size $L$ increases, the Thouless time does not scale with $L$.  $\epsilon=0.1$ is used for data. (d) The scaled SFF with scaled time $t/t_{\textrm{Th}}$ for various different lattice sizes $L=5-10$ from yellow to dark red. The green line is the power-law fit with $\xi=-2.75$.}
\label{SFig7}
\end{figure}

Finally we set the number of atoms to $N=5$ and increase only the lattice size. With this set of parameters, we can faithfully obtain ramp and plateau as seen in Fig.~\ref{SFig7a}. The bump in Fig.~\ref{SFig7b} is significantly smaller compared to the previous cases that we discussed, and data for different $L$ seem to already collapse on top of each other for times $0\ll t \leq \tth$. Consistently with this observation, we do not find a scaling with $L$ for $\tth$ in Fig.~\ref{SFig7c}. To find $\xi$, we also demonstrate the rescaled SFF in Fig.~\ref{SFig7d}.

\newpage
\section*{\label{AppA}Supplementary Note 2: Interferometric measurement scheme for SFF}

We propose a measurement of SFF through an interference scheme in the main text, which utilizes an already developed experimental technology for cold atoms in optical lattices. Here, we detail an alternative measurement scheme which is interferometry with a control atom. This method can measure
\begin{eqnarray}
K^{\text{ex}}(t) &=& \overline{ \Braket{\Bra{\psi}\hat{U}(t)\Ket{\psi}\Bra{\psi'}\hat{U}^{\dagger}(t)\Ket{\psi'}}} \times D^2,  \label{sffExp}
\end{eqnarray}
where $\overline{(\dots)}$ denotes averages over components in \textit{two} independently prepared random states $\ket{\psi} $ and $\ket{\psi'}$, such that no correlation between the two states is introduced. 

(i) Prepare two random initial states, $\Ket{\psi}$ and $\Ket{\psi'}$ to be used in two identical optical lattices which implement the (same) Hamiltonian of interest. If these states are prepared with VE-protocol, then we have to make sure $\Ket{\psi}=\hat{U}_{\text{VE}} \Ket{\mathbf{1}}$ and $\Ket{\psi'}=\hat{U}'_{\text{VE}} \Ket{\mathbf{1}}$ where $\hat{U}_{\text{VE}}\neq \hat{U}'_{\text{VE}}$. Subsequently, initiate two independent control atoms in a superposition state $\Ket{\psi_{c_1}}=(\Ket{0}_{c_1}+\Ket{1}_{c_1})/\sqrt{2}$ and $\Ket{\psi_{c_2}}=(\Ket{0}_{c_2}+\Ket{1}_{c_2})/\sqrt{2}$ where $\Ket{0}$ and $\Ket{1}$ denote the two states of a control atom.

(ii) Couple each control atom to an ensemble of
atoms in the optical lattices and drive the system according to our stroboscopic protocols to prepare the many-body states:
\begin{eqnarray}
\Ket{\psi_1}&=&\frac{1}{\sqrt{2}}\left[\hat{U}(t)\Ket{\psi}\Ket{1}_{c_1} +  \Ket{\psi}\Ket{0}_{c_1}\right],\label{interferometry}\\
\Ket{\psi_2}&=&\frac{1}{\sqrt{2}}\left[\hat{U}(t)\Ket{\psi'}\Ket{1}_{c_2} +\Ket{\psi'}\Ket{0}_{c_2}\right].
\end{eqnarray} 

(iii) Measure the control atoms in $x-$ and $y-$bases to obtain $\Bra{\psi_1}\hat{\sigma}_{c_1}^x\Ket{\psi_1} = \text{Re}\left[\Bra{\psi}\hat{U}(t)\Ket{\psi}\right]$ and $\braket{\hat{\sigma}_{c_1}^y} = \text{Im}\left[\Bra{\psi}\hat{U}(t)\Ket{\psi}\right]$, and similarly for $\Ket{\psi_2}$ followed by conjugation to obtain $\Bra{\psi'}\hat{U}^{\dagger}(t)\Ket{\psi'}$ and eventually calculate $\Bra{\psi}\hat{U}(t)\Ket{\psi}\Bra{\psi'}\hat{U}^{\dagger}(t)\Ket{\psi'}$. 

(iv) Repeat steps (i)-(iii) for a number of different realizations of $\hat{U}$ and with different initial states. Compute Eq.~\eqref{sffExp}.

Step (ii) above could be realized with Rydberg interaction where the states of the control atoms are composed of ground and Rydberg states \cite{PhysRevLett.102.170502}. Such a protocol has been recently proposed to measure the SFF of simulator atoms confined in tweezers that either interact with each other or with the control atom via Van der Waals interactions sequentially \cite{ZollerSFF2020}.
While it is possible to utilize the Rydberg states of atoms confined in an optical lattice to utilize this dipole-blockaded regime, one challenge is the geometry of the setup. Since this method utilizes the Van der Waals interactions, it is important that the control atom is equally distant to simulator atoms \cite{ZollerSFF2020}. In an optical lattice where the atoms are itinerant and extended in 1D-chain, this is hard to achieve. Furthermore, the time evolution of atoms in the optical lattice might be affected by the Van der Waals interaction itself. Perhaps a more optimal implementation of Eq.~\eqref{interferometry} for our systems could be performed through a common cavity mode that couples the control atom and the ensemble of atoms in the optical lattice \cite{Jiang_2008}. In this setup, the states of the control atom $\Ket{0}_{c}$ and $\Ket{1}_{c}$ can be created with hyperfine levels. Both the control and simulator atoms are placed in a cavity, and the cavity photon created conditionally on the control atom would interact with the simulator atoms evolving in the optical lattice.

\begin{figure}[h]
\subfloat[]{\label{SFig1a}\includegraphics[width=0.33\textwidth]{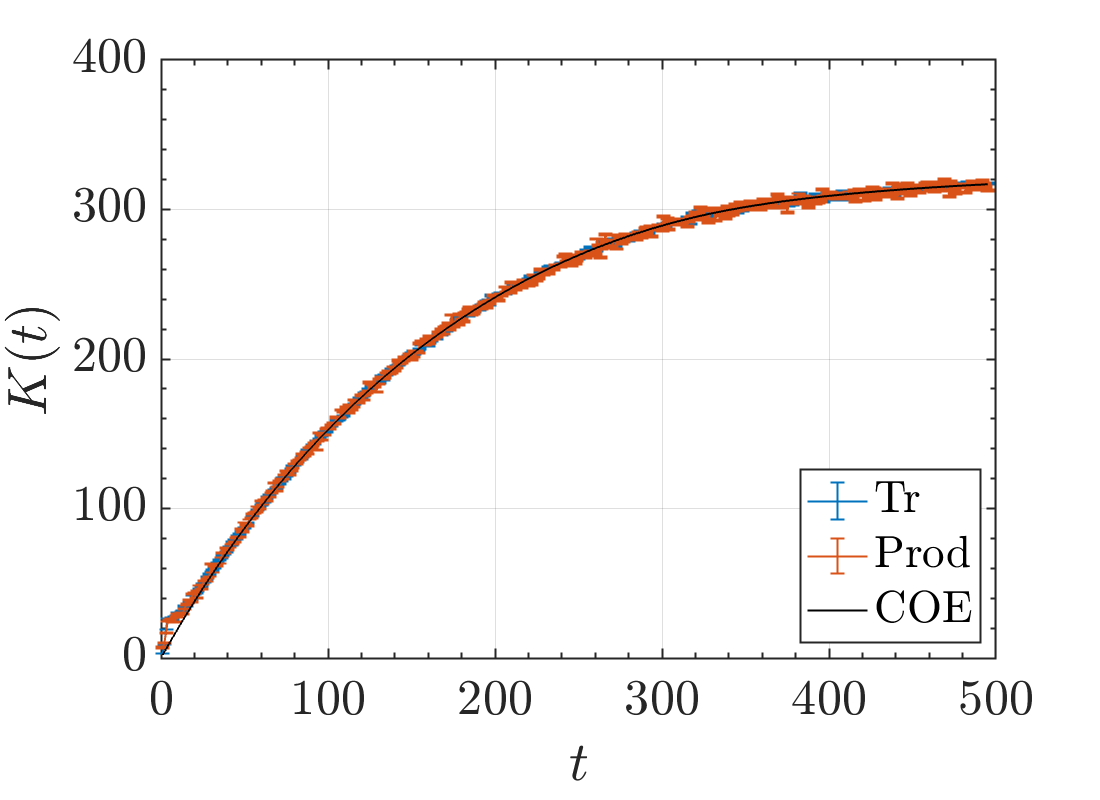}}\hfill
\subfloat[]{\label{SFig1b}\includegraphics[width=0.33\textwidth]{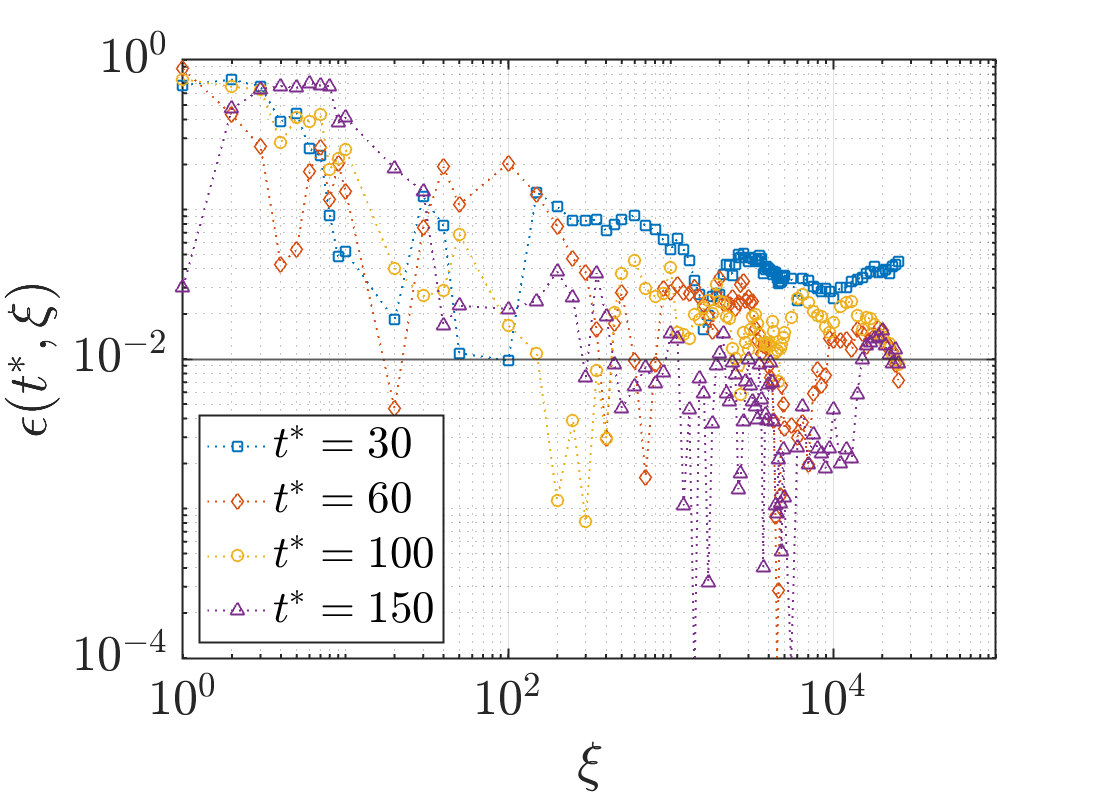}}\hfill
\subfloat[]{\label{SFig1c}\includegraphics[width=0.33\textwidth]{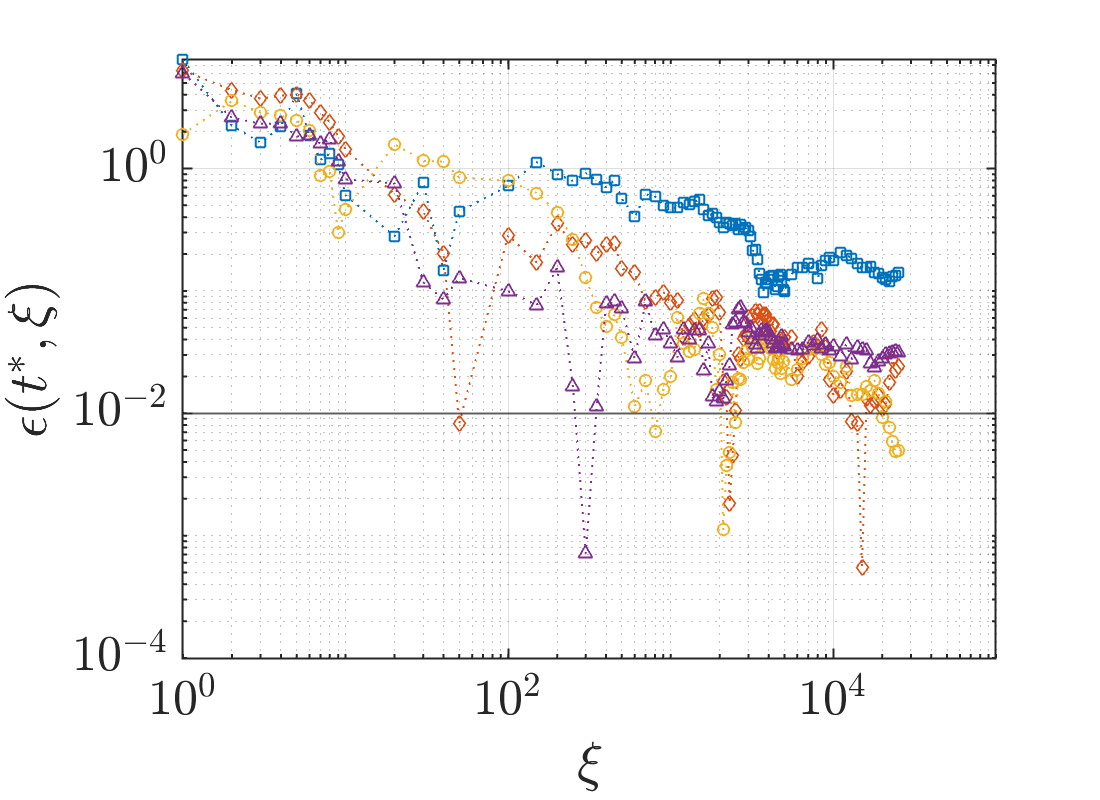}}\hfill
\caption{Spin$-1/2$ BH model with $N=L=4$ and two-step COE protocol. (a) SFF measured with trace (blue) and randomly sampled product states (red) compared to COE prediction (black). No difference is visible. (b) The time-dependent errors at different times during the ramp starting from $t^*=30$ with respect to realization number $\xi$. The error decreases with realization number until $\xi \approx 3\times 10^2$ from where it saturates. Each line is for a different time (see the legend). (c) Same as (b) but with randomly sampled product states. The error decreases with $\xi$ until $\xi \approx 10^3$ slower compared to (b), and they are slightly higher than (a).}
\label{SFig1}
\end{figure}

Let us now elaborate on the initial state preparation. Alternative to generating random states via VE-protocol, we can approximate the trace operation with randomly selected basis states in the Fock space, which has been previously shown to work for quantum chaotic spin systems to compute infinite-temperature out-of-time-order correlator \cite{PhysRevA.99.052322}. In that case, the initial states are set as $\Ket{\psi}=\Ket{\mathbf{n}}$ and $\Ket{\psi'}=\Ket{\mathbf{n'}}$ where $\Ket{\mathbf{n}}$ and $\Ket{\mathbf{n'}}$ are both randomly selected independent Fock basis states at each run of the experiment. To realize a randomly selected Fock basis state in a cold atom simulator, one can first trap the atoms in a Mott insulating regime, and then adiabatically ramp the lattice potential to a shallow lattice which results in an entangled state. By increasing the lattice depth to suppress the tunneling of atoms, one could initiate the system to a randomly-set Fock basis in a controlled fashion. Snapshots of the simulator could be used to confirm that the system is indeed in a Fock basis state. Fig.~\ref{SFig1} systematically compares the SFF calculated with product states to the SFF calculated with trace. 

We calculate the error function $\epsilon(t^*,\xi)$ for selected times $t^* > t_{\textrm{Th}}$. This error helps us see how close the data for the Hamiltonian evolution of a short-range interacting system to random matrix theory prediction gets in Fig.~\ref{SFig1b}. The errors steadily decrease at each shown time and saturates at a value around a realization number $\xi \approx 3\times 10^2$. The error also is smaller later during the ramp (purple-triangle) than the error around the Thouless time (blue-circle), suggesting that the SFF keeps approaching to the RMT prediction. Fig.~\ref{SFig1c} depicts the same physics but with product states randomly sampled from the Fock basis. The latter is experimentally accessible. We observe that the errors still decrease with the realization number, however more slowly, and saturates around $\xi \approx 10^3$. They are also slightly larger than the errors in Fig.~\ref{SFig1b} which implies that the randomly sampled product states indeed only approximate the trace operation.

\newpage
\section*{Supplementary Note 3: Parameter sets relevant to emergence of RMT in driven spin$-1/2$ BH model}

Now we provide numerical evidence for the flexibility of our driving protocol of the spin-$1/2$ BH model. The spinless BH model has a simple driving protocol which alternates between a clean Mott insulating Hamiltonian and a disordered Hamiltonian close to the critical line between a Mott insulator and a superfluid. However, spin-$1/2$ BH model has many more parameters and that enlarges the parameter space that could lead to RMT behavior. Our aim is to find the important parameters 
that determine the path to RMT in spin-$1/2$ BH chains. 

\begin{figure}[h]
\subfloat[]{\label{SFig8a}\includegraphics[width=0.33\textwidth]{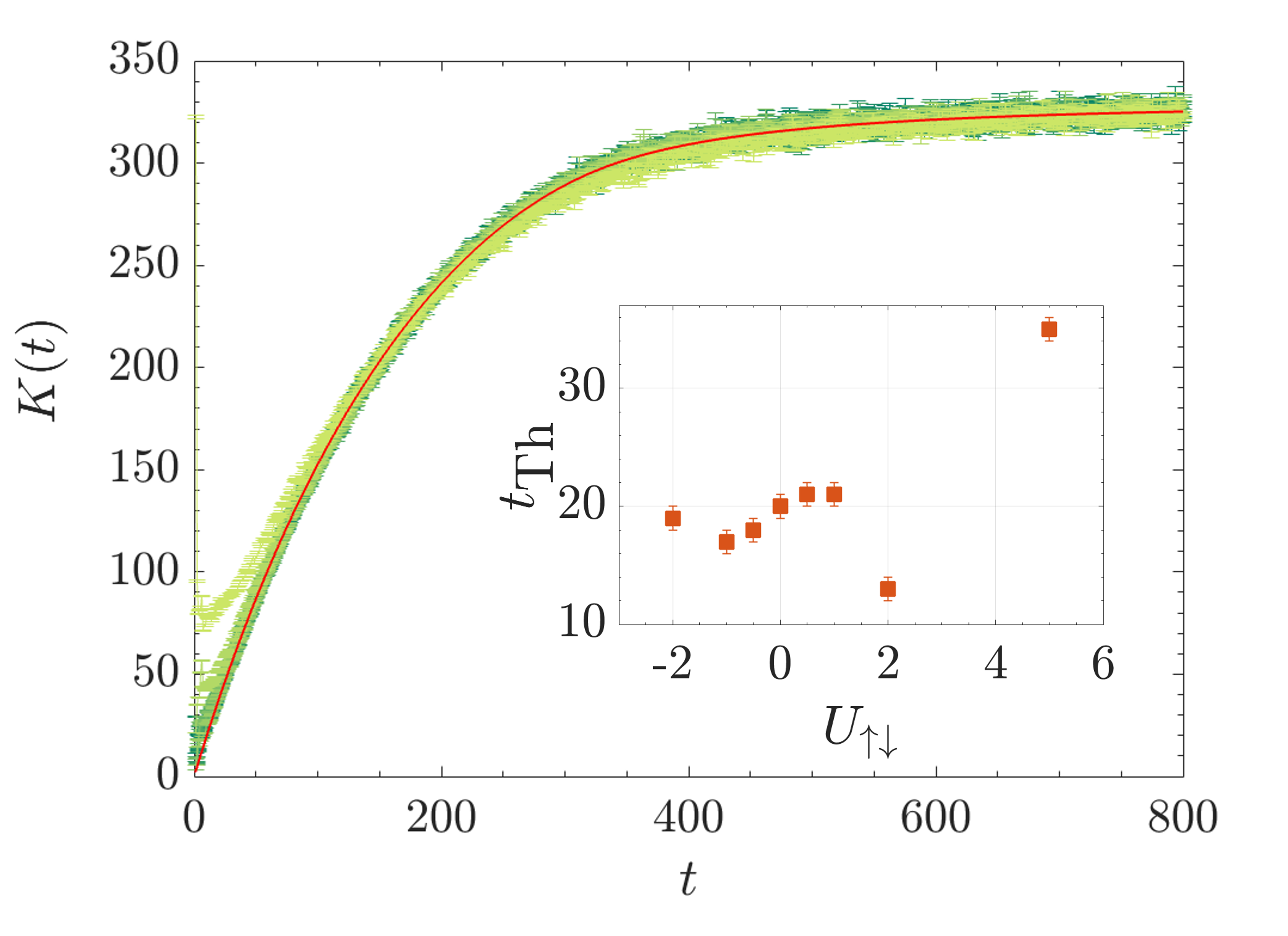}}\hfill
\subfloat[]{\label{SFig8b}\includegraphics[width=0.33\textwidth]{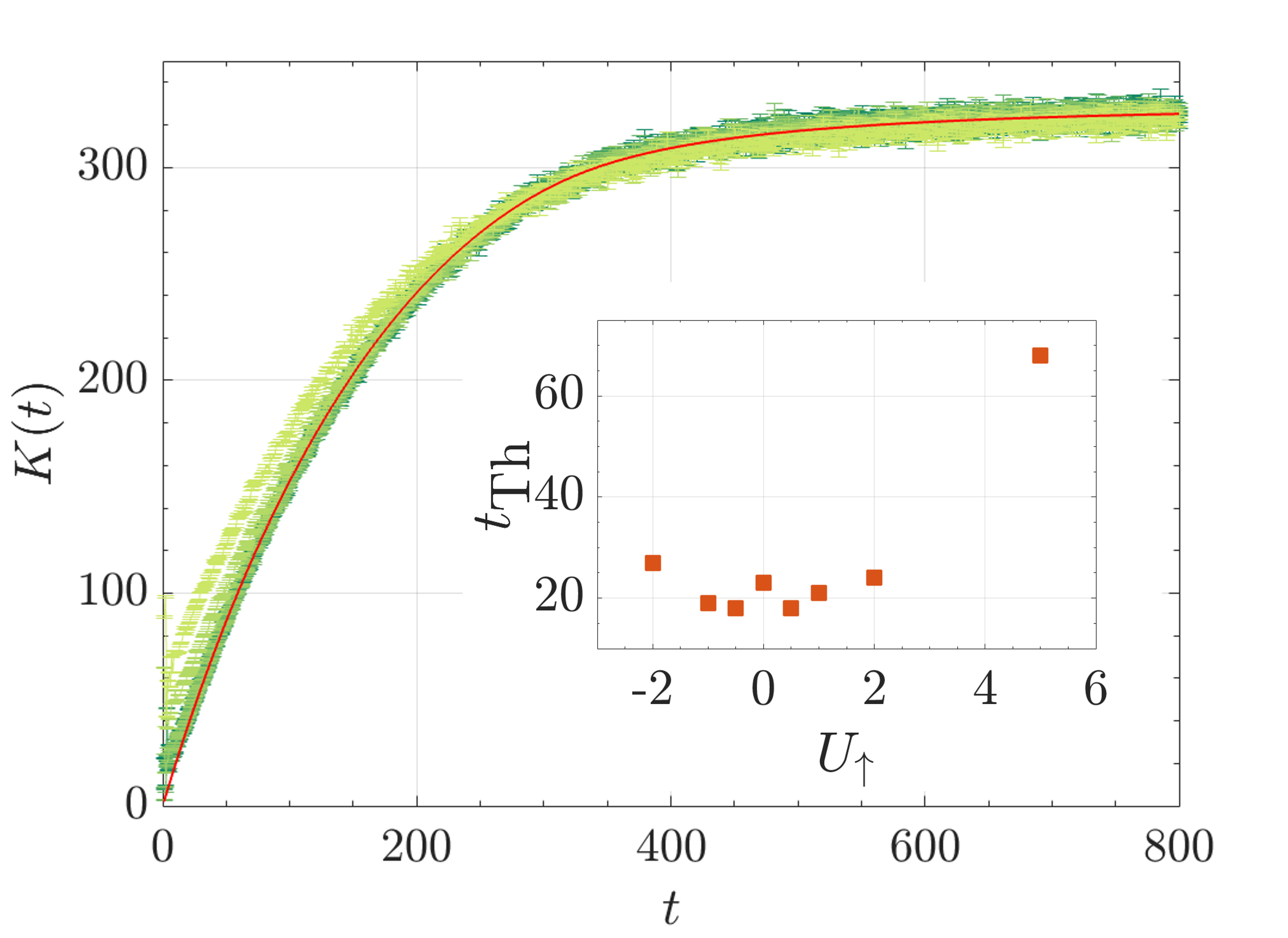}}\hfill
\subfloat[]{\label{SFig8c}\includegraphics[width=0.33\textwidth]{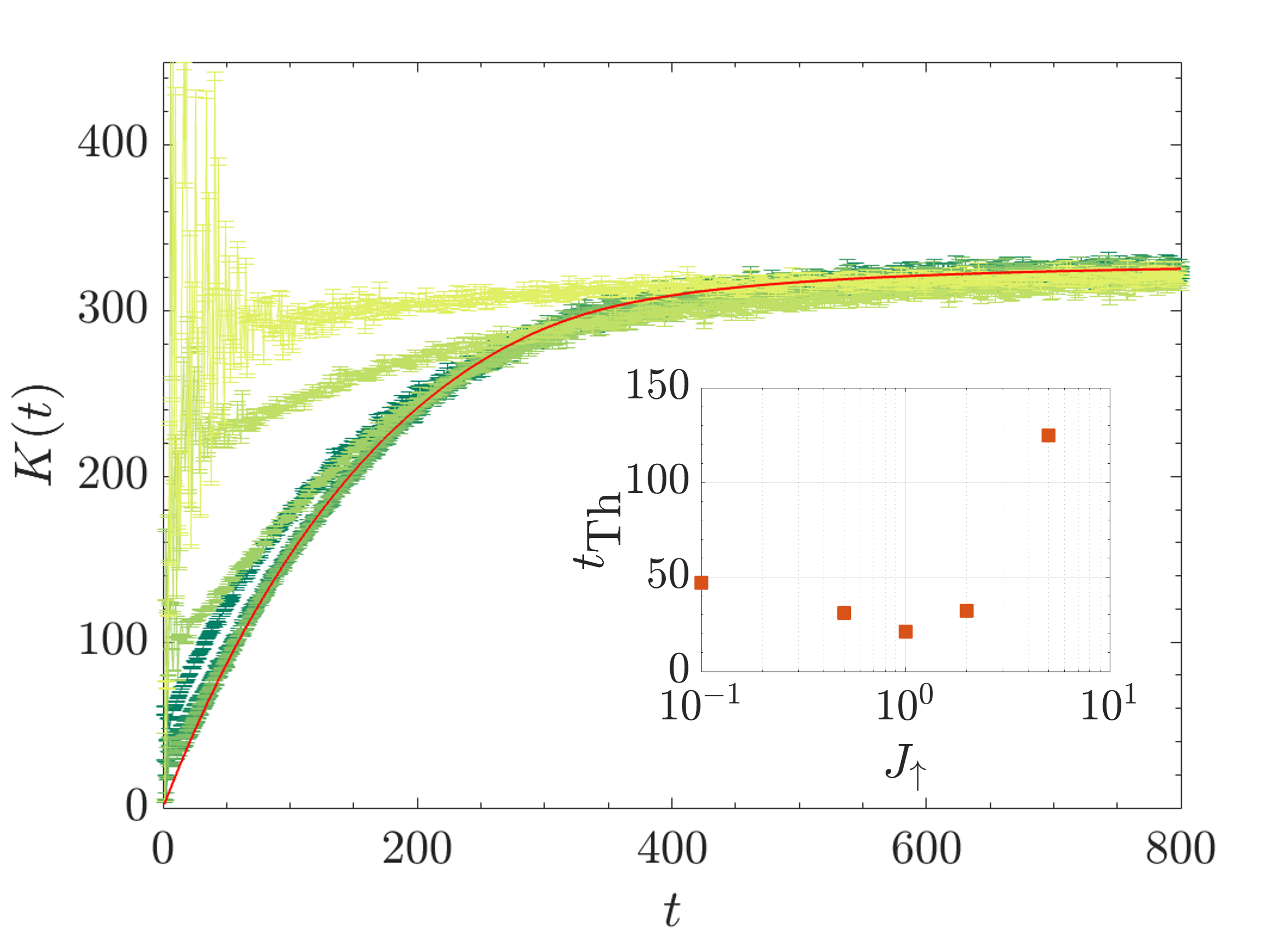}}\hfill
\caption{SFF for spin-$1/2$ BH model with two-step COE protocol at $N=L=4$. (a) SFF of different spin-mixing interactions $u_{\uparrow \downarrow}$ with values $u_{\uparrow \downarrow}=-2,-1,-0.5,0,0.5,1,2,5,10$ from darkest to lightest green in both steps. Inset: Thouless time with respect to spin-mixing interaction when a Thouless time could be defined. (b) SFF of different on-site interactions of one component $u_{\uparrow}$ with values $u_{\uparrow}=-2,-1,-0.5,0,0.5,1,2,5,10$ from darkest to lightest green in both steps. Inset: Thouless time with respect to $u_{\uparrow}$. (c) SFF of different tunneling strengths of one component $J_{\uparrow}$ with values $J_{\uparrow}=0,0.1,0.5,1,2,5,10,20$ from darkest to lightest green in both steps. Inset: Thouless time with respect to tunneling strength $J_{\uparrow}$.}
\label{SFig8}
\end{figure}
\begin{figure}[h]
\subfloat[]{\label{SFig11a}\includegraphics[width=0.33\textwidth]{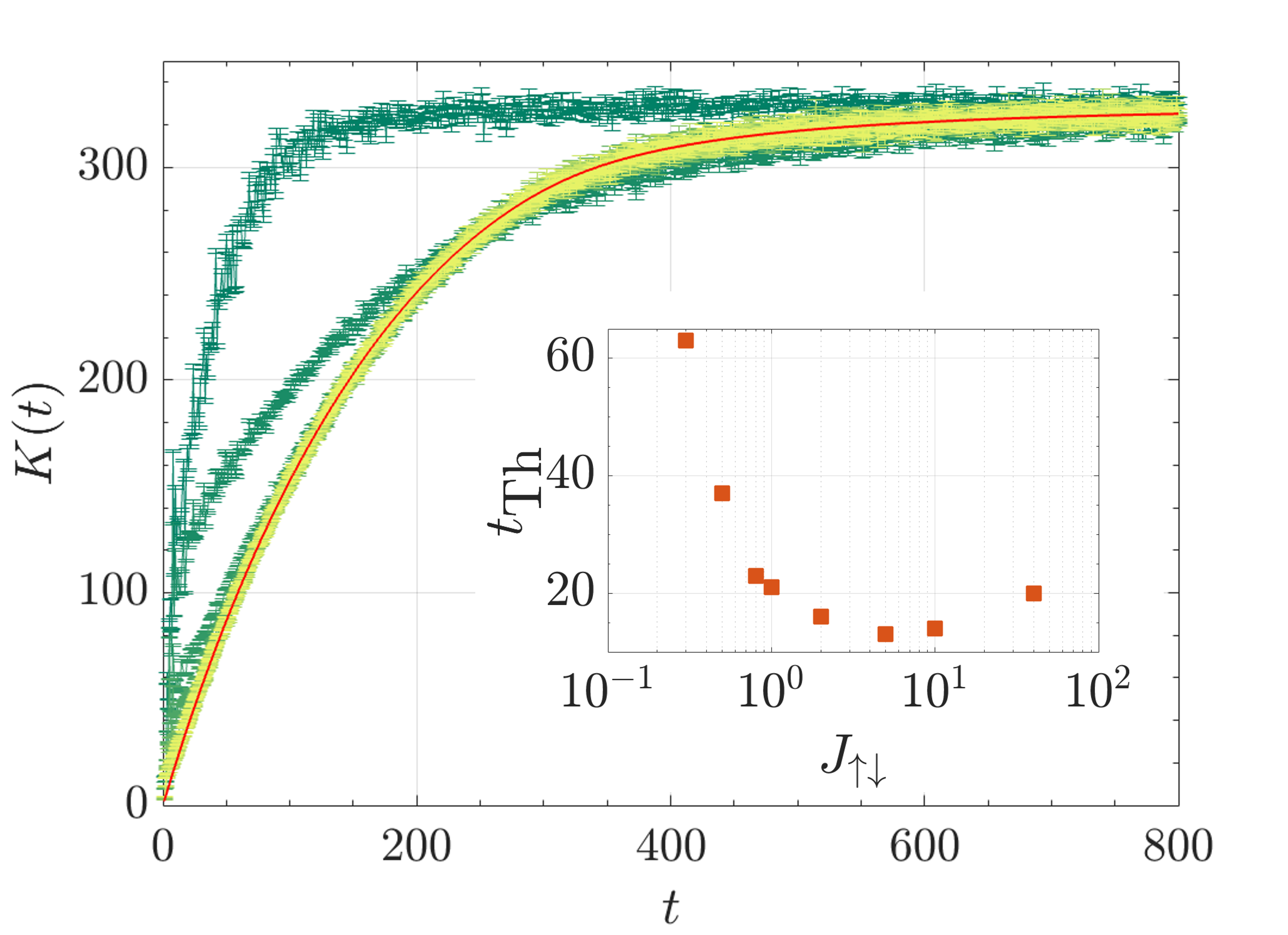}}\hfill
\subfloat[]{\label{SFig11b}\includegraphics[width=0.33\textwidth]{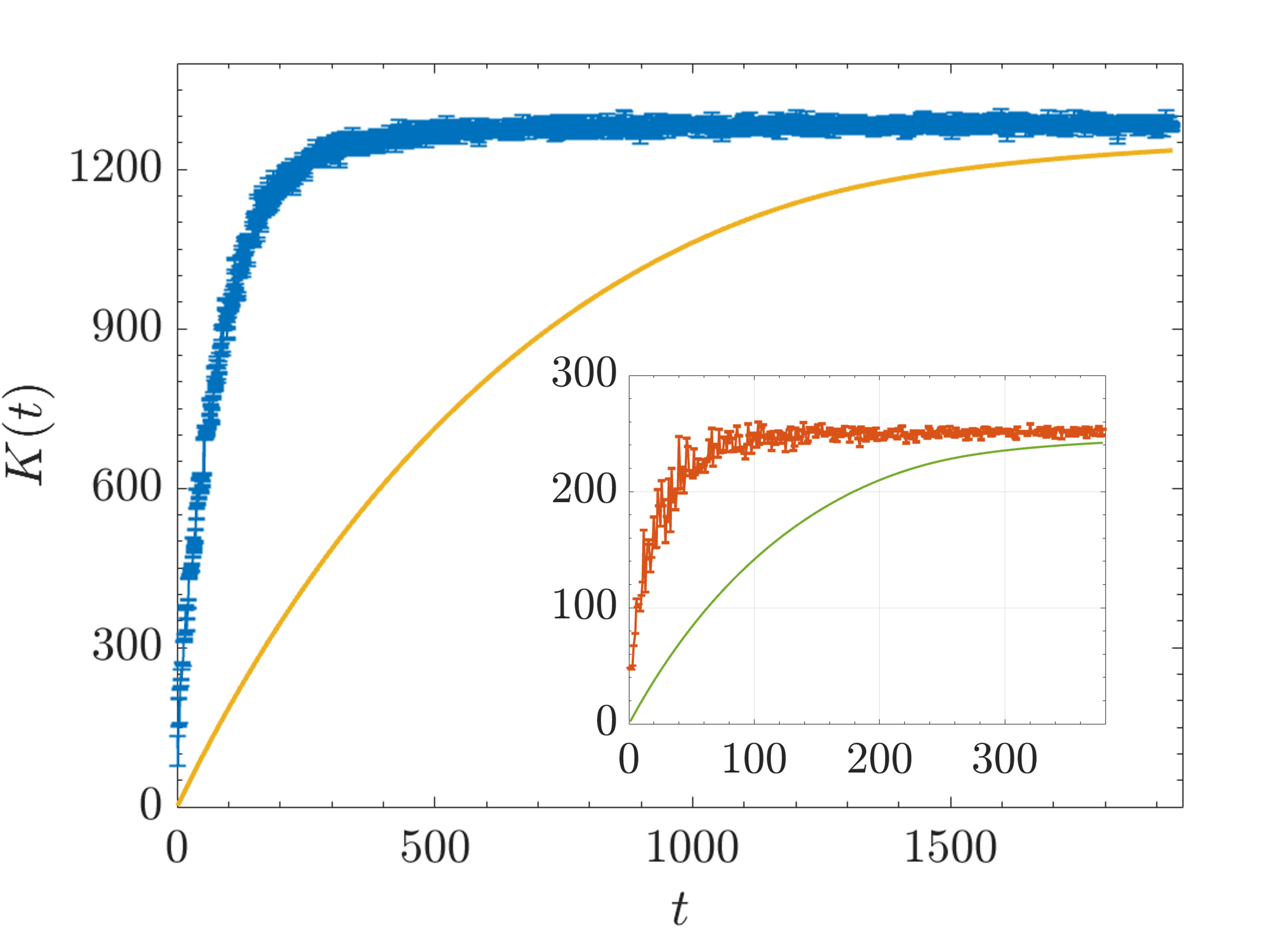}}\hfill
\subfloat[]{\label{SFig11c}\includegraphics[width=0.33\textwidth]{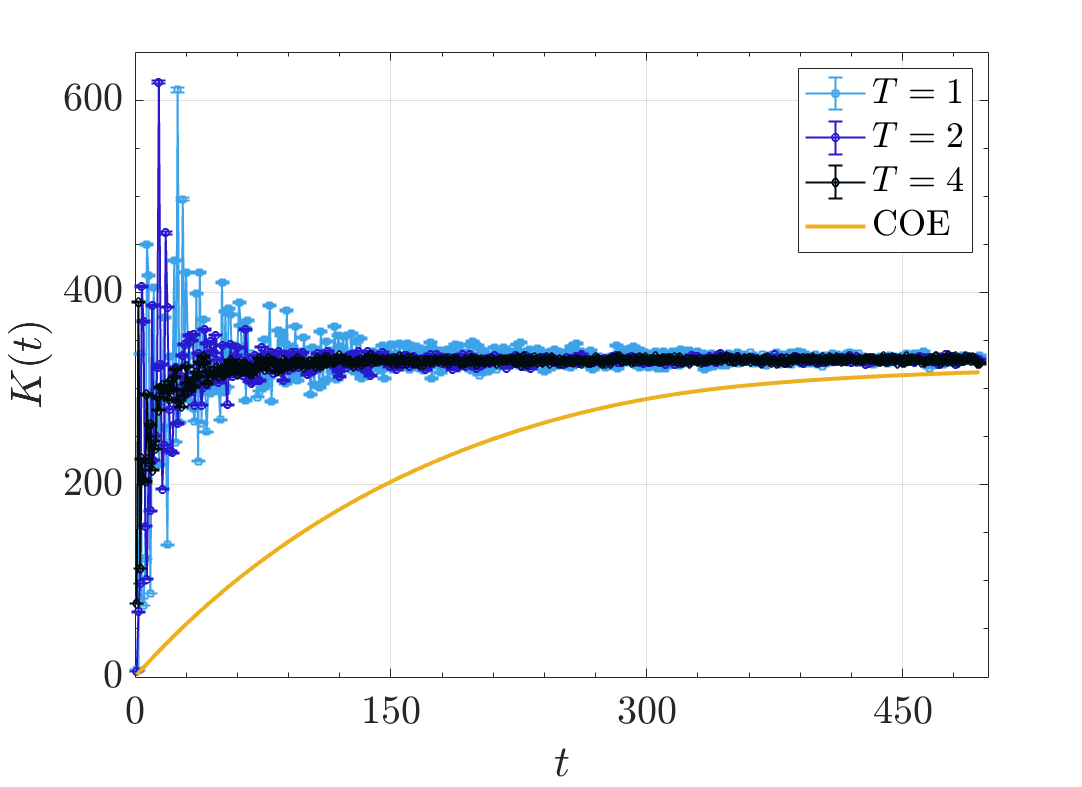}}\hfill
\caption{The SFF for spin-$1/2$ BH model with two-step COE protocol. (a) SFF of different spin-mixing coupling $J_{\uparrow \downarrow}$ with values $J_{\uparrow \downarrow}=0,0.1,0.3,0.5,0.8,1,2,5,10,40$ from darkest to lightest green at $N/L=4/4$ in the second step. Inset: Thouless time with respect to spin-mixing coupling when a Thouless time could be defined $J_{\uparrow \downarrow} \geq 0.3$. (b) The SFF for $J_{\uparrow \downarrow}=0$ in both steps with different $D$, $N/L=8/3$ in the main panel and $N/L=4/3$ in the inset. The lack of RMT behavior is not a finite-size effect. (c) The SFF when two spin sectors are completely decoupled with $u_{\uparrow \downarrow } = J_{\uparrow \downarrow}=0$ in both steps for different driving periods (see legend).}
\label{SFig11}
\end{figure}

\begin{figure}[h]
\subfloat[]{\label{SFig12a}\includegraphics[width=0.33\textwidth]{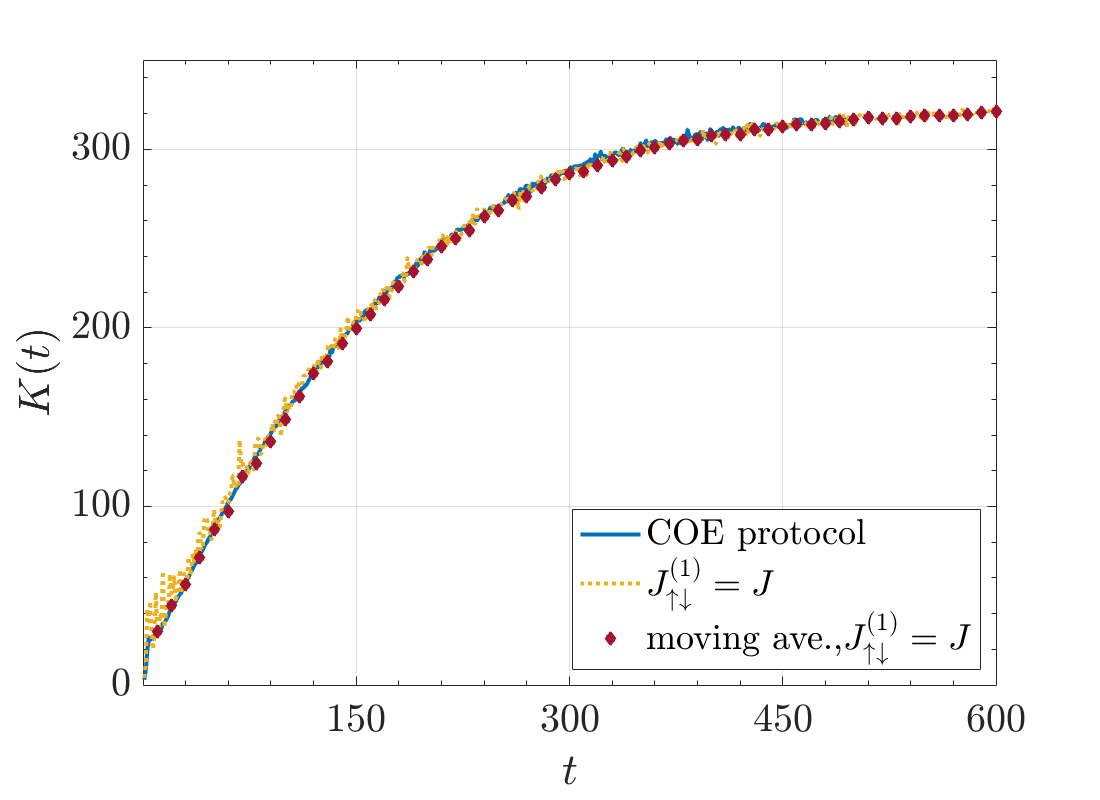}}\hfill
\subfloat[]{\label{SFig12b}\includegraphics[width=0.33\textwidth]{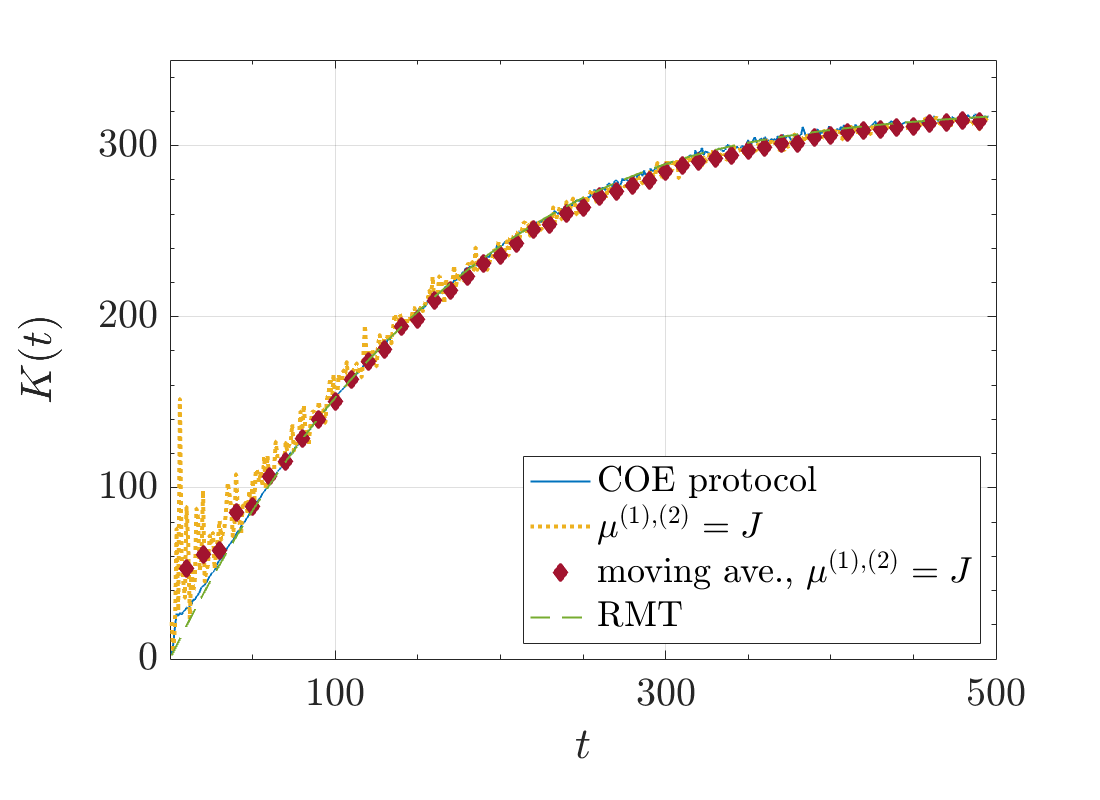}}\hfill
\subfloat[]{\label{SFig12c}\includegraphics[width=0.33\textwidth]{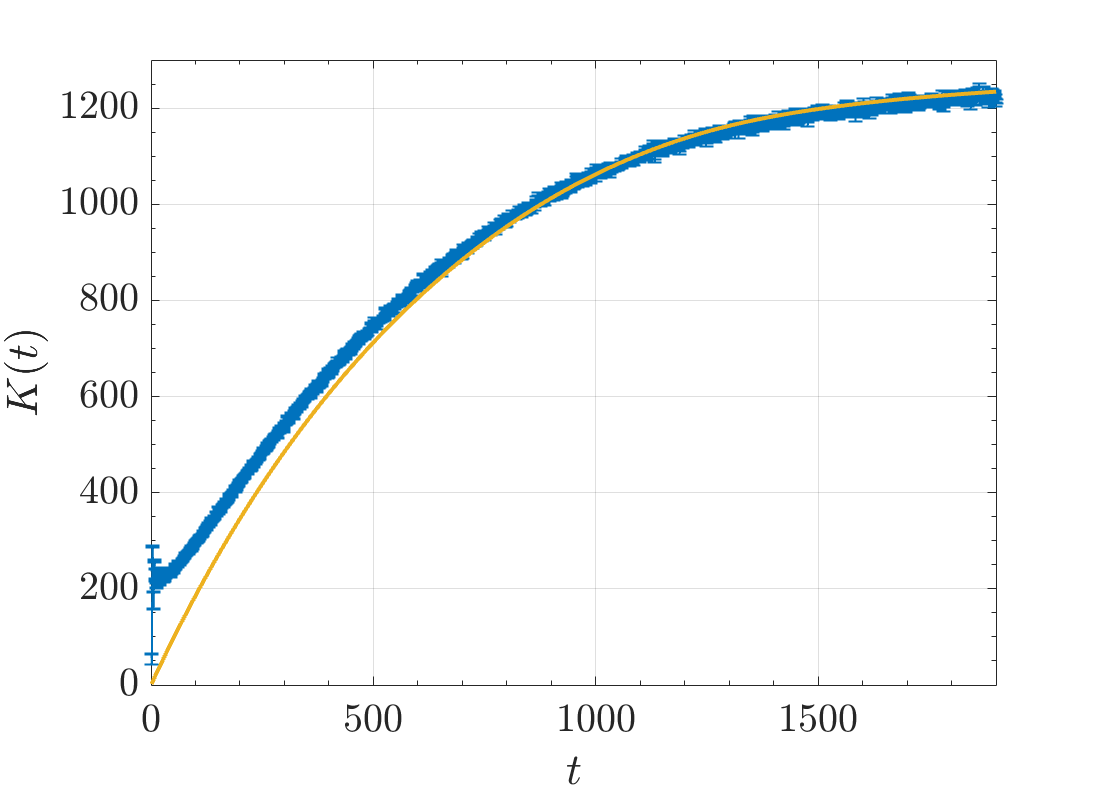}}\hfill
\caption{The SFF for spin-$1/2$ BH model at $N/L=4/4$ with the COE two-step protocol stated in the main text (blue-solid in all panels). (a) The spin-mixing coupling in the second step is set to be constant at $J$ instead of random (yellow-dotted). Randomness due to random disorder potential in one Floquet step is sufficient for the RMT to emerge up to oscillations which can be averaged over time (red-diamonds). (b) The potential is set to be constant at $J$ throughout the time evolution instead of randomly disordered only in the first step (yellow-dotted). Randomness due to random spin-mixing coupling in one Floquet is sufficient for the RMT to emerge up to oscillations which can be averaged over time (red-diamonds). COE analytical curve is given for comparison (dashed-green). Fixing the potential constant throughout the time evolution also decreases the number of simultaneous pulses in Floquet driving to two. (c) The SFF for spin-$1/2$ BH model at $N/L=8/3$ with the two-step protocol when only the randomized terms switch on and off, i.e.,~$J_{\downarrow}=J_{\uparrow}=J$ in both steps. Two simultaneous pulses are sufficient.}
\label{SFig12}
\end{figure}

Fig.~\ref{SFig8} tests how tweaking the two-step protocol presented in main text would affect SFF. In panel (a), we change the inter-species interaction $u_{\uparrow \downarrow}$ sweeping from negative to positive values. Unless we crank it up too large, we see perfect CUE behavior. The corresponding $\tth$ values are given in the inset. Simply setting this interaction strength to $0$ does not affect the chaos. For very large values, e.g.,~$u_{\uparrow \downarrow} = 10$, the resulting SFF suggests an increased sensitivity to finite size effects. This means that, if we increase either $N$ or $L$, possibly we will recover the CUE prediction well. Similar results also hold when we sweep either of the intra-species interactions. Fig.~\ref{SFig8b} depicts particularly the case when $u_{\uparrow}$ changes between $-2$ and $10$. Fig.~\ref{SFig8c} shows the SFF driven according to the two-step protocol and when we sweep the tunneling of spin-$\uparrow$ between $0-20$. We observe in this case that the driven system becomes less chaotic when $J_{\uparrow}$ increases, and eventually becomes integrable when $J_{\uparrow}=20$, i.e.,~the SFF becomes flat at a value equal to $\sim D$. This suggests that having one component in superfluid regime pushes the driven system away from RMT.

Next we aim to understand the impact of spin-mixing tunneling, $J_{\uparrow \downarrow}$. Fig.~\ref{SFig11a} shows SFF for a spin$-1/2$ gas driven according to the two-step protocol but with different spin-mixing tunneling amplitudes ranging between $J_{\uparrow \downarrow}=0-40$. We observe that turning off $J_{\uparrow \downarrow}$ completely pushes away the system from COE statistics. Although the SFF still dips down, it quickly ramps up to the dimension of the Hilbert space in a different trend than what COE dictates. This is likely a consequence of SU$(2)$ symmetry which is present when $J_{\uparrow \downarrow}=0$, even though the symmetry subspaces are chaotic. Cranking up $J_{\uparrow \downarrow}$ does not change the statistics, but merely affects $\tth$ (inset). We focus on the $J_{\uparrow\downarrow}=0$ case in Fig.~\ref{SFig11b} with a larger $D$, and observe that quick ramp-up is not a finite-size effect. We conclude that unprojected SFF for a two-step Floquet unitary in the presence of SU$(2)$ symmetry cannot simulate the COE. We further remove the inter-component interactions $u_{\uparrow \downarrow}$ in Fig.~\ref{SFig11c} to completely decouple two spin components. In this case, the SFF further moves away from the COE prediction and this behavior does not change for different driving periods. Let us note that the three-step protocol could be more helpful to simulate CUE and understand the behavior of SFF in the presence of SU$(2)$ symmetry, because the ramp and plateau are very well separated in CUE statistics.  

Although we randomize each step in Floquet unitary in the two-step protocol, randomness at only one step is sufficient to simulate COE statistics. Fig.~\ref{SFig12a} shows how COE-like statistics could be obtained even when we assume spin-mixing tunneling with constant (nonrandom) amplitude $J_{\uparrow \downarrow}=J$. Alternatively, one could obtain statistically similar systems by only randomizing the amplitude of $J_{\uparrow \downarrow}$ with a constant potential landscape for the 1D optical lattice (Fig.~\ref{SFig12b}). Let us note that, Fig.~\ref{SFig12b} also shows the emergence of COE statistics when the number of simultaneous pulses is two. Similarly, Fig.~\ref{SFig12c} shows COE statistics even when we do not change the tunneling of one spin component, i.e.,~$J_{\downarrow}=J$ throughout. 

\newpage
\section*{\label{AppB}Supplementary Note 4: Further results on the BH models driven according to two- or three-step protocol}

\begin{figure}[h]
\subfloat[]{\label{SFig15a}\includegraphics[width=0.4\textwidth]{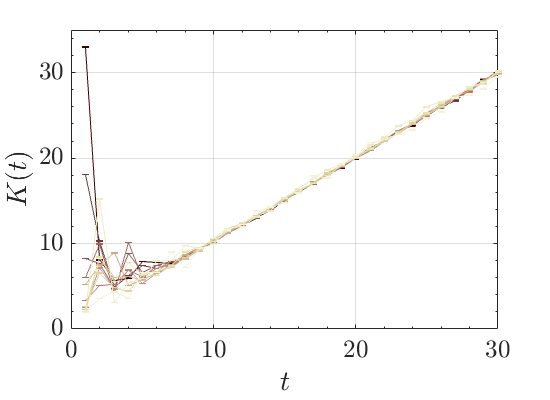}}
\subfloat[]{\label{SFig15b}\includegraphics[width=0.4\textwidth]{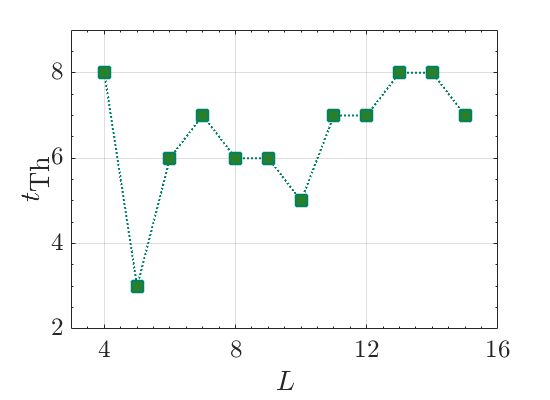}}
\caption{Results for three-step protocol applied to BH model in an optical lattice with $N=4$ number of atoms. (a) SFF for various lattice sizes $L=4-15$ from light yellow to dark red exhibiting CUE statistics and showcasing the bump. (b) The Thouless time in $L$ fluctuates. $\epsilon=0.1$ is used for data.}
\label{SFig15}
\end{figure}

Fig.~\ref{SFig15a} shows the SFF for the spinless BH model with four atoms (instead of five, which is the case discussed in the main text) for different $L$ and driven according to the three-step protocol. The bump is present in early time SFF. Fig.~\ref{SFig15b} presents the extracted $\tth$ for this data set. For $N=5$, we could not identify any scaling behavior in $L$ due to computational limitations that prevents us from reaching larger lattice sizes. Here for $N=4$, we can simulate up to $L=15$. Although the extracted $\tth$ fluctuates, it still suggests a slowly increasing overall trend with $L$. 

\begin{figure}[h]
\subfloat[]{\label{SFig13a}\includegraphics[width=0.4\textwidth]{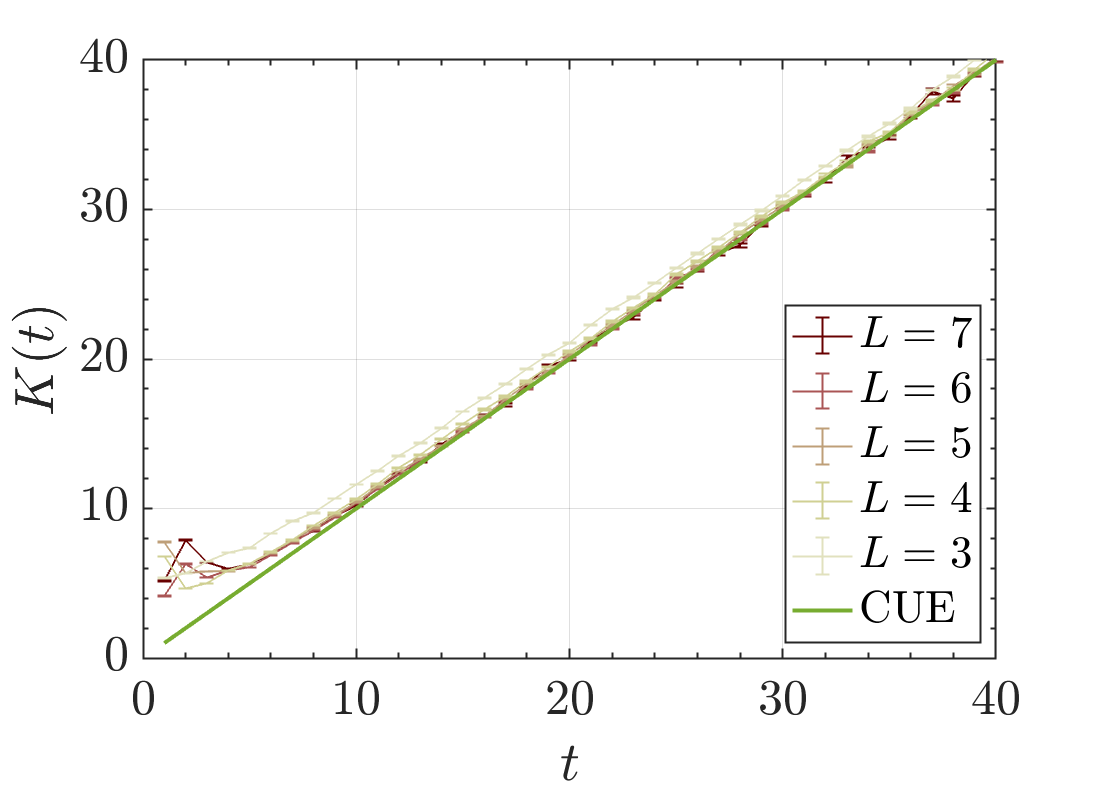}}
\subfloat[]{\label{SFig13b}\includegraphics[width=0.4\textwidth]{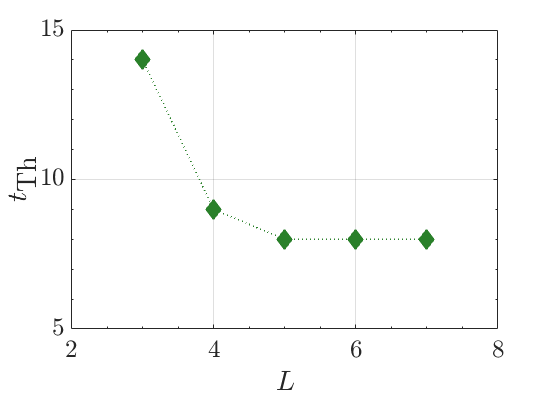}}\hfill
\caption{(a) SFF for three-step CUE protocol with $T=3$ applied to spin-$1/2$ BH model in an optical lattice with $N=4$ atoms and varying sizes $L=3-7$ from light yellow to dark red. (b) Thouless time with respect to lattice size $L$ for $\epsilon=0.1$. We do not detect an increase in $L$.}
\label{SFig13}
\end{figure}

Fig.~\ref{SFig13} shows the SFF and the associated $\tth$ for the spin$-1/2$ BH model with four atoms (instead of three, which is discussed in the main text) and varying chain sizes $L=3-7$. Although we observe a bump feature, we do not find an increasing trend with $L$. This is likely a result of small sizes and very slow scaling of $\tth$ in lattice size.

\begin{figure}[h]
\subfloat[]{\label{SFig4b}\includegraphics[width=0.4\textwidth]{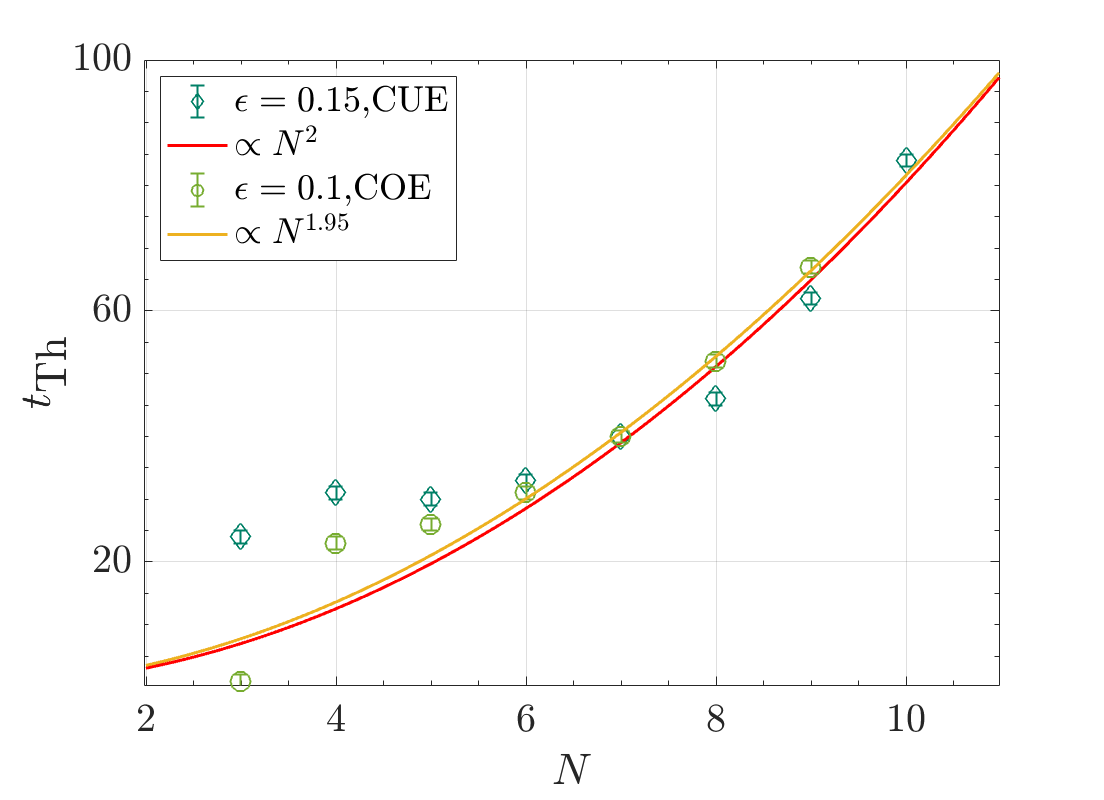}}
\subfloat[]{\label{SFig4c}\includegraphics[width=0.4\textwidth]{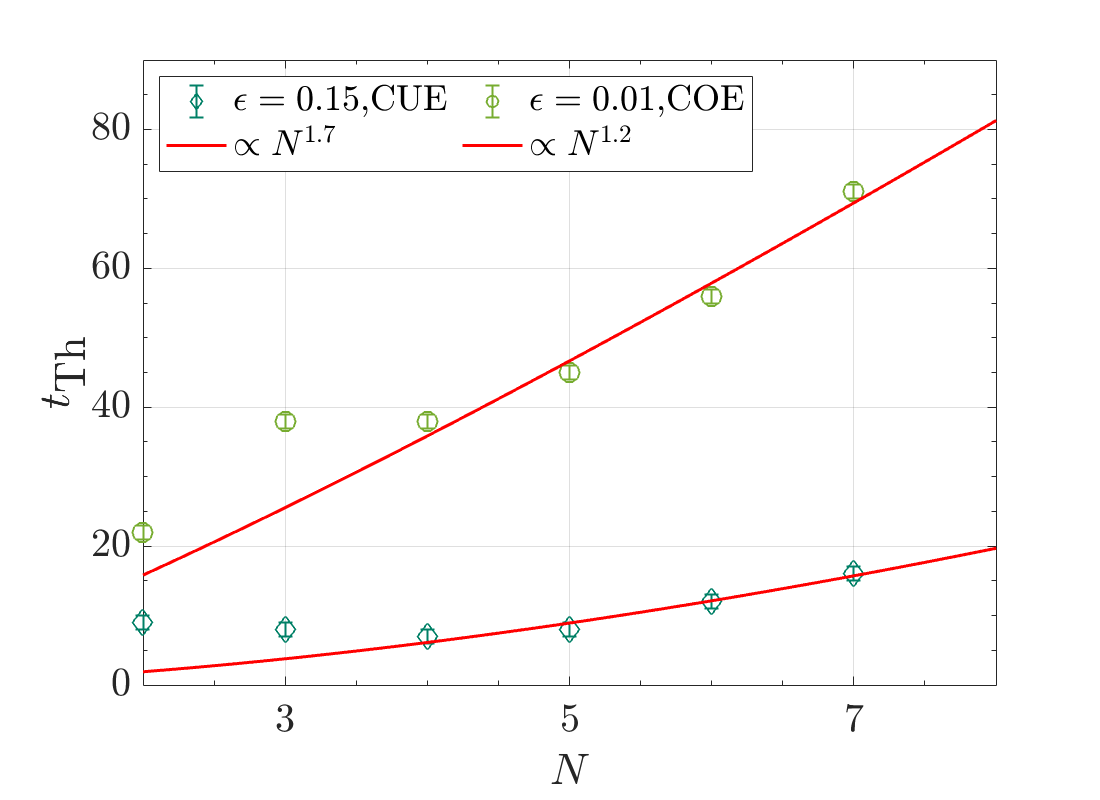}}\hfill
\caption{Thouless time scaling 
(a) with atom number $N$ for a spin-$1/2$ BH model either with COE ($T=1$) or CUE ($T=1.5$) in a triple well, and (b) with atom number $N$ for a spin-$1/2$ BH model either with COE ($T=1$) or CUE ($T=3$) in $L=4$ number of wells. We exclude the first three data points in fits of (a) and the first two data points in (b). $\epsilon$ in the legends shows the threshold of choice to determine the Thouless time.}
\label{SFig4}
\end{figure}

Now we provide further details on the fits to obtain a scaling function for $\tth(N,L)$ in $N$ and $L$. All panels in  Fig.~\ref{SFig4} compare CUE and COE scalings. Fig.~\ref{SFig4b} shows $\tth$ scaling in $N$ when the lattice size is fixed to $L=3$. $\tth(N,L=3)$ scaling for the three-step protocol fits a
power-law in $N$ with an exponent $2.0282\pm 0.8385$. Similarly for the two-step protocol, we obtain a power-law exponent of $1.9543\pm 0.3505$. We find very close power-law exponents in $N$ for both two- and three-step protocols. For these data sets, linear and logarithmic trends can be discarded. 


In Fig.~\ref{SFig4c}, we again compare $\tth$ scaling for two- and three-step protocols in $N$ both for $L=4$. We simulate up to $N=7$ which gives us scaling exponents smaller than what we have found for $L=3$. We attribute this to finite size effects. One could notice the increase in the scaling exponent in Fig.~\ref{SFig4b} for $L=3$ as one passes further beyond $N=7$. This also suggests that our finite-size analysis is indeed restricted and cannot provide conclusive scaling exponents.

\begin{figure}
\subfloat{\label{SFig3b}\includegraphics[width=0.4\textwidth]{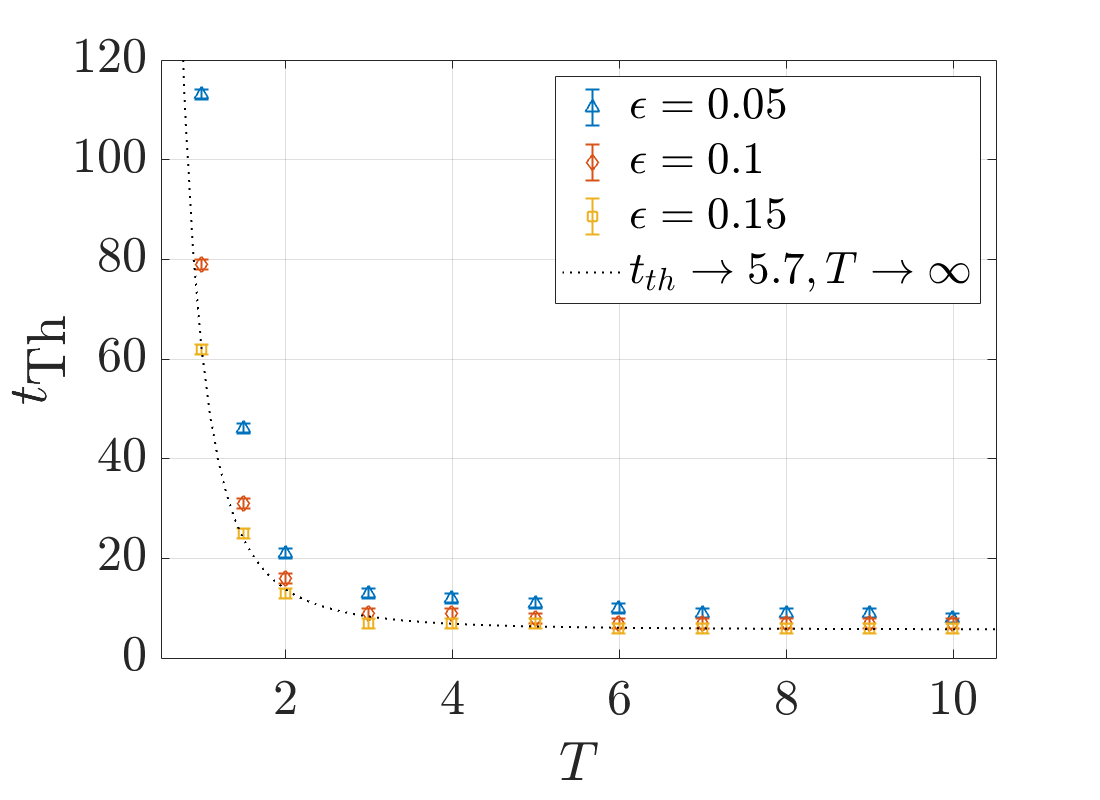}}\hfill
\caption{$\tth$ values decrease as a power-law in driving period and approach to $\tth \sim 5.7$ in $T\rightarrow \infty$ limit. Parameter set is $N=L=4$ with two-step driving protocol. $\epsilon$ in the legend shows the threshold of choice to determine the Thouless time.}
\end{figure}

Fig.~\ref{SFig3b} shows how $\tth$ changes with increasing driving period when we set $N/L=4/4$ with two-step driving protocol. We observe a power-law decrease to $\tth \sim 5.7$ for $T \gg 1$. This suggests that the RMT behavior lingers although we significantly increase the driving period and this is consistent with the observations in the literature \cite{PhysRevX.4.041048}. Interestingly, although $\tth$ decreases, it remains nonzero. On the other side, $\tth$ increases as the driving frequency becomes very large. An interesting question to answer in the future is whether this observation is in any way related to the vanishing of heating effect with fast Floquet driving. 

\newpage
\section*{\label{AppC}Supplementary Note 5: SFF of spinor condensate $H_{\text{SC}}$ with no additional parameters}

\begin{figure}[h]
\subfloat[]{\label{SFig14a}\includegraphics[width=0.45\textwidth]{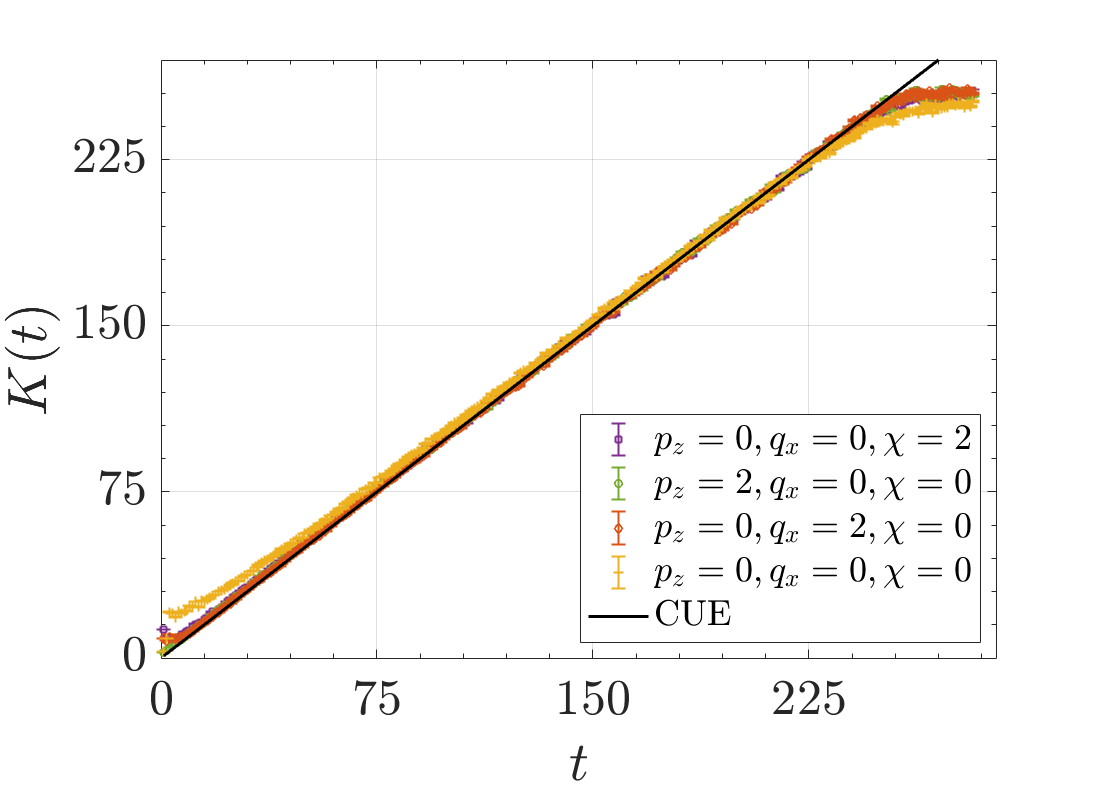}}
\subfloat[]{\label{SFig14b}\includegraphics[width=0.45\textwidth]{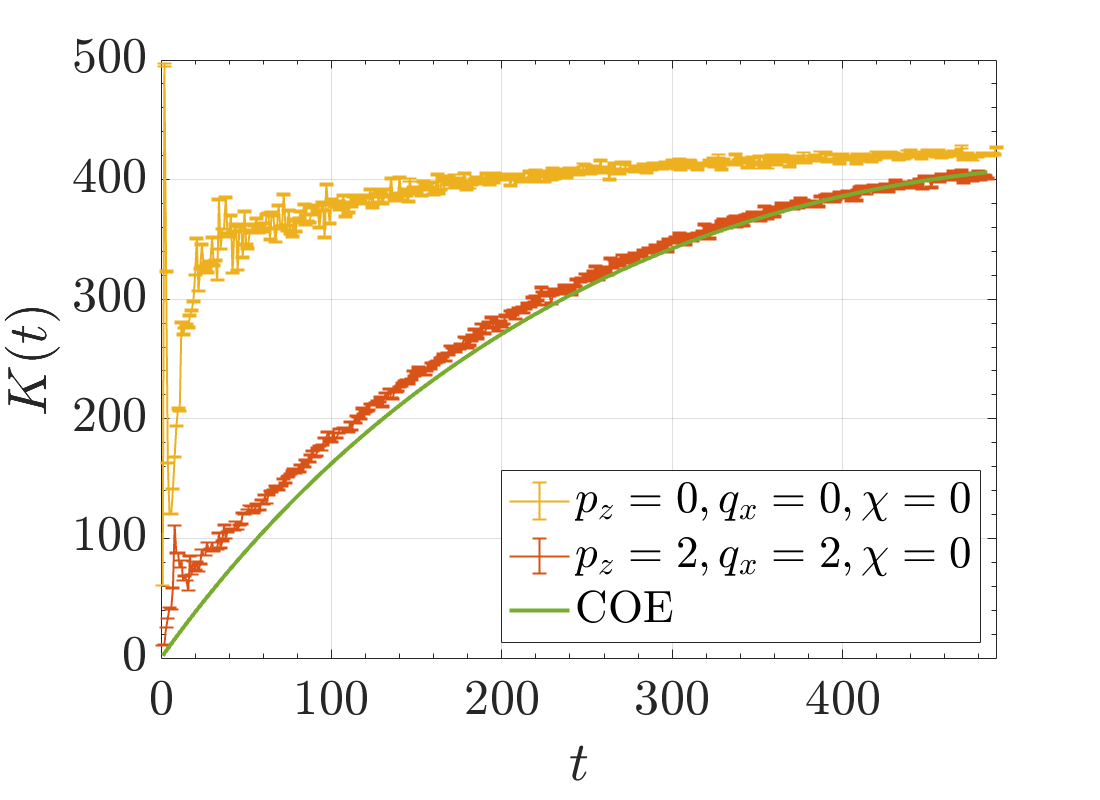}}\hfill
\caption{SFF of chaotic spinor condensates driven according to (a) three-step and with an atom number of $N=30$, (b) two-step protocol and with an atom number of $N=40$. (a) When all additional parameters are zero, and we only break the SO$(2)$ symmetry of the gas as in $H_{\text{SC}}$, the strength of the level repulsion is not maximal, i.e.,~the kink between the ramp and plateau is not as sharp as the other cases (see legend). (b) $H_{\text{SC}}$ without additional parameters cannot simulate COE statistics, in contrast to a case where we switch on $p_z$ and $q_x$.}
\label{SFig14}
\end{figure}

We test whether $H_{\text{SC}}$ with no additional parameters, e.g.,~$p_z=q_x=\chi=0$ could simulate COE and CUE. Fig.~\ref{SFig14b} shows that it cannot simulate COE, which is compared to a case where we switch on $p_z=q_x=2$. Fig.~\ref{SFig14a} shows $H_{\text{SC}}$ can simulate CUE up to a degree where the level repulsion and spectral rigidity are not as strong as the other cases where either of $p_z$, $q_x$ or $\chi$ is turned on. These results are obtained after a projection to even-parity symmetry subspace. 

\newpage
\section*{\label{AppE}Supplementary Note 6: SFF of spin-$1/2$ BH model in its Mott insulating phase $u\gg J$}

\begin{figure}[h]
\subfloat[]{\label{SFig9a}\includegraphics[width=0.4\textwidth]{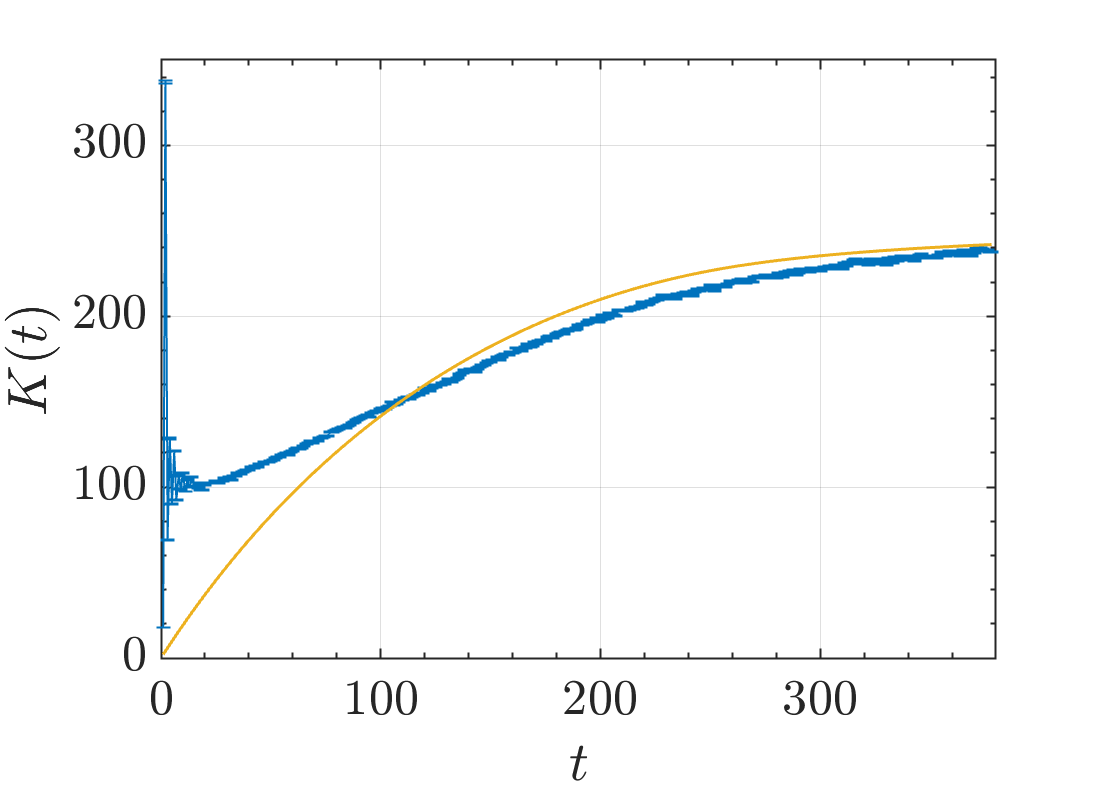}}
\subfloat[]{\label{SFig9b}\includegraphics[width=0.4\textwidth]{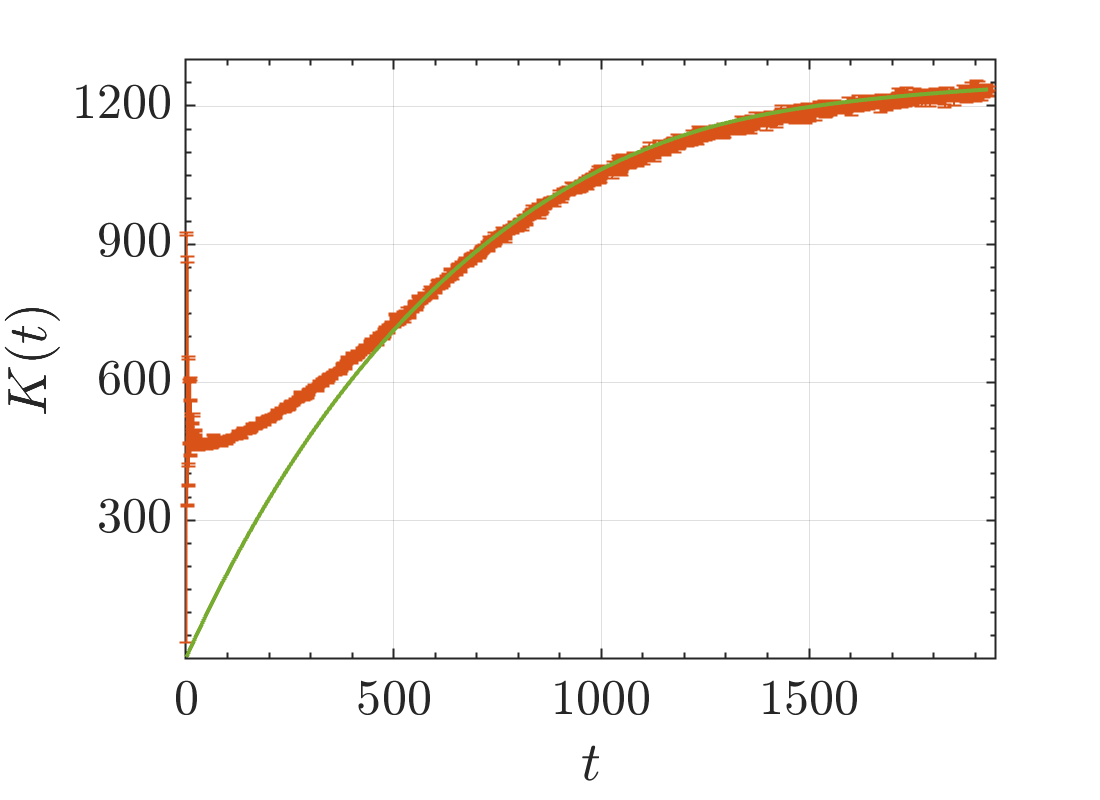}}\\
\subfloat[]{\label{SFig9c}\includegraphics[width=0.4\textwidth]{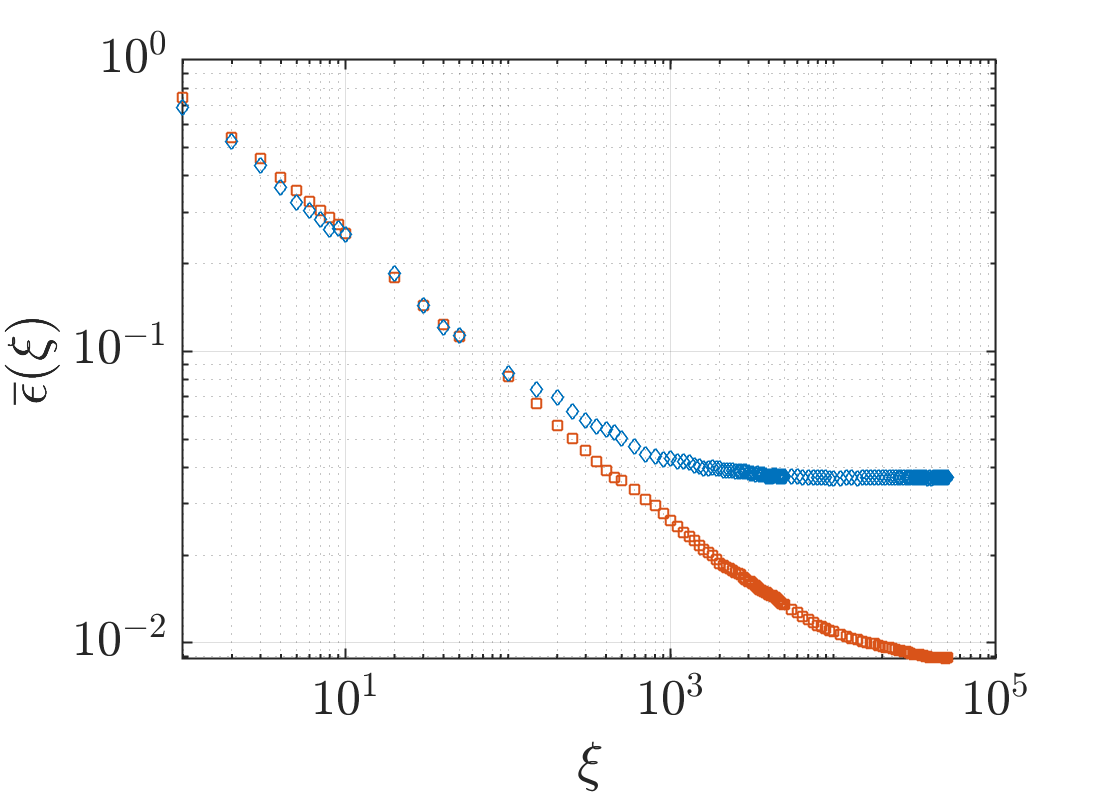}}
\subfloat[]{\label{SFig9d}\includegraphics[width=0.4\textwidth]{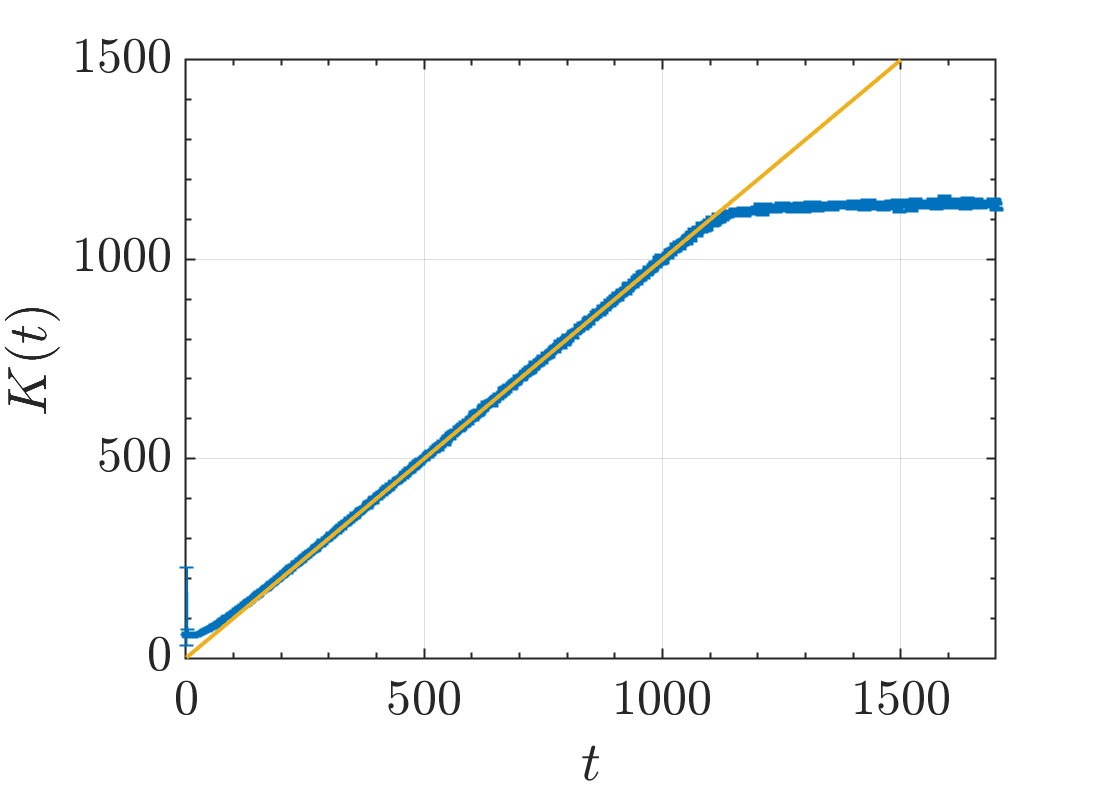}}
\caption{(a-c) SFF of spin-$1/2$ BH model driven according to three-step protocol in a triple well $L=3$ in its Mott insulating regime where $u_{\uparrow}=u_{\downarrow}=u_{\uparrow \downarrow}=u=10J$ with (a) $N=5$ atoms and (b) $N=8$ atoms. As the number of atoms increase, SFF of the BH model approaches to the RMT. (c) The time-averaged error between the Thouless time and the simulation time for $N=5$ (blue-diamonds) and $N=8$ (red-squares) atoms. (d) SFF of spin-$1/2$ BH model driven according to three-step protocol with three atoms $N=3$ in a 1D lattice with nine sites $L=9$ in its Mott insulating regime where $u_{\uparrow}=u_{\downarrow}=u_{\uparrow \downarrow}=u=10J$. In all panels, the driving period is set to $T=3$.}
\label{SFig9}
\end{figure}

After sweeping through the parameter space of spin$-1/2$ BH model, we can state that so long as any of the spin components is not in superfluid regime, the system will likely exhibit some degree of RMT. In Fig.~\ref{SFig9} we focus on the case when the system is deep in the Mott insulator regime after driven according to the three-step protocol. We find two distinct behaviors depending on whether the filling factor is $\nu>1$ [panels (a-c)] and $\nu < 1$ [panel (d)]. In panel (d) where we have three atoms in nine sites, we recover the CUE ``ramp-plateau'' feature with a clear bump behavior at $\tth \sim 100$ which is much larger than $\tth$ for $u=1$. Additionally, we notice a larger bump. The same behavior holds true for $\nu > 1$ in panels (a) and (b). Interestingly however, the system statistics diverge from CUE-like to COE-like. Although we do not find an ideal match between the simulated dynamics and the COE in panel (a), we increase the atom number in panel (b), and we find a better match with COE. Panel (c) shows that the difference between the simulated result and the analytical COE behavior, $\bar{\epsilon}(\xi)$ (Eq.~\eqref{timeAveragedError}) indeed decreases with increasing atom number. In this case too, the predicted $\tth$ is much larger than what we found for $u=1$, and the bump is much larger too. 

\newpage
\section*{Supplementary Note 7: Time-evolution of the spin populations and heating times}

\begin{figure}[h]
\subfloat[]{\label{SFig16a}\includegraphics[width=0.45\textwidth]{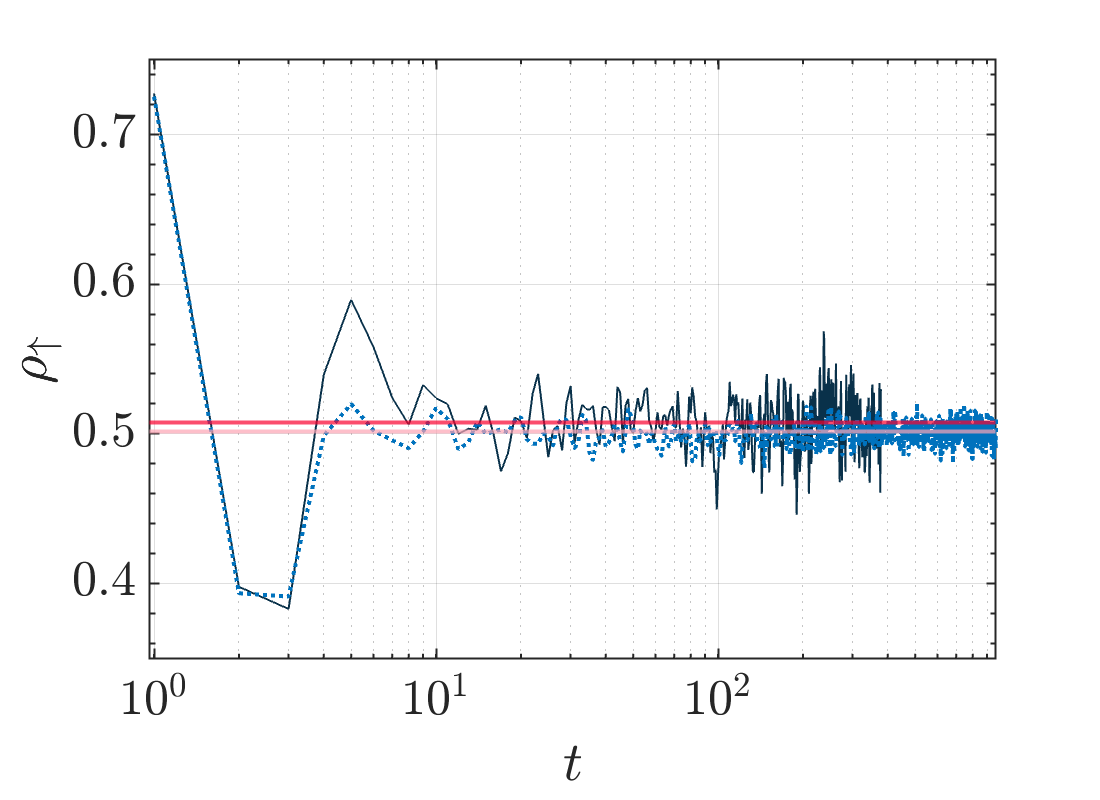}}
\subfloat[]{\label{SFig16b}\includegraphics[width=0.45\textwidth]{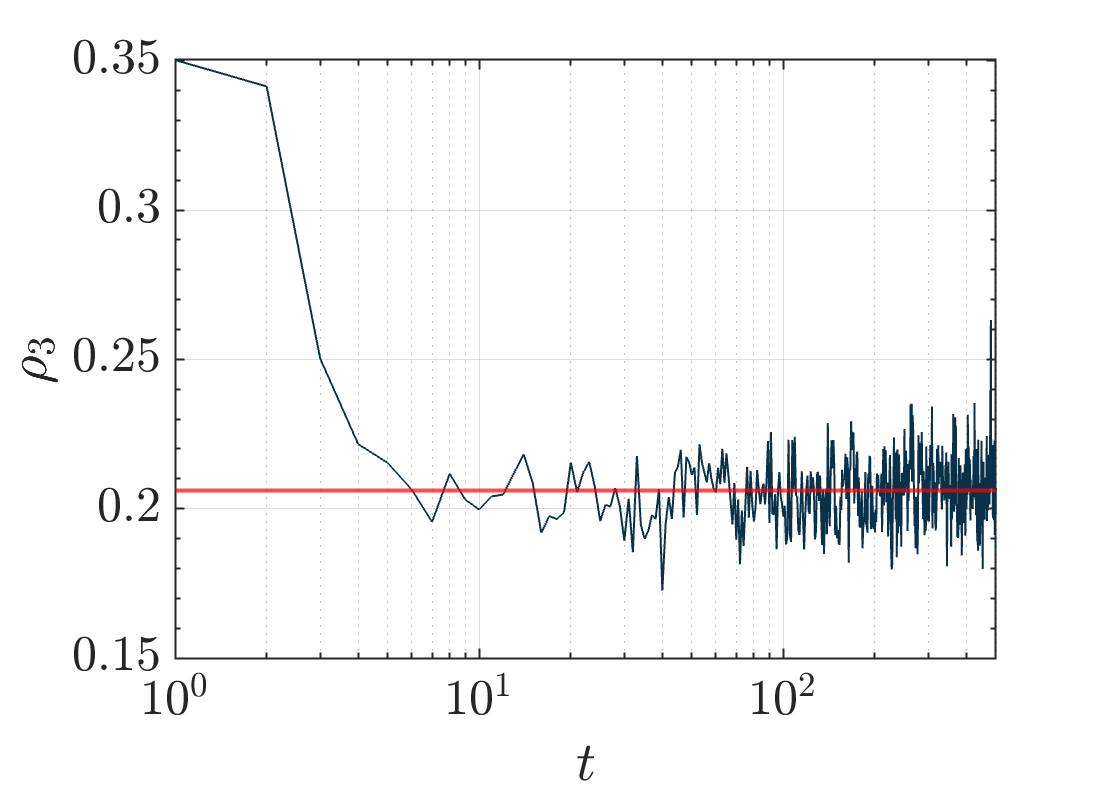}} \\
\subfloat[]{\label{SFig16c}\includegraphics[width=0.33\textwidth]{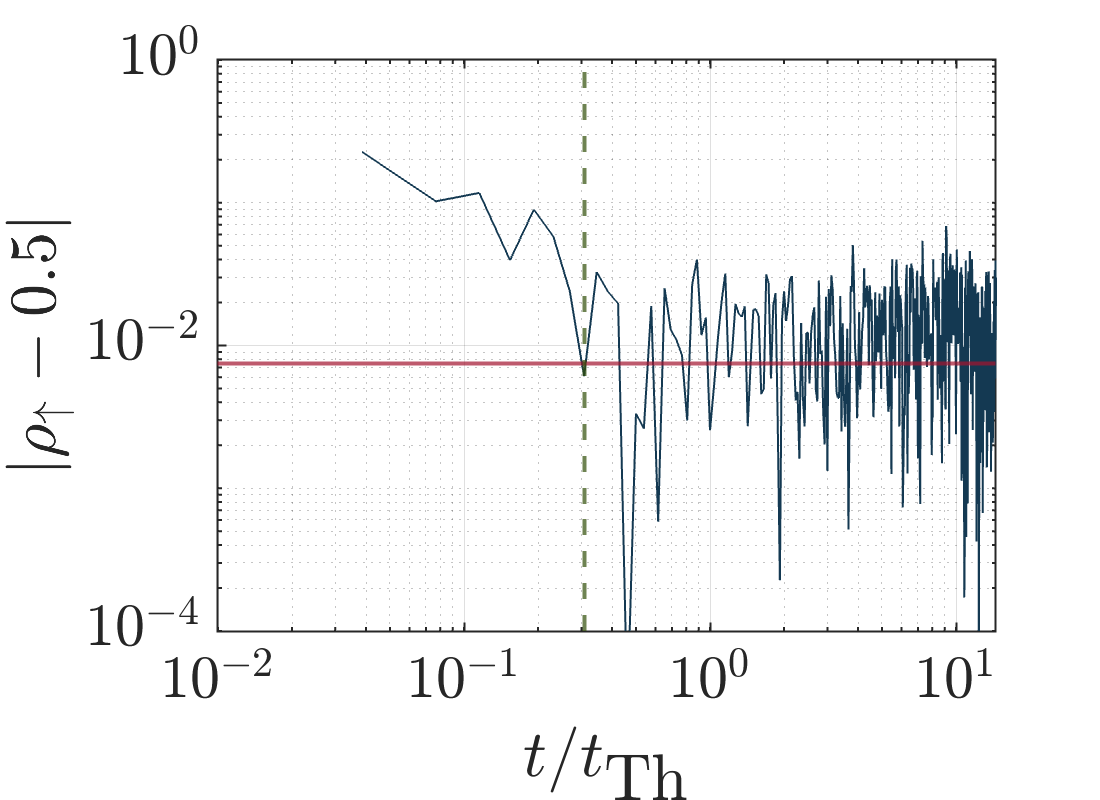}}
\subfloat[]{\label{SFig16d}\includegraphics[width=0.33\textwidth]{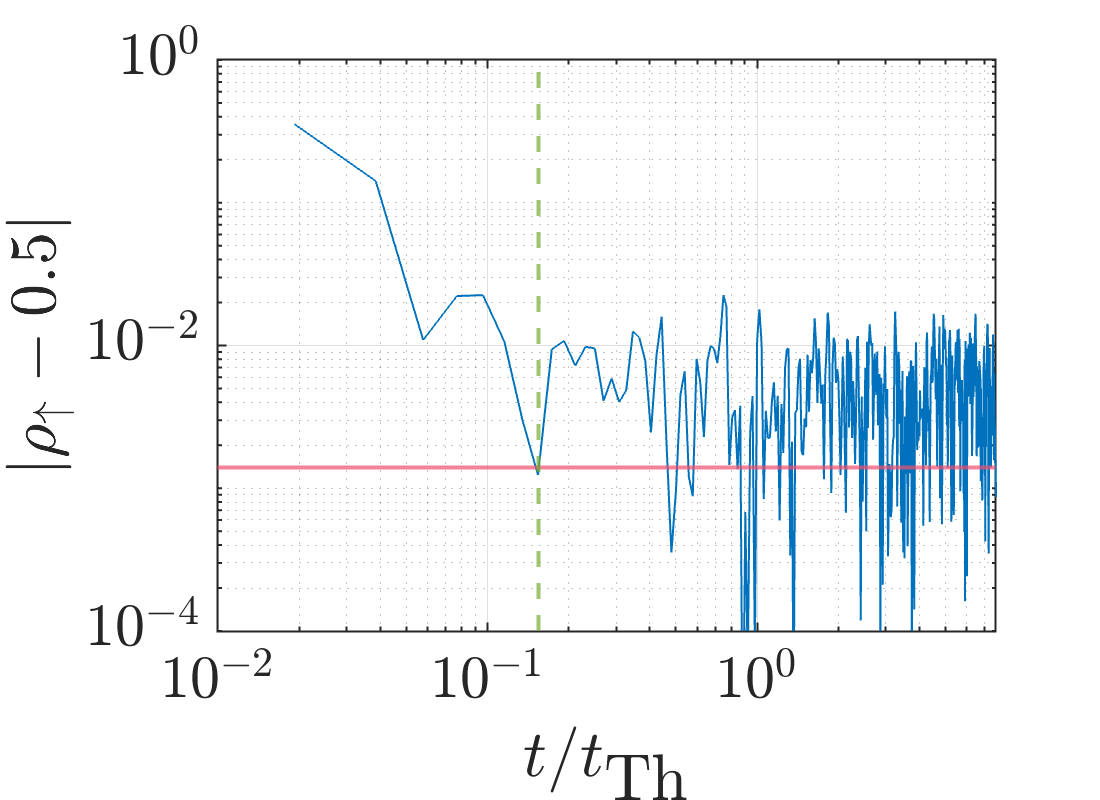}}
\subfloat[]{\label{SFig16e}\includegraphics[width=0.33\textwidth]{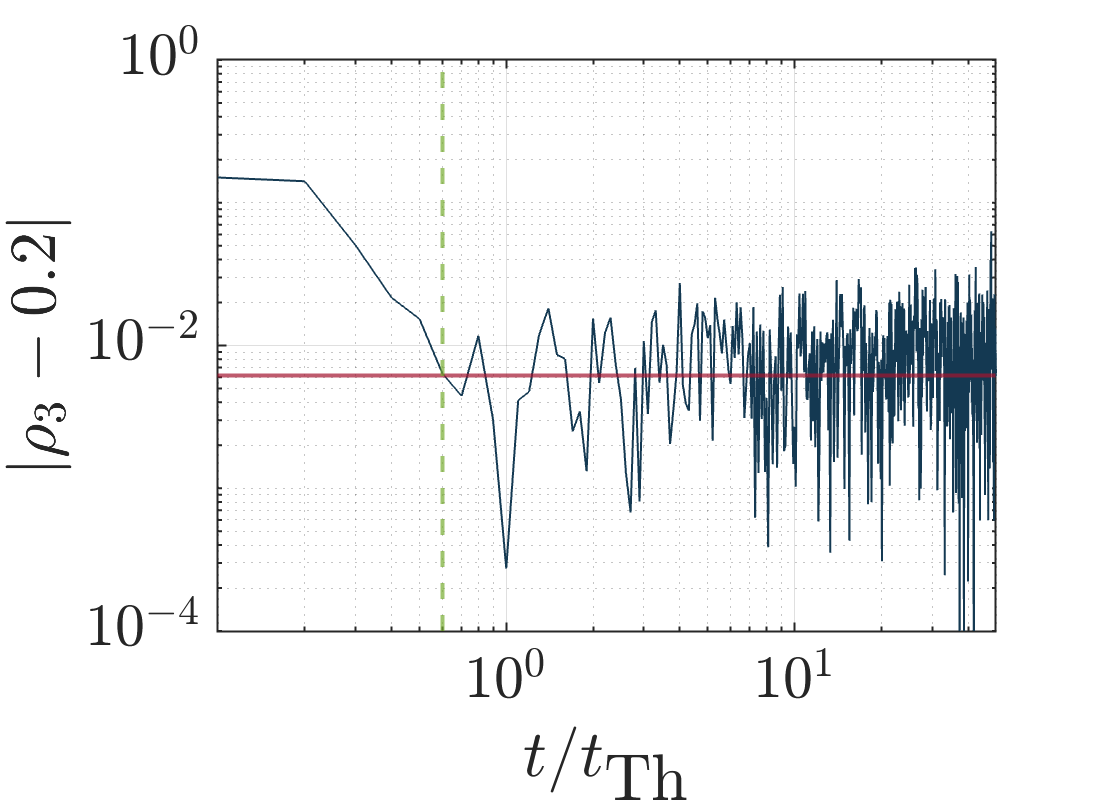}}
\caption{(a) The density of atoms with respect to the normalized time $t=t'/T$ with spin-$\uparrow$, $\rho_{\uparrow}$, in a spin$-1/2$ BH model with $L=3$, $N=5$ (solid-dark line) and $N=8$ (light-dotted line) after a two-step Floquet protocol initialized in a spin-$\uparrow$ polarized state. The horizontal lines are the long time average of $\rho_{\uparrow}$ respectively for $N=5$ (red) and $N=8$ (pink). (b) The density of atoms in the middle well with respect to time  in a spinless BH model with $L=5$ and $N=8$ initialized in a product state with atom numbers $\Ket{0,0,N,0,0}$ and subjected to a three-step Floquet protocol. The horizontal line is the long time average of $\rho_{3}$. The driving periods are (a) $T=1$ and (b) $T=3$. The spin-$1/2$ BH model with (c) $N=5$ and (d) $N=8$, and the spinless BH model with (e) $N=8$ in logarithmic density scale to show when the populations approach the long time average (solid-horizontal lines). The time axes are rescaled with the corresponding Thouless times to emphasize the timescale difference between $\tth$ and the heating times (marked with dashed vertical lines).}
\label{SFig16}
\end{figure}

The local observables in Floquet-driven systems could become featureless after a \emph{heating time} which strongly depends on the driving period \cite{PhysRevE.90.012110,PhysRevX.4.041048}. Hence, an indirect evidence for  identifying the RMT behavior can be obtained by examining the time evolution of local observables. Fig.~\ref{SFig16a} presents the dynamics of the atom density with spin-$\uparrow$ in the entire lattice, $\rho_{\uparrow}$, for a spin$-1/2$ BH model in a triple well $L=3$ with $N=5$ and $N=8$ initialized in a spin-$\uparrow$ polarized state and following a two-step Floquet protocol. As it can be seen a rapid decay towards $\rho_{\uparrow} \sim 0.5$ occurs signalling that the system approaches a featureless state at long evolution times. The horizontal lines are the long time average of $\rho_{\uparrow}$, namely  $\Braket{\rho_{\uparrow}(t)}_{t\rightarrow \infty}$, respectively for $N=5$ (red line) and $N=8$ (pink line). We observe that as the atom number increases, $\Braket{\rho_{\uparrow}(t)}_{t\rightarrow \infty}$ approaches $0.5$, as expected since the individual lattice sites tend to acquire a uniform distribution over the spin states. Fig.~\ref{SFig16b} shows the atom density in the middle well for a spinless BH model after a three-step Floquet protocol initialized in a product state with atom numbers $\Ket{0,0,N,0,0}$. The featureless long-time state for $L=5$ is expected to give for all sites $i$, $\rho_i=1/L=0.2$ in the $N\rightarrow \infty$ limit, and we indeed observe $\Braket{\rho_3(t)}_{t\rightarrow \infty} \sim 0.2$ for $N=8$. Slight deviations from this limit are attributed to finite size effects.

For the spin-$1/2$ BH model, we define an empirical heating time as 
\begin{eqnarray}
\tau_{\text{heating}} = \text{min}_t \left( |\rho_{\uparrow}(t)-0.5| \leq \Braket{\rho_{\uparrow}(t)-0.5}_{t\rightarrow \infty} \right).
\end{eqnarray}
We identify $\tau_{\text{heating}}=8$ in both cases in Fig.~\ref{SFig16a} which are much smaller than $\tth=26$ and $\tth=52$, respectively for $N=5$ and $N=8$ with an error threshold of $\epsilon=0.1$ in SFF (see Eq.~\eqref{errorFunc} for the definition of $\epsilon$). To explicitly show the difference between timescales of $\tth$ and $\tau_{\text{heating}}$, we plot the same data in logarithmic y-axis in Figs.~\ref{SFig16c}-\ref{SFig16d} together with rescaled time $t/\tth$. The heating times are marked with dashed vertical lines and can be seen to differ from $t/\tth=1$. Let us also note that we find $t_b=5$, as the time that the bump region collapse starts, for both data resulting in the interesting observation of $t_b \sim \tau_{\text{heating}} < \tth$. 

For the spinless BH model, we apply the same analysis with the following definition for the heating time,
\begin{eqnarray}
\tau_{\text{heating}} = \text{min}_t \left( |\rho_{3}(t)-0.2| \leq \Braket{\rho_{3}(t)-0.2}_{t\rightarrow \infty} \right).
\end{eqnarray}
This gives $\tau_{\text{heating}}=6 < \tth = 10$ for $\epsilon=0.1$, where $\tau_{\text{heating}}$ is marked with dashed vertical line in Fig.~\ref{SFig16e}. Let us however note that we find $t_b = 1$ for this data set. 

These observations may demand an in-depth investigation on the timescale hierarchy of quantum chaos, and the possible connections between the heating times and the SFF timescales. Furthermore, the observation that the local observables are already featureless at $\tth$ could be an indirect physical evidence that the universality of quantum chaos in many-body systems indeed extends beyond the RMT ramp into the bump regime. 

\section*{Supplementary Note 8: Survival probability and spectral form factor}

In this section, we discuss the relationship between survival probability and spectral form factor, and investigate the survival probability of randomly sampled initial states.
The survival probability for the system of interest with initial state $\ket{\psi}$ and time evolution operator $\hat{U}$ is given by Eq.~(10) in the main text. When the state $\ket{\psi}$ is chosen to be the equal-amplitude superposition of all eigenstates of $\hat{U}^t$, then 
\begin{equation}
P\left( \left\{\ket{\psi} =  \sum_{n} \ket{\theta_n}, \hat{U} \right\},t \right)=
K_{\hat{U}}(t).
\end{equation}
Next, we consider the survival probability of two randomly sampled initial states, namely, 
\begin{enumerate}
    \item \textit{Globally randomly rotated states}, i.e. \begin{equation}
      \ket{\psi} = \hat{W}\ket{\Psi} , \quad \; \; \hat{V} \in \mathrm{CUE}(D)= \mathrm{CUE}(q^L) \label{eq:globalRotated} 
    \end{equation} 
    where $\hat{W}$ is a random unitary drawn from the $D$-by-$D$ CUE, $D$ and $q$ are the Hilbert space dimension and the on-site Hilbert space dimension respectively. $\ket{\Psi}$ is any pure state including product states.
    
    \item \textit{Locally randomly rotated states}, i.e. 
    \begin{equation}
     \ket{\psi}=   \left( \bigotimes^{L}_i \hat{v}_i \right)  \ket{\Psi}, \quad  \;\;  \hat{v}_i \in \mathrm{CUE}(q).\label{eq:localRotated} 
    \end{equation}
\end{enumerate}

Using the two-design formula for $m$-by-$m$ random unitaries $\hat{W}$ from CUE, 
\begin{equation}
\begin{aligned}
\mathbb{E}[ \hat{W}_{a_1 a'_1}\hat{W}_{a_2 a'_2} 
\hat{W}^{*}_{b_1 b'_1}
\hat{W}^{*}_{b_2 b'_2}
]_{\hat{W}} & = \; \frac{1}{m^2 -1}  \left(
\delta_{a_1 b_1} \delta_{a'_1 b'_1}
\delta_{a_2 b_2} \delta_{a'_2 b'_2}
+
\delta_{a_1 b_2} \delta_{a'_1 b'_2}
\delta_{a_2 b_1} \delta_{a'_2 b'_1}
\right)
\\
&- 
\frac{1}{m(m^2-1)}
\left(
\delta_{a_1 b_2} \delta_{a'_1 b'_1}
\delta_{a_2 b_1} \delta_{a'_2 b'_2}
+
\delta_{a_1 b_1} \delta_{a'_1 b'_2}
\delta_{a_2 b_2} \delta_{a'_2 b'_1}
\right),
\end{aligned}
\end{equation}
one can compute Eq.~(13) in the main text for both cases in Eqs.~\eqref{eq:globalRotated} and~\eqref{eq:localRotated}.

\bibliographystyle{apsrev4-1}